%% file: SL-WPmain.tex
\documentclass[pdflatex]{sn-jnl}

\usepackage{graphicx}%
\usepackage{verbatim}
\usepackage{multirow}%
\usepackage{amsmath,amssymb,amsfonts}%
\usepackage{amsthm}%
\usepackage[title]{appendix}%
\usepackage{xcolor}%
\usepackage{textcomp}%
\usepackage{manyfoot}%
\usepackage{booktabs}%
\usepackage{algorithm}%
\usepackage{algorithmicx}%
\usepackage{algpseudocode}%
\usepackage{listings}%
\usepackage{slashed}
\usepackage[utf8]{inputenc}
\usepackage{textgreek}
\usepackage[normalem]{ulem}
\definecolor{DarkGreen}{rgb}{0,0.5,0}

\RequirePackage[mcite,style=numeric-comp,eprint=true,giveninits=true,maxbibnames=4,backend=biber,defernumbers=true,sorting=none,labelnumber=true]{biblatex}
\bibliography{sn-bibliography}
\DeclareBibliographyCategory{reviews}
\addtocategory{reviews}{FlavourLatticeAveragingGroupFLAG:2024oxs,ParticleDataGroup:2024cfk}

\DeclareSourcemap{
  \maps[datatype=bibtex]{
    \map{
      \step[fieldsource=entrykey, match=FlavourLatticeAveragingGroupFLAG:2024oxs, final]
      \step[fieldset=shorthand, fieldvalue={FLAG}]
    }
    \map{
      \step[fieldsource=entrykey, match=ParticleDataGroup:2024cfk, final]
      \step[fieldset=shorthand, fieldvalue={PDG}]
    }
  }
}

\DeclareUnicodeCharacter{202F}{,}

\newcommand{\smallcap}[1]{{\mathchoice
  {\scalebox{1.1}{$\scriptstyle \mathsf{#1}$}}
  {\scalebox{1.1}{$\scriptstyle \mathsf{#1}$}}
  {\scalebox{1.1}{$\scriptscriptstyle \mathsf{#1}$}}
  {\scalebox{1.1}{$\scriptscriptstyle \mathsf{#1}$}}
}}

\usepackage{tocloft}

\begin{document}

\title[Article Title]{Electroweak Transitions of Hyperons and Charmed Baryons: \\ Opportunities and Prospects}

\author[1,2]{\fnm{Patrik} \sur{Adlarson}}

\author[3,4]{\fnm{Constantia} \sur{Alexandrou}}

\author[4]{\fnm{Simone} \sur{Bacchio}}

\author[2,5]{\fnm{Varvara} \sur{Batozskaya}}

\author[6]{\fnm{Johan} \sur{Bijnens}}

\author[7]{\fnm{Alexandre} \sur{Brea Rodríguez}}

\author[8]{\fnm{Gernot} \sur{Eichmann}}

\author[3,4]{\fnm{Andreas} \sur{Konstantinou}}

\author[1,2]{\fnm{Andrzej} \sur{Kupsc}}

\author[1]{\fnm{Stefan} \sur{Leupold}}

\author[9]{\fnm{Piotr} \sur{Salabura}}

\author[10]{\fnm{Christoph} \sur{Schwanda}}

\author[9]{\fnm{Marcin} \sur{Zielinski}}

\affil[1]{\orgdiv{Department of
  Physics and Astronomy}, \orgname{Uppsala University}, \orgaddress{\street{Box 516}, \postcode{SE-75120}, \city{Uppsala},  \country{Sweden}}}

\affil[2]{\orgname{National Centre for Nuclear Research}, \orgaddress{\street{Pasteura 7}, \city{Warsaw}, \postcode{02-093}, \country{Poland}}}

\affil[3]{\orgdiv{Department of Physics}, \orgname{University of Cyprus}, \orgaddress{P.O. Box 20537, 1678 Nicosia, Cyprus}}

\affil[4]{\orgdiv{Computation-based Science and Technology Research Center}, \orgname{The Cyprus Institute}, \orgaddress{20 Kavafi Str., 2121  Nicosia,  Cyprus}}

\affil[5]{\orgname{Institute of High Energy Physics}, \orgaddress{\postcode{100049}, \city{Beijing}, \country{People's Republic of China}}}

\affil[6]{\orgname{Division of Particle and Nuclear Physics, Department of Physics, Lund University,
Box 118, \\SE 221 00 Lund, Sweden}}

\affil[7]{\orgdiv{LPHE}, \orgname{École Polytechnique Fédérale de Lausanne}, \orgaddress{\street{Route de la Sorge}, \city{Écublens}, \postcode{1015}, \state{Vaud}, \country{Switzerland}}}

\affil[8]{University of Graz, NAWI Graz, Institute of Physics, Universitätsplatz 5, 8010 Graz, Austria}

\affil[9]{   Jagiellonian University,
        Marian Smoluchowski Institute of Physics,
        Łojasiewicza 11,
        30-348
        Kraków, Poland}

\affil[10]{Marietta Blau Institute for Particle Physics, Austrian Academy of Sciences, Dominikanerbastei 16, 1010
Vienna, Austria}        

\abstract{\unboldmath
We present a phenomenology-driven overview of opportunities and outstanding challenges in baryon electroweak transitions, focusing on hyperon and charmed-baryon processes for which lattice-QCD results are available, including semileptonic decays and selected radiative, dileptonic, and neutrino-induced channels. We identify semileptonic decays of strange and charmed baryons as a particularly promising domain for synergy between phenomenology, experiment, and lattice QCD. Particular emphasis is placed on the determination of vector and axial form factors, the precision and model-independent extraction of $|V_{us}|$ from hyperon semileptonic decays including radiative corrections, and on searches for non-standard interactions through lepton-flavor- and CP-violating observables. We examine the momentum dependence of form factors in ground-state transitions, highlight the emerging $\Xi_c \to \Xi$ semileptonic puzzle stemming from tensions between lattice-QCD predictions and experimental measurements, discuss inclusive decay rates, and outline strategies for extending these studies to transitions involving excited baryon states. For completeness, we review also results involving beauty-baryon decays.
}

\maketitle
\clearpage

\tableofcontents


\section{Introduction}\label{sec:intro}

Electroweak transitions among low-lying strange and charmed baryons have emerged as a promising frontier for precision studies of the Standard Model (SM), offering unique opportunities for close interplay between experiment, phenomenology, and lattice QCD (LQCD). Recent measurements and forthcoming high-statistics datasets from BESIII, Belle II, and LHCb are transforming baryon semileptonic decays into precision probes of baryon structure, the Cabibbo–Kobayashi–Maskawa (CKM) matrix, SU(3) flavor-symmetry breaking, and baryon–meson dynamics. These channels probe SM currents in a regime complementary to meson decays, where a richer spin structure provides enhanced and qualitatively distinct sensitivity to hadron form factors, as well as possible contributions from physics beyond the Standard Model (BSM).  

These experimental developments have been accompanied by rapid progress in lattice QCD. Advances in algorithms,  improved control over systematic uncertainties, new theoretical developments, and simulations at or near physical quark masses have significantly increased the precision and reliability of lattice calculations of baryon transition matrix elements.  For several benchmark observables, lattice QCD is now reaching a level of accuracy that both complements and guides experimental measurements, enabling stringent tests of the Standard Model and increasingly precise determinations of its fundamental parameters.

Despite this remarkable progress, baryon weak decays remain considerably more challenging to study than their mesonic counterparts due to the exponentially poorer signal-to-noise ratio in baryon correlation functions and the larger number of independent form factors.
As noted in the review 50 Years of Quantum Chromodynamics \cite{Gross:2022hyw} 
\textit{``[...]lattice QCD calculation of form factors for weak decays of baryons is still in its infancy because of the extra challenges provided by the poorer signal-to-noise. [...]This is clearly a developing area for the future.''}

Building on these developments, our aim with this white paper is to identify the most promising opportunities at the interface  of experiment, phenomenology, and lattice QCD, and to provide a roadmap for coordinated progress over the coming decade.  On the experimental side, we  assess the  near-term  physics potential of semileptonic baryon decays, outlining realistic precision targets that can be achieved using forthcoming datasets from LHCb, BESIII, and Belle~II.  We also discuss opportunities for measuring rare leptonic decays of hyperons at the FAIR facility, where   the new CBM detector is expected to come into operation by the end of this decade. On the lattice QCD side, we review  the current  status  of calculations of baryon form factors and other hadronic matrix elements relevant to weak decays and identify the theoretical and computational advances required to fully capitalize on the anticipated experimental program. Particular emphasis is placed on improving the control of systematic uncertainties, extending the range of accessible observables, and achieving the precision needed for meaningful comparisons with experiments. Finally, on the side of functional methods, we sketch the status of electromagnetic and axial elastic and transition form factor calculations of baryons and provide a roadmap
for future studies of electroweak transitions.

Effective field theories (EFTs) based on hadronic degrees of freedom are complementary to ``microscopic'' methods based on quarks and gluons (lattice QCD, functional methods, quark models, \ldots). The strength of EFTs lies in the possibility to systematically order the importance of various contributions to a given observable: pion-loop effects, flavor-symmetry breaking, isospin breaking, and radiative corrections, to name a few. The price to pay is typically a growing number of low-energy constants that come with the demand to increase precision, i.e., to push the power expansion in small scales to higher orders. An important opportunity is, therefore, the determination of low-energy constants using microscopic methods rather than relying exclusively on phenomenological fits. In addition, the convergence of the power expansion can become an issue, particularly if not all relevant small scales have been properly identified. In practice, this can give rise to several competing EFT approaches and combinations with phenomenological models, resummation schemes, or dispersion theory. It is also true that the application of EFTs to mesonic systems \cite{Gasser:1984gg,Cirigliano:2011ny,Bijnens:2014lea,Colangelo:2018jxw,Colangelo:2025ivq} is further developed than for baryons. 

Throughout this white paper, we highlight the distinctive role of baryon transitions as complementary probes of the Standard Model.  Compared with mesonic decays, they offer richer spin and helicity structures, enhanced sensitivity to the underlying operator structure, access to complementary kinematic regimes, and valuable cross-checks of theoretical and experimental systematic uncertainties. 

The remainder of the paper is organized as follows: We begin by outlining the theoretical framework,  including first-principles approaches to the non-perturbative regime of QCD based on lattice simulations, effective field theory, and functional methods. We then review the relevant experimental programs,  with  particular emphasis on  BESIII, STCF experiments, LHCb, Belle and Belle~II, HADES, and CBM, highlighting their respective capabilities and projected sensitivities. The core of the white paper is devoted to a systematic discussion of weak processes, organized by classes of baryon transitions.

For each of the most promising channels, we provide: (a) an assessment of the experimental outlook, including the current status, anticipated datasets, dominant sources of systematic uncertainty, and  expected precision; (b)  a review of the corresponding lattice QCD calculations, identifying   the required  matrix elements, kinematic coverage of form factors, treatment of multi-hadron states, control of lattice systematic uncertainties, and realistic precision targets; and (c) an evaluation of the broader physics impact, including  implications  for Standard Model tests, searches for physics beyond the Standard Model, and the understanding of non-perturbative QCD.

Throughout the white paper, we distinguish between opportunities and prospects: By \textit{opportunities}, we refer to processes that have not yet been measured or are currently known with limited precision but are expected to become experimentally  or theoretically accessible through new data sets,  methodologies, or computational advances. By \textit{prospects}, we refer to channels that are already established experimentally  but for which substantial improvements in precision, theoretical control, or physics reach can realistically be achieved in the coming years.

Rather than providing a comprehensive catalog of all electroweak baryon decays, this white paper focuses on those channels where coordinated advances in experiment, phenomenology, and lattice QCD are expected to have the greatest scientific impact. It aims to identify the key priorities and benchmarks that will shape the field over the next decade.

\section{ Theoretical Framework }\label{sec:theory}
\subsection{Formalism}
Semileptonic baryon decays are mediated by the charged–current weak interaction. Owing to the large mass of the $W$ boson, its propagator reduces to a point-like interaction at the quark level, so that a local four-fermion description provides an excellent approximation over the entire kinematic regime of interest. The corresponding low-energy dynamics are therefore encoded in the effective Lagrangian
\begin{equation}
\mathcal{L}_{\text{eff}} = -\frac{G_F}{\sqrt{2}}\, V_{qq'}\, \bigl(\bar q' \gamma^\mu(1-\gamma_5) q \bigr)\cdot \bigl(\bar\ell \gamma_\mu(1-\gamma_5) \nu_\ell\bigr)\,,    
\end{equation}
where $G_F$ denotes the Fermi constant, $V_{qq'}$ the appropriate Cabibbo–Kobayashi–Maskawa (CKM) matrix element, using the conventions of Bjorken and Drell \cite{Bjorken:1965zz}. The chiral structure $\gamma^\mu(1-\gamma_5)$ reflects the purely left-handed nature of charged weak currents in the Standard Model. Generalizations of this operator basis and their implications for semileptonic transitions are discussed below.

The Fermi constant $G_F$ is one of the most precisely determined parameters of the Standard Model. It is extracted from the muon lifetime, which is dominated by the purely leptonic decay $\mu^- \to e^- \bar\nu_e \nu_\mu$. At tree level, the decay rate is proportional to $G_F^2 m_\mu^5$, but achieving the current level of precision requires incorporating higher-order electroweak corrections, phase-space effects from the nonzero electron mass, and QED radiative contributions. The resulting value, $G_F = 1.1663787(6)\times 10^{-5}\,\mathrm{GeV}^{-2}$, is known with sub-ppm accuracy. For applications to semileptonic processes, this definition must be refined to account for process-dependent radiative effects. In particular, Sirlin~\cite{Sirlin:1967zza} identified universal short-distance electroweak corrections that can be absorbed into an effective normalization of the four-fermion interaction, along with additional long-distance contributions that depend on the hadronic structure. This separation ensures that $G_F$, as extracted from muon decay, can be consistently employed in semileptonic amplitudes once the appropriate radiative corrections are included.

\begin{wrapfigure}[15]{r}{0.39\linewidth}\vspace*{-0.3cm}
    \centering
    \includegraphics[width=\linewidth]{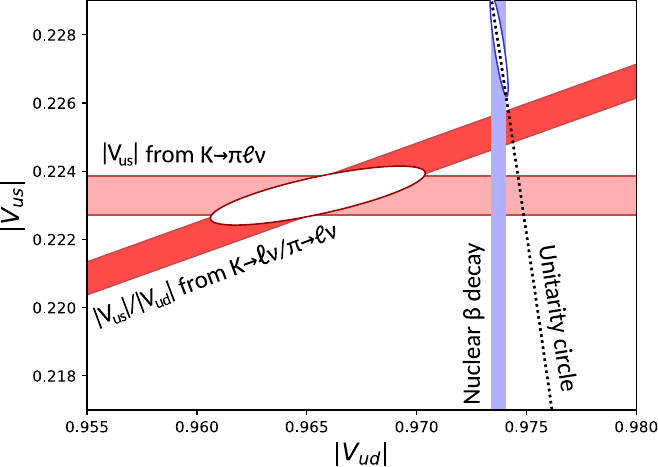}\vspace*{-0.5cm}
    \caption{Determination of $|V_{ud}|$ from nuclear $\beta$ decay, $|V_{us}|$ from kaon semileptonic decay and $|V_{us}|/|V_{ud}|$ from the ratio of pion-to-kaon leptonic decays. The dashed line represents the unitary circle where $|V_{ud}|^2+|V_{us}|^2=1-|V_{ub}|^2$. Details on these determinations can be found either in~\cite{ParticleDataGroup:2024cfk} or~\cite{FlavourLatticeAveragingGroupFLAG:2024oxs}.}
    \label{fig:VudVus}
\end{wrapfigure}
Flavor-changing transitions are governed by the CKM matrix, which parametrizes the misalignment between the weak-interaction and mass eigenbases of quarks. It is a unitary $3\times 3$ matrix, implying a set of normalization and orthogonality conditions among its elements, 
\begin{equation}   
V_{\smallcap{CKM}} =
\begin{pmatrix}
V_{ud} & V_{us} & V_{ub} \\
V_{cd} & V_{cs} & V_{cb} \\
V_{td} & V_{ts} & V_{tb}
\end{pmatrix}\quad\text{with}\quad   
|V_{\smallcap{CKM}}| \approx
\begin{pmatrix}
0.974 & 0.225 & 0.004 \\
0.225 & 0.973& 0.042 \\
0.009 & 0.041 & 0.999
\end{pmatrix}
\end{equation}
with relative uncertainties (in percent)
\[\frac{\delta |V_\smallcap{CKM}|}{|V_\smallcap{CKM}|}\;(\%)\approx\left(\begin{array}{lll}0.016 & 0.30 & 3.0\\ 
0.30 & 0.016 & 2.0\\
2.3 & 2.0 & 0.003
\end{array}\right)\,,\]
where the values are taken from global fits to SM~\cite{ParticleDataGroup:2024cfk}.

Unitarity plays a central role in testing the consistency of the Standard Model, most notably through the closure of the unitarity triangles. The first-row relation $|V_{ud}|^2 + |V_{us}|^2 + |V_{ub}|^2 = 1$ is currently the most precisely tested, with $|V_{ud}|$ determined from superallowed nuclear beta decays, $|V_{us}|$ from kaon and $\tau$ decays, and $|V_{ub}|$ from inclusive and exclusive $B$ decays. While $|V_{ud}|$ is known at the $10^{-4}$ level,  determinations of $|V_{us}|$ with per-mile accuracy show tensions at the percent level when extracted from different processes. This leads to deviations from exact unitarity at the level of a few standard deviations, as shown in Fig.~\ref{fig:VudVus}. The second and third rows are less precisely constrained, with notable discrepancies between inclusive and exclusive determinations of $|V_{cb}|$ and $|V_{ub}|$. Overall, the CKM framework remains remarkably successful, but the tensions continue to motivate the development of improved theoretical methods and higher-precision measurements across multiple processes.

The decay rate of a semileptonic transition is given by
\begin{equation}\label{eq:decay_rate}
\Gamma = \frac{1}{2M_1} \int d\Phi_3 \, \big| \mathcal{M}(B_1 \to B_2  \ell \nu) \big|^2 \quad\text{with}\quad \mathcal{M}(B_1 \to B_2  \ell \nu) = \frac{G_F}{\sqrt{2}} \, V_{qq'} \sum_\mu\, \langle B_2,\ell,\nu_\ell\ | J^{q'q}_\mu J^{\ell\nu_\ell}_\mu | B_1 \rangle
\end{equation}
where $M_1$ is the mass of the initial state,  $J^{q^\prime q}=\bar{q}^\prime \gamma_\mu(1-\gamma_{5})q$, $J^{l\nu_l }=\bar{l} \gamma_\mu(1-\gamma_{5})\nu_l$; $d\Phi_3$ is the Lorentz-invariant measure that spans the kinematically allowed configurations of the outgoing lepton, neutrino, and baryon, with differential observables obtained by imposing angular or energy constraints.
\begin{wrapfigure}[11]{r}{0.4\linewidth}\vspace*{-0.3cm}
    \centering
    \includegraphics[width=\linewidth]{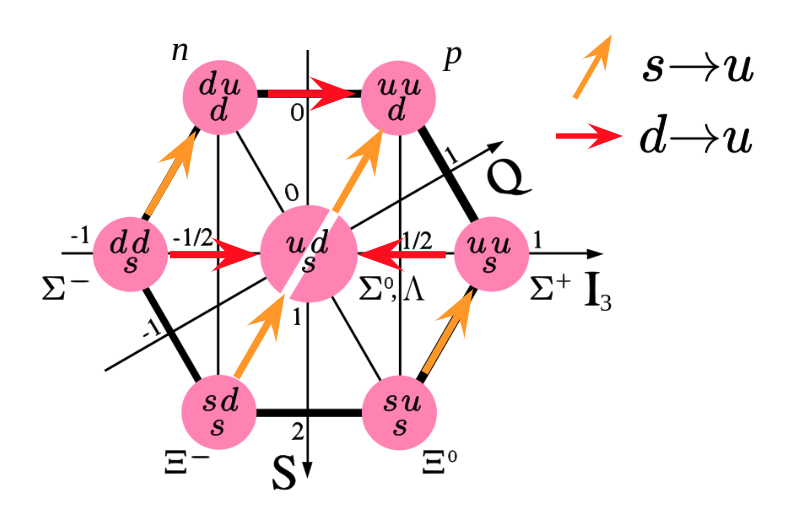}\vspace*{-0.3cm}
    \caption{Possible semileptonic decays within the baryon octet.}
    \label{fig:octet}
\end{wrapfigure}
Let us consider a semileptonic decay of a 1/2$^+$ hyperon $B_1$ into a 1/2$^+$ baryon $B_2$, as shown in Fig.~\ref{fig:octet}, and an off-shell $W^-$-boson decaying to the lepton pair $l^-\bar\nu_l$, denoted as  $B_1(p_1,M_1)\to B_2(p_2,M_2)+l^-(p_l,m_l)+\bar\nu_l(p_{\nu},0)$, where  baryon $B_1$ ($B_2$)  has momentum $p_1$ ($p_2$) and mass $M_1$ ($M_2$). The matrix elements of the vector $J^V_{\mu}$ and axial-vector $J^A_{\mu}$ currents in the notation of Ref.~\cite{Kadeer:2005aq} are:
\begin{equation}\label{eq:matrixelem_semil}
 \begin{aligned}
 \langle B_2|J^{V}_{\mu}-J^{A}_{\mu}|B_1\rangle 
& = \bar{u}(p_2)\left[\gamma_{\mu}\left(F^V_1(q^2)-F^A_1(q^2)\gamma_5\right)-\frac{i\sigma_{\mu\nu}q^{\nu}}{M_1}\left(F^V_2(q^2)\right.\right.\\
&-\left.F^A_2(q^2)\gamma_5\right)+\left.\frac{q_{\mu}}{M_1}\left(F^V_3(q^2)-F^A_3(q^2)\gamma_5\right)\right]u(p_1) \ ,
 \end{aligned}
\end{equation}
where $q_{\mu}=(p_1-p_2)_{\mu}=(p_l+p_\nu)_\mu$ is the four-momentum transfer. The four-momentum transfer squared $q^2$ ranges from $m_l^2$ to $(M_1-M_2)^2$. One often represents the phase space in terms of the parameter $\delta=(M_1-M_2)/M_1$. Since the parameter {$\delta$} is small for $u-d$ and $s-u$ transitions, it is useful to order approximations for the decay distributions by their powers. The hadronic matrix elements are expressed in terms of Lorentz-invariant form factors and thus depend only on the momentum transfer $q^2$. They provide a general decomposition consistent with the symmetries of QCD.
Furthermore, the helicity form factors $F_{1,2,3}^{V, A}(q^2)$ are complex  functions of $q^2$ that describe hadronic effects in the transition.  Neglecting possible CP-odd weak phases, the corresponding form factors are the same for the $(l^-,\bar{\nu}_l)$ and $(l^+,\nu_l)$ transitions. 
To fully determine the hadronic part of a semileptonic decay, the six form factors involved should be determined as a function of $q^2$. 
 An alternative parametrization is provided by Weinberg's form factors~\cite{Weinberg:1958ut}, which are standard in phenomenological analyses. However,  the helicity~\cite{Feldmann:2011xf} form factors are better suited for computing the decay distributions as they map directly onto angular distributions and polarization observables.

The form factors $F^V_{1,2,3}(q^2)$ originate from the vector current $V_{\mu}=\bar{u}\gamma_{\mu}s$, while $F^A_{1,2,3}(q^2)$ arise from the axial-vector current $A_{\mu}=\bar{u}\gamma_{\mu}\gamma_5s$. $F^V_1(q^2)$ and $F^V_2(q^2)$ are known as the Dirac and Pauli form factors, and $F^A_1(q^2)$ and $F^A_3(q^2)$ are the axial and induced pseudoscalar form factors -- structures also present in diagonal matrix elements. In contrast, $F^V_3(q^2)$ and $F^A_2(q^2)$ are second-class form factors~\cite{Weinberg:1958ut} that are proportional to the mass splitting and vanish in the exact SU(3)-flavor symmetry limit, where the quark masses are degenerate and $M_1 = M_2$. Additionally, both $F^V_3(q^2)$ and $F^A_3(q^2)$ contribute to the decay rate only through terms suppressed by $m^2_l$, where $m_l$ is the lepton mass. As a result, $F^V_3(q^2)$ is negligible in all channels at the targeted precision, while $F^A_3(q^2)$ becomes relevant only in the muonic mode, since $m_{\mu}$ is similar in size to the mass splitting.

Since the $q^2$ range is limited for hyperon decays, the dependence of the form factors on $q^2$ can be  neglected at first order, and we use their values at $q^2=0$.
To this NLO, the decay width is given by
\begin{equation}
\Gamma_{\rm NLO} = \frac{G_F^2 |V_{ij}|^2 M_1^5\delta^5}{60 \pi^3}\left(1-\frac{3}{2}\delta\right)\bigg(f_1^2+3g_1^2\bigg) \ ,
\label{eq:GNLO}
\end{equation}
where $V_{ij}$ is the corresponding CKM matrix element and the coupling constants are  $f_i= F^V_i(0)$, $g_i= F^A_i(0)$.
A better approximation is to assume the effective range expansion for the form factors given by 
\begin{equation}\label{eq:range}
F_i(q^2) \approx F_i(0)\left(1+\frac{\langle{r_i^2}\rangle}{6}q^2\right)\ ,    
\end{equation}
where $\langle{r_i^2}\rangle$ are the corresponding root mean square (r.m.s.) radii. The decay rate, thus, depends on the couplings and the r.m.s. radii of the contributing form factors. 
 These parameters can be extracted from the observed distributions in experiments or computed  from theory.
The form factors are usually parameterized by the axial-vector to vector coupling $g_{av}$, the weak-magnetism coupling $g_w$, and the pseudoscalar coupling  $g_{av3}$. They are defined by normalizing their values to  the $f_1$ value as
\begin{equation}\label{eq:g_q2}
 g_{av}=\frac{g_1}{f_1}\ , \qquad g_w=\frac{f_2}{f_1}\ , \qquad g_{av2}=\frac{g_2}{f_1}\ , \qquad
 g_{av3}=\frac{g_3}{f_1}\ .
\end{equation}

In  general, for a transition of the form $B_1\to B_2$, the ratio of decay rates of the $\ell_1$ to $\ell_2$ mode is
\[
R_{\ell_1\ell_2}^{B_1\to B_2} \equiv \frac{\Gamma(B_1\to B_2\ell_1\bar{\nu})}{\Gamma(B_1\to B_2 \ell_2\bar{\nu})}
\]
which, at NLO in $\delta$, is not dependent on the SM form factors.

The decomposition of matrix elements of transitions between the decuplet and octet baryons  $3/2^+ \to 1/2^+$ reads \cite{Korner:1976hv,Salone:2021bvx}
\begin{eqnarray}
  \label{eq:def-transFF13}
  \langle B_2(1/2^+) \vert J_\mu^V - J_\mu^A \vert  B_1(3/2^+) \rangle = 
  \bar u(p_2) \, \Gamma_{\mu\nu}(p_1-p_2) \, u^\nu(p_1)
\end{eqnarray}
with
\begin{eqnarray}
  \label{eq:def-TFF-Gam}
  \Gamma_{\mu\nu}(q) &:=& \frac{\gamma_\mu q_\nu}{M_1} \left( F_1^V(q^2) \gamma_5 - F_1^A(q^2) \right) 
  - \frac{i \sigma_{\mu\alpha} q^\alpha q_\nu}{M_1^2} \left( F_2^V(q^2) \gamma_5 - F_2^A(q^2) \right) \nonumber \\
  && {} + \frac{q_\mu q_\nu}{M_1^2} \left( F_3^V(q^2) \gamma_5 - F_3^A(q^2) \right) 
  + g_{\mu\nu} \left( F_4^V(q^2) \gamma_5 - F_4^A(q^2) \right) \,.
\end{eqnarray}
Note that the ``natural'' parity of a spin-3/2 state is negative. Therefore, the appearance of $\gamma_5$ is reversed compared to the transition $1/2^+ \to 1/2^+$ provided in Eq.~(\ref{eq:matrixelem_semil}). 

The decomposition of matrix elements of transitions among  the decuplet baryons  $3/2^+ \to 3/2^+$ is given by
\begin{eqnarray}
  \label{eq:def-transFF33}
  \langle B_2(3/2^+) \vert J_\mu^V - J_\mu^A \vert  B_1(3/2^+) \rangle = 
  \bar u^\alpha(p_2) \, \Gamma_{\mu\alpha\beta}(p_1-p_2) \, u^\beta(p_1)
\end{eqnarray}
with
\begin{eqnarray}
  \label{eq:def-TFF-Gam33}
  \Gamma_{\mu\alpha\beta}(q) &:=& \gamma_\mu g_{\alpha\beta} \left( F_1^V(q^2) - F_1^A(q^2) \gamma_5 \right) 
  - \frac{i \sigma_{\mu\nu} q^\nu g_{\alpha\beta}}{M_1} \left( F_2^V(q^2) - F_2^A(q^2) \gamma_5 \right) \nonumber \\
  && {} + \frac{q_\mu g_{\alpha\beta}}{M_1} \left( F_3^V(q^2) - F_3^A(q^2) \gamma_5 \right) 
     +\frac{\gamma_\mu q_\alpha q_\beta}{M_1^2} \left( F_4^V(q^2) - F_4^A(q^2) \gamma_5 \right) \nonumber \\
   && {}  - \frac{i \sigma_{\mu\nu} q^\nu q_\alpha q_\beta}{M_1^3} \left( F_5^V(q^2) - F_5^A(q^2) \gamma_5 \right) \nonumber \\
   && {} + \frac{q_\mu q_\alpha q_\beta}{M_1^3} \left( F_6^V(q^2) - F_6^A(q^2) \gamma_5 \right) 
  + \frac{q_\beta g_{\mu \alpha}}{M_1} \left( F_7^V(q^2) - F_7^A(q^2) \gamma_5 \right)  \,.  \nonumber \\ &&
\end{eqnarray}
Note that one can use the equations of motion to show that the structures $\bar u^\alpha(p_2) \, q_\alpha g_{\mu \beta} \, (\gamma_5) \, u^\beta(p_1)$ are redundant.

In general, the number of transition form factors can be obtained from the following considerations. The currents $J^V_\mu$ and $J^A_\mu$ are not conserved. Their divergence is proportional to the difference or sum of the masses of the quarks that participate in the flavor change. Therefore, the four-vectors have spin-0 and spin-1 components. All that is needed now is to count the number of baryon helicity amplitudes that can combine such that the spin of the quark current is 0 or 1 and that the corresponding magnetic quantum number fits. For instance, if we consider the transition $3/2 \to 3/2$, then there are in total $4 \times 4$ helicity combinations. However, six of them lead to magnetic quantum numbers with a modulus larger than 1. This leaves 10 helicity amplitudes where the spin of the quark current can be 1. Following the same logic, there are in total 4 helicity combinations to obtain spin 0. The corresponding combinations of transition form factors constitute induced scalar and pseudoscalar form factors. In total, one obtains $10+4$ transition form factors, which split up into 7 vector and 7 axial-vector form factors. 
\begin{figure}
    \centering
    \includegraphics[width=0.8\linewidth]{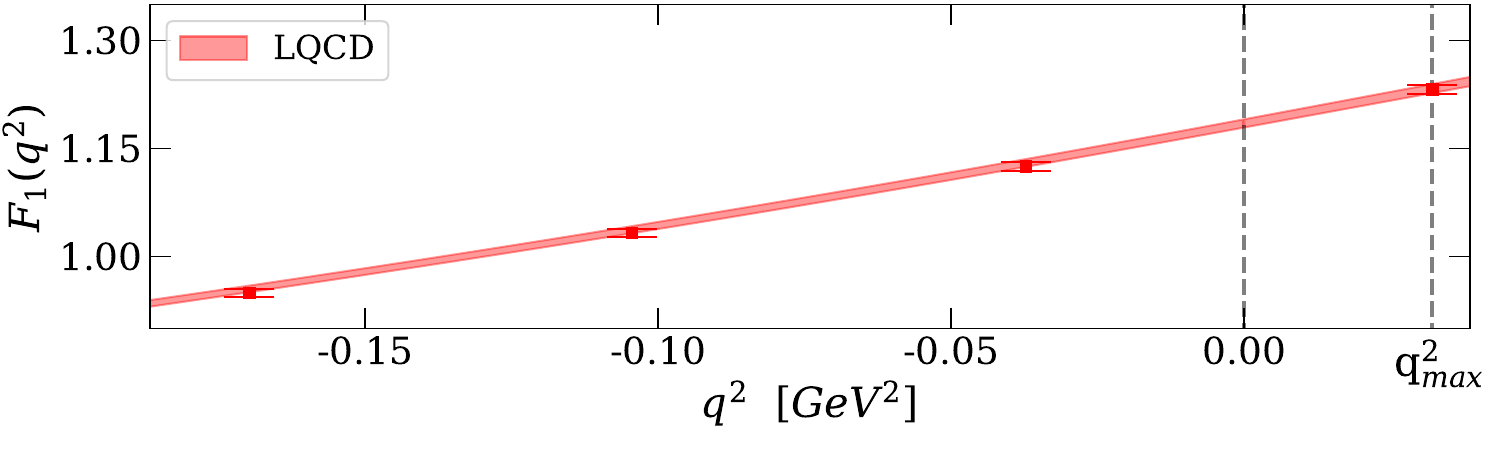}
    \caption{Ilustration of a lattice results for $\Lambda\to p$ $F_1(q^2)$ form factor~\cite{Bacchio:2025auj}. The region between the dashed lines is the kinematical range $m_e^2<q^2<(M_\Lambda-M_p)^2$ in the semileptinic decay $\Lambda\to p e^-\bar\nu_e$. Experimental results could be parameterized with high accuracy using \eqref{eq:range}, but one needs models to describe the full range obtained in lattice QCD calculations that include spacelike region. }
    \label{fig:F1illustration}
\end{figure}

\subsubsection{\texorpdfstring{$q^2$}{q2}-dependence of form factors}
For transitions with a large phase space, such as $c-s/d$ or $b-c/u$, one parametrizes the $q^2$-dependence of the form factors using a model. The same holds for lattice QCD calculations, which obtain form factors in the range $q^2<(M_1-M_2)^2$ down to the timelike region. This is illustrated in Fig.~\ref{fig:F1illustration} using results of the $F_1(q^2)$ $\Lambda\to p$ form factor from~\cite{Bacchio:2025auj}. Sparse $q^2$ bins of the LQCD results require proper interpolation, whereas for the experiment, the effective range approximation is fully sufficient.

The standard parametrization results from including one or multiple meson poles that have the correct quantum numbers to mix with the $W$ boson and have masses close to the $q^2$ range of the decay. Traditionally, one pole is explicitly included together with  an effective contribution from other poles~\cite{HFLAV:2019otj},
such as in the  Becirevic--Kaidalov (BK)~\cite{Becirevic:1999kt} parameterization given by
\begin{equation}\label{eq:beckai}
    F_i(q^2)=
    \frac{F_i(0)}{M^2-{q^2}}~\frac{M^4}{M^2-\alpha_{\text{BK}}{q^2}}\ ,
\end{equation}
where the dominant pole mass $M$ is outside the kinematic region, and the parameter $\alpha_{\text{BK}}$ represents an effective contribution from the meson poles with higher masses. The case $\alpha_{\text{BK}}=0$ represents the dominant pole contribution. This parameterization gives real-valued form factors. If more data is available, one or more extra parameters can be added to describe the $q^2$-dependence. Another parameterization of the form factors is the model-independent expansion after conformal transformation into $z$-variable~\cite{Hill:2010yb}, defined by
\begin{equation}
    F(q^2)=\sum_{k=0}^{k_{max}}a_kz^k \quad{\rm with}\quad z=\frac{\sqrt{t_{cut}-q^2}-\sqrt{t_{cut}}}{\sqrt{t_{cut}-q^2}+\sqrt{t_{cut}}},
\end{equation}
where $t_{cut}$ is the mass squared at the start of the branch cut, e.g., for hyperon decays $t_{cut}=(m_{\pi}+m_K)^2$ with $m_{\pi}$ and $m_K$ denoting the
pion and kaon masses, respectively. It is customary~\cite{Hill:2010yb} to constrain the $z$-expansion to vanish in the large-$Q^2=-q^2$ limit by imposing $a_{k_{max}} =-\sum^{k_{max}-1}_{k=0} a_k$.

\subsubsection{Neutrino vs Antineutrino mode}\label{sec:(anti)neutrino}

Semileptonic decays with a neutrino or an antineutrino in the final state exhibit opposite signs in the $V\!A$ interference terms. This follows from comparing the two leptonic currents,
$L_\alpha = \bar\ell\,\gamma_\alpha(1-\gamma_5)\nu$ (neutrino mode) and
$L_\alpha = \bar\nu\,\gamma_\alpha(1-\gamma_5)\ell$ (antineutrino mode),
which leads, upon squaring the amplitude, to the leptonic tensors

\begin{equation}
L_{\alpha\beta}=
\begin{cases}
\mathrm{Tr}\!\left[
(\slashed{k}_\ell-m_\ell)\,\gamma_\alpha(1+\gamma_5)\,\slashed{k}_\nu\,\gamma_\beta(1+\gamma_5)
\right] & (\text{neutrino})\\
\mathrm{Tr}\!\left[
(\slashed{k}_\ell+m_\ell)\,\gamma_\alpha(1-\gamma_5)\,\slashed{k}_\nu\,\gamma_\beta(1-\gamma_5)
\right] & (\text{antineutrino})
\end{cases}
\end{equation}
The only difference is the ordering of the spinors in the trace, which is not symmetric under this exchange because the $\gamma_5$ term generates an antisymmetric tensor. A standard evaluation with $\epsilon_{0123}=1$ yields
\begin{equation}
L_{\alpha\beta}
= 8k_{\nu\alpha}k_{\ell\beta}+8k_{\ell\alpha}k_{\nu\beta}-8(\frac{q^2-m_\ell^2}{2}) g_{\alpha\beta}
\begin{cases}
-8\,i\,\varepsilon_{\alpha\beta\rho\sigma}\,k_\nu^\rho k_\ell^\sigma & (\text{neutrino})\\
+8\,i\,\varepsilon_{\alpha\beta\rho\sigma}\,k_\nu^\rho k_\ell^\sigma & (\text{antineutrino})
\end{cases}
\end{equation}
so that, after contracting with $H^{\alpha\beta}$, the symmetric terms, $VV$ and $AA$, remain identical, while the antisymmetric part, $V\!A$, changes sign.

For the absolute decay rate, the $V\!A$ interference term does not survive since spin averaging and full angular integration eliminate the triple–product structures to which the antisymmetric piece contributes. The $V\!A$ term survives only if at least one of the following conditions is satisfied:
\begin{itemize}
    \item angular distributions are retained, i.e., the lepton directions are not integrated out, producing forward–backward or $T$-odd correlations;
    \item  the initial and final states are differently polarized, so that  $S^\mu \neq 0$  prevents the triple product from averaging to zero;
    \item a $T$-odd observable is constructed, for example, $(\vec p_\ell \times \vec p_B)\cdot\vec S_{B'}$.
\end{itemize}
In such cases, the interference term changes sign between neutrino and antineutrino modes, leading to measurable differences in angular and polarization observables.

\subsubsection{Radiative corrections}

Radiative corrections are required for a precise determination of decay rates in semileptonic baryon decays and for the extraction of CKM matrix elements. The decay width including radiative corrections can be written as~\cite{Flores-Mendieta:1996vgq}
\begin{equation}
\Gamma^{c}=\Gamma\cdot(1+C_r),    
\end{equation}
where $\Gamma$ denotes the tree-level decay width and $C_r$ is the radiative correction factor. For example, for the decay $\Lambda\to p e^-\bar{\nu}_e$, the radiative correction has been calculated to be $C_r=(1.6\pm0.5)\%$~\cite{Flores-Mendieta:1996vgq}. This correction corresponds to the model-independent component of the radiative corrections derived in Ref.~\cite{Flores-Mendieta:1996vgq}.

For the determination of CKM matrix elements, electroweak radiative corrections are incorporated through an effective modification of the Fermi coupling constant. The exact implementation depends on the theoretical framework used to determine the vector form factor.

The revised universal electroweak correction~\cite{Czarnecki:2019mwq,Czarnecki:2004cw} fr the $\Lambda$ hyperon decay is,
\begin{equation}
\Delta_R^V=\frac{\alpha}{2\pi}\left[3\ln\left(\frac{M_Z}{M_p}\right)+\ln\left(\frac{M_Z}{M_A}\right)+2C+A_g\right]=0.0191,   
\end{equation}
where $M_Z=91.188$~GeV/$c^2$, $M_p=0.938$~GeV/$c^2$, $M_A=1.2$~GeV/$c^2$, which corresponds to the axial-vector meson $a_1(1260)$, $C\simeq-0.629$~\cite{Marciano:1985pd,Marciano:2005ec,Czarnecki:2019mwq}, and $A_g\simeq-0.34$. This correction is applied through a modification of the Fermi coupling constant:
\begin{equation}
G_F^2 \equiv G_F^2(1+\Delta_R^V)\,.    
\end{equation}

\subsubsection{Non-standard interactions}
\label{sec:BMS}

Semileptonic hyperon decays are part of a broader class of charged-current semileptonic processes mediated by quark transitions of the form $u_i \leftrightarrow d_j$, with $u_i=(u,c,t)$ and $d_j=(d,s,b)$.
A model-independent description of possible heavy new physics is naturally organized in the Standard Model Effective Field Theory (SMEFT) above the electroweak scale~\cite{Grzadkowski:2010es} and in the low-energy effective field theory (LEFT), or weak effective theory, below the electroweak scale \cite{Jenkins:2017jig, Dekens:2019ept}.

At energies below the electroweak scale, and for processes in which the top quark is not an external degree of freedom, possible non-standard charged-current interactions can be described by a low-energy effective Lagrangian \cite{Cirigliano:2012ab, Cirigliano:2013xha, GonzalezAlonso:2016etj, Chang:2014iba}
\begin{align}
{\cal L}_{\rm eff}^{ij\ell}
=
-\frac{G_F V_{ij}}{\sqrt{2}}
\Big[&
(1+\epsilon_L^{ij\ell})
\bar\ell\gamma_\mu(1-\gamma_5)\nu_\ell\,
\bar u_i\gamma^\mu(1-\gamma_5)d_j
+\epsilon_R^{ij\ell}
\bar\ell\gamma_\mu(1-\gamma_5)\nu_\ell\,
\bar u_i\gamma^\mu(1+\gamma_5)d_j
\nonumber\\
&+\bar\ell(1-\gamma_5)\nu_\ell\,
\bar u_i\left(\epsilon_S^{ij\ell}
-\epsilon_P^{ij\ell}\gamma_5\right)d_j
+\epsilon_T^{ij\ell}
\bar\ell\sigma_{\mu\nu}(1-\gamma_5)\nu_\ell\,
\bar u_i\sigma^{\mu\nu}(1-\gamma_5)d_j
\Big]
+{\rm h.c.} .
\end{align}
Here \(V_{ij}\) is the corresponding CKM matrix element, and the Wilson coefficients \(\epsilon_X^{ij\ell}\), with \(X=L,R,S,P,T\), parameterize short-distance physics beyond the Standard Model. In a general analysis, nuclear and neutron beta decays probe mostly \(\epsilon_X^{ud\ell}\) \cite{Gonzalez-Alonso:2018omy}, kaon and hyperon decays probe \(\epsilon_X^{us\ell}\) \cite{GonzalezAlonso:2016etj, Chang:2014iba, Alexandrou:2026noh}, charm decays probe \(\epsilon_X^{cd\ell}\) and \(\epsilon_X^{cs\ell}\) \cite{Fajfer:2015mia, Becirevic:2026tle}, while bottom decays probe \(\epsilon_X^{ub\ell}\) and \(\epsilon_X^{cb\ell}\) \cite{Fajfer:2012vx, Sakaki:2013bfa, Leljak:2023gna}. Top charged-current interactions, involving \(td\), \(ts\), and \(tb\), are usually treated directly in SMEFT or in anomalous \(Wtb\)-type couplings rather than in a purely low-energy hadronic EFT \cite{Aguilar-Saavedra:2008quj}.

The same operator basis, therefore, applies across a wide range of semileptonic charged-current processes. However, the Wilson coefficients are not universal numbers independent of flavor. They depend on the quark transition and, in general, on the charged-lepton flavor. Moreover, the numerical sensitivity of a given observable to these coefficients is process dependent because it depends on phase space, lepton-mass effects, QCD running, and the relevant hadronic matrix elements. Lepton-flavor-universality ratios 
$R_{a\to\beta}^{\ell_2\ell_1}$ are especially useful because they are $V_{ij}$ independent. 

At  first order in the non-standard Wilson coefficients, we have,
\begin{equation}
\frac{R_{a\to\beta}^{\ell_2\ell_1}}
{R_{{a\to\beta},{\rm SM}}^{\ell_2\ell_1}}
=
1+
\sum_X
\left[
r_{a\to\beta,X}^{ij\ell_2}\epsilon_X^{ij\ell_2}
-
r_{a\to\beta,X}^{ij\ell_1}\epsilon_X^{ij\ell_1}
\right]
+\cdots .
\end{equation}

In what follows, we specialize this general framework to the strangeness-changing sector relevant for semileptonic hyperon decays. Using the same notation as above, this corresponds to \((i,j)=(u,s)\), so the relevant Wilson coefficients are
\(\epsilon_X^{us\ell}\). 
A particularly important point, emphasized in Ref.~\cite{Chang:2014iba}, is that semileptonic hyperon decays (SHDs) are governed by a small SU(3)-breaking parameter, which allows a simultaneous SU(3)-breaking and recoil expansion of the decay amplitudes. As a result, several observables admit accurate SM predictions with reduced dependence on hadronic form factors, making SHDs an especially clean laboratory in which to search for deviations from the \((V-A)\) structure of the weak interaction.

In the effective-field-theory description adopted in Ref.~\cite{Chang:2014iba}, the low-energy charged-current Lagrangian is extended by non-standard scalar, pseudoscalar, and tensor operators, parameterized by Wilson coefficients $\epsilon_i$. Within this framework, SHDs are sensitive to new physics in a way that is complementary to kaon decays and collider searches. Namely, low-energy observables probe the interference between the Standard Model  and non-standard amplitudes, while collider observables are typically sensitive only to quadratic combinations of the same couplings. This makes improved SHD measurements particularly interesting in a broad SMEFT interpretation of $s\to u$ transitions. 

A central result of Ref.~\cite{Chang:2014iba} is that the lepton-flavor universality ratio
$R_{\mu e}$ is especially clean from a theoretical standpoint. At next-to-leading order in the SU(3) breaking expansion, the SM prediction is independent of hadronic form factors up to relative corrections of $\mathcal{O}(\delta^2)$. This is a remarkable feature, since it means that deviations in $R_{\mu e}$ can be interpreted with unusually small hadronic uncertainty compared to many other semileptonic observables. 

Ref.~\cite{Chang:2014iba} also explains why  \emph{muonic} channels are much more sensitive to non-standard scalar and tensor interactions compared to  their electronic counterparts. At first order in the Wilson coefficients, these new contributions enter through interference with the SM amplitude and are suppressed by a factor proportional to $m_\ell/\sqrt{q^2}$. As a consequence, the effect is strongly suppressed for electrons but significantly enhanced for muons. In practice, this means that electronic modes are best suited to determine normalization factors and standard form-factor information, whereas muonic modes provide the strongest direct sensitivity to scalar and tensor operators. Pseudoscalar effects are additionally kinematically suppressed in SHDs, so the dominant non-standard sensitivity comes from scalar and tensor interactions. 

Using the then-available data on $R_{\mu e}$ in channels such as $\Lambda\to p\ell\bar{\nu}$, $\Sigma^-\to n\ell\bar{\nu}$, $\Xi^0\to\Sigma^+\ell\bar{\nu}$, and $\Xi^-\to\Lambda\ell\bar{\nu}$, Ref.~\cite{Chang:2014iba} found agreement with the Standard Model within uncertainties and translated this into constraints on scalar and tensor Wilson coefficients. An important conclusion was not only the numerical bound itself but also the observation that sensitivity is strongly channel-dependent, as different hyperon transitions probe different combinations of charges and form factors. This makes a broad experimental program across several SHD modes particularly valuable. A future measurement of $R_{\Xi}^{\mu e}$ and $R_{\Sigma}^{\mu e}$ therefore would not simply improve statistics but would add a differently oriented constraint in the $(\epsilon_S,\epsilon_T)$ plane, helping to disentangle scalar from tensor effects.

\begin{figure}
    \centering
    \includegraphics[width=0.45\linewidth]{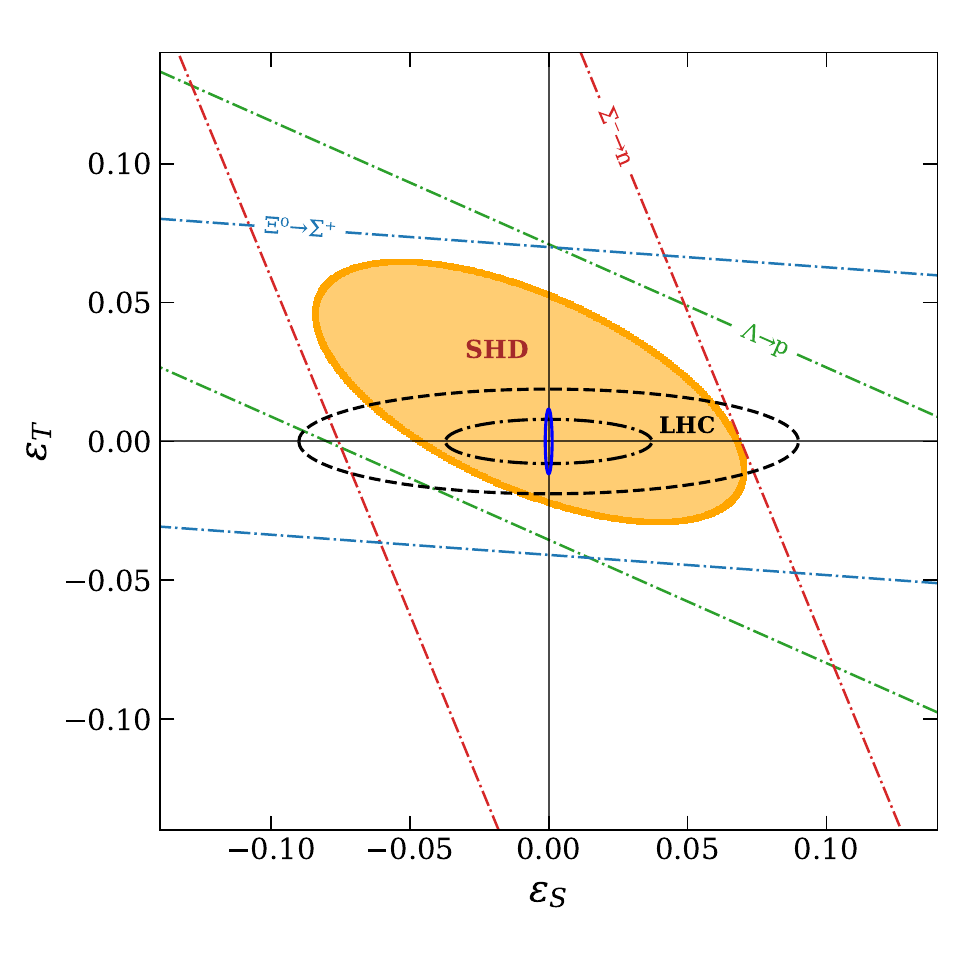}
    \caption{Constraints in the $(\epsilon_S,\epsilon_T)$ plane from semileptonic hyperon decays. The individual bands correspond to different hyperon transitions and illustrate their complementary sensitivities to scalar and tensor interactions, while the filled SHD region denotes their combined constraint. For the $\Lambda\to p$ channel, the scalar and tensor sensitivities are obtained from the lattice-QCD form factors of Ref.~\cite{Alexandrou:2026noh}. The black dashed contours show the high-$p_T$ collider constraints on non-standard $us\to e\nu$ interactions, with the outer and inner contours corresponding to $\sqrt{s}=7$ and $8$~TeV ~\cite{Cirigliano:2012ab,Chang:2014iba}, respectively. The blue contour shows the constraint from semileptonic kaon decays ~\cite{GonzalezAlonso:2016etj}. The Wilson coefficients are defined in the $\overline{\rm MS}$ scheme at $\mu=2~{\rm GeV}$.}
    \label{Fig:WC}
    \end{figure}

In Fig.~\ref{Fig:WC}, we show the impact of the recent LHCb measurement of $R_{\Lambda}^{\mu e}$ using the lattice-QCD determination of the scalar and tensor $\Lambda\to p$ form factors from Ref.~\cite{Alexandrou:2026noh}. For a given SHD channel, the lepton-flavor-universality (LFU) ratio can be expressed as
\begin{equation}
\frac{R_{\mu e}}{R_{\mu e}^{\rm SM}}
=
1+r_S\,\epsilon_S+r_T\,\epsilon_T ,
\end{equation}
where $\epsilon_S$ and $\epsilon_T$ are the scalar and tensor Wilson coefficients, while the coefficients $r_S$ and $r_T$ are channel dependent. For $\Lambda\to p\ell\bar{\nu}$, the lattice-QCD analysis of Ref.~\cite{Alexandrou:2026noh} gives $r_S=1.271(29)$ and $r_T=2.975(34)$. Using the LHCb result $R_{\Lambda}^{\mu e}=0.175(12)$~\cite{LHCb:2025wld} and the corresponding lattice-QCD Standard Model prediction $R_{\Lambda,{\rm SM}}^{\mu e}=0.1735(98)$~\cite{Alexandrou:2026noh},
one obtains
\begin{equation}
1.271(29)\,\epsilon_S+2.975(34)\,\epsilon_T=0.009(90).
\end{equation}

The principal advantage of combining several SHD modes is that the different channels probe different linear combinations of $\epsilon_S$ and $\epsilon_T$. In the updated analysis of Ref.~\cite{Alexandrou:2026noh}, the lattice-QCD input is used for the $\Lambda\to p$ channel, while phenomenological and experimental inputs, together with the SU(3)-breaking expansion, are retained for the other hyperon channels. Their combination gives
\begin{equation}
\epsilon_S=-0.009(36),
\qquad
\epsilon_T=0.011(21),
\end{equation}
with a correlation coefficient $\rho=-0.61$ in the
$\overline{\rm MS}$ scheme at $\mu=2~{\rm GeV}$.
Accordingly, Fig.~\ref{Fig:WC} displays an allowed band in the $(\epsilon_S,\epsilon_T)$ plane rather than a closed contour, since a single channel constrains only one linear combination of Wilson coefficients. This illustrates both the constraining power already reached by modern measurements in the $\Lambda$ channel and the strong motivation for a broader SHD program including several hyperon transitions, which would provide complementary directions in the $(\epsilon_S,\epsilon_T)$ plane.

From the perspective of the present paper, the main message is therefore that SHDs are not simply an alternative way to repeat kaon tests. Rather, they provide a distinct and theoretically clean handle on non-standard $s\to u$ charged-current interactions, with muonic modes playing a special role because of their enhanced first order sensitivity to scalar and tensor operators. This provides a strong motivation for precise measurements of $\Lambda\to p\mu^-\bar{\nu}_\mu$, $\Xi^-\to\Lambda\mu^-\bar{\nu}_\mu$, and their corresponding LFU ratios. 

\subsection{Chiral Perturbation Theory}

For the transitions of strange (or down) to up quarks in baryon decays, one can identify hard and soft scales in the strong-interaction part,
which suggests the applicability of an effective field theory (EFT). The hard scale is set by the (average) baryon mass. For those baryons where weak processes are not completely overshadowed by strong or electromagnetic decays, the mass difference between the initial and final baryon is relatively small. This entails that the available phase space for the decay is also small, i.e.,the momentum of the outgoing baryon and the invariant mass of the dilepton (virtual $W$) are small compared to the baryon mass. For weak decays of ground-state (spin-1/2) hyperons, the smallness of the baryon mass difference can be traced back to the fact that QCD has an approximate chiral $SU(3)$-flavor symmetry. If the up, down, and strange quark masses are neglected relative to the dynamical scale of $\Lambda_{\rm QCD}$, the hyperons become degenerate with the nucleons. 

Chiral perturbation theory is the EFT framework that deals with quark mass effects and low momenta in a systematic way.\footnote{We note in passing that it does not make sense to neglect the charm quark mass relative to $\Lambda_{\rm QCD}$. Also, the momenta of the final states emerging from the semi-leptonic decay of a charm baryon are not small. Therefore, such an EFT would not be meaningful for the decays of charm baryons.} It is suggestive that chiral perturbation theory can then be used to obtain a unifying picture of the semi-leptonic hyperon decays and to provide an understanding of the size of $SU(3)$ flavor breaking. In particular, the spontaneous breaking of (three-flavor) chiral symmetry gives rise to Goldstone bosons (pions, kaons, and $\eta$). The long-distance effects in the loop corrections caused by corresponding chiral logarithms might dominate the flavor breaking. Definitely, they point to physics effects that cannot be neglected. In addition to the semi-leptonic decays themselves, an EFT can be very valuable for the determination of radiative corrections. 

On the other hand, when it comes to quantitative considerations, the strange-quark mass is not negligibly small compared to $\Lambda_{\rm QCD}$ (while the masses of the up and down quarks are). This can present a challenge for the convergence properties of an EFT. Indeed, the quality of three-flavor chiral perturbation theory depends on the chosen observable and on the details of the methods (heavy-baryon, infrared regularization, extended on mass shell, \ldots).

And there is yet another small scale, specific to flavor-changing processes and unrelated to the chiral limit of massless quarks. In every EFT, one must check the relevant degrees of freedom, i.e., the states that sit close to the energy/mass range of interest. Let us consider the mass splitting between ground-state baryons where strangeness differs by one unit. It is essentially proportional to the strange-quark mass $m_s$. Quantitatively, that mass splitting is in the range of 120 to 250 MeV. This should be compared to the typical baryon mass of 1 GeV, i.e., the typical ratio of soft and hard scales is about 25\% (or lower). The scale of the chiral loops is set by the square of the kaon mass (which is also proportional to $m_s$ on account of the Gell-Mann--Oakes--Renner relation \cite{Gell-Mann:1968hlm}). Relative to (the square of) 1 GeV, this is again a 25\% effect. In decays, the momenta of the outgoing states will also scale with some power of $m_s$. All this suggests that it makes sense to expand in powers of the strange-quark mass. But one must check if there are other states close to the hyperons (and with a strangeness difference of one unit) that need to be taken into account. Indeed, this criterion applies at least to the spin-3/2 states of the lowest-lying decuplet. The $\Delta(1232)$ is very close in mass to the (ground-state) $\Sigma$ hyperon. The sibling $\Sigma(1385)$ is rather close to the (ground-state) $\Xi$. But even beyond the decuplet, there is the Roper state $N(1440)$ which is only about 250 MeV away from the (ground-state) $\Sigma$. Note that the mass difference between the spin-1/2 ground state baryons and other states (resonances) does not vanish in the chiral limit. Taking such resonance states into account is not purely an expansion around the chiral limit. This makes a systematic treatment challenging, and various schemes have been proposed on how to order the various soft scales relative to each other. So far, the decuplet states, but no other resonances, have been taken into account for EFT calculations of semi-leptonic decays of hyperons. 

While the inclusion or disregard of resonance states in the loop corrections might be regarded as optional for decays of spin-1/2 hyperons, it is trivially clear that for the decays of decuplet states, it is necessary to include decuplet states as relevant degrees of freedom. For semi-leptonic decays, this concerns the processes $\Omega^- \to \Xi^0 \, \ell^- \bar\nu_\ell$ and $\Omega^- \to \Xi^0(1530) \, \ell^- \bar\nu_\ell$, where so far only the former process has been observed.

The $q^2$ range needed for $d\to u$ and $s\to u$ transitions is small, so one would expect that Chiral Perturbation Theory (ChPT) would provide an excellent framework to discuss SHDs. The lowest order was worked out by Cabibbo in the sixties \cite{Cabibbo:1963yz} and the chiral logarithms were later added in \cite{Bijnens:1985kj}. The corrections to the lowest order were very large, 100\% or more. Adding the decuplet degrees of freedom did improve matters \cite{Jenkins:1991es}, but the corrections are still large. Later work found that the small pionic chiral logarithms improved matters somewhat \cite{Luty:1993gi}.

However, the chiral logarithms are only part of the story; one really needs to add all the higher-order low-energy constant contributions as well. The problem now becomes two-fold: there are too many parameters compared to the data, and it is difficult to find reasonable parameters that give convergence. Baryon ChPT exists in a number of different versions, a nonrelativistic scheme, HBChPT, and two different relativistic schemes, infra-red regularization and Extended-On-Mass-Shell (EOMS), all of them with and without the decuplet degrees of freedom. The convergence of the expansions in these schemes is very different, but never very good. The HBChPT calculation can be found in \cite{Villadoro:2006nj}, and the infra-red  regularization (IRChPT) in \cite{Lacour:2007wm} for the pure octet part. At this level, we have 15 relevant parameters.

The decuplet was added in the work of \cite{Geng:2009ik}, introducing several more unknown parameters. Finally, the EOMS scheme calculation, including the octet and decuplet, was performed in \cite{Sauerwein:2021jxb}. 

The ChPT results in all schemes are known to third order, and the pure octet in some schemes is known to fourth order.
The corrections are shown in Table~\ref{tab:chpt}, where one can judge the convergence.
\begin{table}
\caption{The corrections to $f_1$ (in percentage).
Errors are estimated by varying $\mu$ from 0.7 to 1.3 GeV. Top: octet contributions, bottom decuplet contributions, Table adapted from \cite{Geng:2009ik}.}
\label{tab:chpt}
  \begin{tabular}{c|ccc|c|c}
  \hline\hline
 Octet &\multicolumn{3}{c|}{EOMS} & HBChPT & IRChPT\\\hline
& $\delta^{(2)}$&  $\delta^{(3)}$ & $\delta^{(2)}+\delta^{(3)}$& $\delta^{(2)}+\delta^{(3)}$& $\delta^{(2)}+\delta^{(3)}$\\
\hline
$\Lambda\rightarrow N$ & $-3.8$  &  $\phantom{-}0.2^{+1.2}_{-0.9}$ & $-3.6^{+1.2}_{-0.9}$  & $2.7$ & $-5.7\pm2.1$ \\
$\Sigma\rightarrow N$  & $-0.8$ &  $\phantom{-}4.7^{+3.8}_{-2.8}$ & $\phantom{-}3.9^{+3.8}_{-2.8}$   & $4.1$ & $\phantom{-}2.8\pm0.2$\\
$\Xi\rightarrow\Lambda$& $-2.9$  &  $\phantom{-}1.7^{+2.4}_{-1.8}$ & $-1.2^{+2.4}_{-1.8}$ & $4.3$ & $-1.1\pm1.7$\\
   $\Xi\rightarrow\Sigma$ & $-3.7$  &  $-1.3^{+0.3}_{-0.2}$ & $-5.0^{+0.3}_{-0.2}$  & $0.9$ & $-5.6\pm1.6$\\
  \end{tabular}
\begin{tabular}{c|ccc|ccc}
\hline\hline
Decuplet&\multicolumn{3}{c}{EOMS}&\multicolumn{3}{c}{HBChPT}\\\hline
& $\delta^{(2)}$&  $\delta^{(3)}$ & $\delta^{(2)}+\delta^{(3)}$  & $\delta^{(2)}$&  $\delta^{(3)}$ & $\delta^{(2)}+\delta^{(3)}$\\
   \hline
$\Lambda\rightarrow N$ &    $\phantom{-}0.7$ & $3.0^{+0.1}_{-0.1}$ & $3.7^{+0.1}_{-0.1}$ & $\phantom{-}1.8$ & $\phantom{-}1.3$ & $3.1$ \\
$\Sigma\rightarrow N$  & $-1.4$ & $6.2^{+0.4}_{-0.3}$ & $4.8^{+0.4}_{-0.3}$& $-3.6$ & $\phantom{-}8.8$ & $5.2$\\
$\Xi\rightarrow\Lambda$&  $-0.02$ & $5.2^{+0.4}_{-0.3}$ & $5.2^{+0.4}_{-0.3}$ &$-0.05$ & $\phantom{-}4.2$ & $4.1$ \\
$\Xi\rightarrow\Sigma$ &  $\phantom{-}0.7$ & $6.0^{+1.9}_{-1.4}$ & $6.7^{+1.9}_{-1.4}$ &$\phantom{-}1.9$ & $-0.2$ & $1.7$\\
\hline\hline
\end{tabular}
\end{table}

In conclusion, pure ChPT can best be used as a guide for SHDs in understanding lattice and functional methods results, as well as for estimates of the uncertainties. It is also very relevant for estimating the radiative corrections, as discussed earlier.

\subsection{Lattice QCD methods}

Lattice QCD provides a systematically improvable first-principles framework for studying strongly interacting hadrons and their interactions~\cite{Wilson:1974sk,Gattringer:2010zz}. By formulating QCD on a discretized space-time lattice, it furnishes a non-perturbative regularization of the theory that, after Wick rotation to Euclidean time and in a finite space-time volume, becomes amenable to numerical simulation. This enables the computation of Euclidean correlation functions from which hadron masses, matrix elements, and transition amplitudes relevant to the weak and electromagnetic processes discussed in this white paper can be determined.

The extraction of transition matrix elements is based on the calculation of Euclidean two- and three-point correlation functions, from which the desired initial and final hadronic states are isolated~\cite{Martinelli:1988rr}. For low-lying single-particle states, this procedure is well established: at sufficiently large Euclidean time separations, the ground-state contribution dominates, and excited-state contamination can be systematically controlled~\cite{Alexandrou:2017hac}. The situation is considerably more challenging for resonances and processes involving multi-hadron states, where finite-volume effects become essential and dedicated theoretical formalisms are required to relate lattice observables to physical scattering amplitudes and transition matrix elements~\cite{Luscher:1986pf,Luscher:1990ux,Lellouch:2000pv}.

Compared with mesons, lattice calculations involving baryons face an additional and fundamental challenge: baryonic correlation functions suffer from an exponentially deteriorating signal-to-noise ratio as the Euclidean time separation increases~\cite{Parisi:1983ae,Lepage:1989hd}. Since weak baryon decays require precise determinations of multiple form factors over a range of momentum transfers, overcoming this limitation has driven significant methodological developments in recent years, including improved interpolating operators, optimized analysis strategies, and advanced variance-reduction techniques. These advances have substantially expanded the scope and precision of lattice calculations, making baryon electroweak transitions an increasingly important area of contemporary lattice QCD.

Early lattice investigations of hyperon semileptonic decays often relied on the Feynman–Hellmann theorem to access matrix elements indirectly through energy shifts induced by external currents~\cite{Bouchard:2016heu}. Although well suited for determining forward matrix elements, this approach becomes increasingly impractical when multiple momentum transfers and different current structures are required, as needed for the complete $q^2$ determination of semileptonic form factors. Consequently, early studies were largely restricted to the calculation of vector charges, corresponding to the zero-momentum-transfer value of the leading form factor $f_1$, for processes such as $\Sigma \to N$ and $\Xi \to \Sigma$~\cite{Shanahan:2015dka,Sasaki:2012ne,Sasaki:2017jue}.

The use of three-point correlation functions provides a significantly more versatile framework, as it enables the direct extraction of matrix elements for arbitrary local currents and at non-zero momentum transfers. By constructing suitable ratios of two- and three-point functions, one can isolate the full set of form factors entering the decomposition of matrix elements of hadronic weak currents, thereby accessing not only the charges at  $q^2=0$ but also the complete $q^2$-dependence relevant for phenomenology. This methodology has been extensively employed in the nucleon sector for the study of nucleon structure, including electromagnetic~\cite{Alexandrou:2025vto,Djukanovic:2023jag} and axial~\cite{Alexandrou:2023qbg,Djukanovic:2022wru} form factors, and has become the standard approach for computing transition matrix elements between single baryon states that are stable under the strong interaction.

From a methodological perspective, lattice QCD calculations of weak decays currently follow two complementary strategies. The first, and by far the most widely developed, is based on \emph{exclusive} determinations using three-point correlation functions, in which both the initial and final hadronic states are explicitly isolated, and the relevant matrix elements are extracted as functions of the momentum transfer. This approach provides direct access to the complete set of form factors governing the decay and constitutes the primary framework for the semileptonic baryon transitions discussed throughout this white paper.

A second, more recent strategy aims to compute \emph{inclusive} decay rates directly from four-point correlation functions~\cite{Hansen:2017mnd,Gambino:2020crt,Hansen:2019idp,ExtendedTwistedMass:2024myu,DeSantis:2025yfm}. In this formulation, only the initial hadron is isolated, while the sum over all possible hadronic final states is implemented through time-ordered products of the relevant weak currents. Although computationally and theoretically more demanding, this approach offers a first-principles determination of total inclusive decay rates without requiring the reconstruction of individual final states, providing a complementary avenue for studying weak decays on the lattice.

At present, most lattice calculations are performed in isospin-symmetric QCD (isoQCD), where the up and down quark masses are taken to be equal, and electromagnetic interactions are neglected. Within this approximation, one obtains the leading-order weak decay amplitudes in the absence of radiative corrections. Achieving the precision demanded by current and future experimental programs, however, requires going beyond this approximation. Since strong isospin-breaking effects and QED corrections both enter at leading order in isospin violation, they must ultimately be incorporated consistently through combined QCD and QED simulations. Such calculations are becoming increasingly important for percent-level determinations of CKM matrix elements, precision tests of flavor symmetries, and stringent comparisons between lattice QCD and experiment.

\subsection{Functional methods}\label{sec:function-methods}

Functional methods are another non-perturbative approach to compute electroweak transitions between baryons. Here, the central quantities of interest are QCD's $n$-point correlation functions, which can be computed via Dyson-Schwinger and Bethe-Salpeter equations (DSEs and BSEs)~\cite{Roberts:1994dr,Alkofer:2000wg,Fischer:2006ub,Bashir:2012fs,Cloet:2013jya,Eichmann:2016yit,Huber:2018ned,Fischer:2018sdj} 
or the functional renormalization group (FRG)~\cite{Berges:2000ew,Pawlowski:2005xe,Dupuis:2020fhh}.
Functional methods are complementary to lattice QCD:
While on the lattice one calculates the correlation functions directly from the QCD partition function,
the strategy with functional methods is to calculate them  from \textit{each other} using 
equations derived from the partition function.
This amounts to coupled integral or integral-differential equations, which in practice must be truncated.
The truncations employed in the literature so far cover a broad landscape ranging  
from simple models to ab-initio  calculations without any model input, see e.g.~\cite{Eichmann:2025wgs,Huber:2025cbd} for recent pedagogical introductions.

    \begin{figure}[h!]
	\centering
 \includegraphics[width=0.9\textwidth]{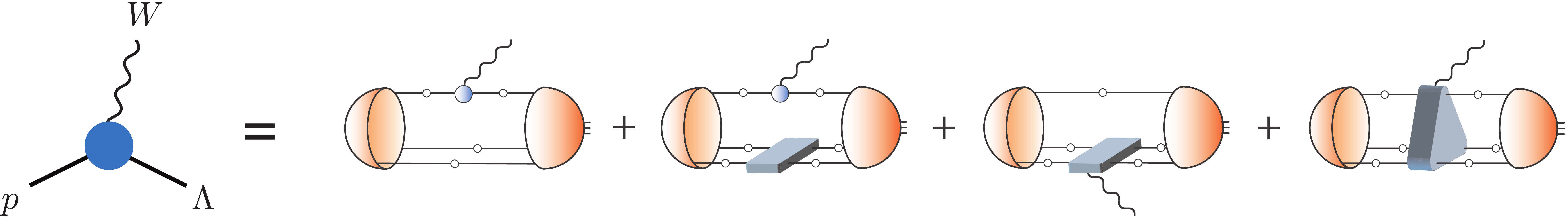}
    \caption{Decomposition of a current matrix element with (vector, axial-vector, pseudoscalar, \dots) currents between the same baryon or describing the transition between different baryons,
    here exemplified for the weak transition $\Lambda \to p e^- \bar{\nu}_e$.}
    \label{fig:fm-current}
    \vspace{0mm}
\end{figure} 

Regarding weak transitions, several studies are available for mesons~\cite{Ivanov:1997yg,Ji:2001pj,Chen:2012txa,Yao:2020vef,Xu:2021iwv,Yao:2021pyf,Yao:2021pdy,Xing:2022sor}, 
while  in the baryon sector, significant foundational progress has been made in recent years regarding semileptonic transitions.
Electroweak transitions between baryons are encoded in the current matrix element shown in Fig.~\ref{fig:fm-current}.
Here, the internal lines represent the quarks, whose propagators for different flavors can be computed from their DSEs.
The half-circles on the left and right are the Faddeev amplitudes, either of the same baryon or different baryons, which 
are obtained along with the baryon spectrum by solving the three-quark covariant Faddeev equation. The external $q\bar{q}$ current that couples to the quark 
can be of any type, such as vector, axial-vector, pseudoscalar, scalar, etc. Therefore, it can also describe
$W$ or $Z$ bosons and mediate flavor transitions. The vertex that describes the coupling to the quark satisfies
a BSE, from which it can  be computed self-consistently. The two- and three-body kernels 
 include all possible interactions between the quarks.
The derivation of the current-matrix element in Fig.~\ref{fig:fm-current} requires no assumptions~\cite{Eichmann:2011vu,Eichmann:2016yit}, so this is \textit{a priori} an exact expression.
The truncations are then made at the level of the two- and three-body kernels.

So far, the approach with the decomposition in Fig.~\ref{fig:fm-current} has been mainly used to compute spacelike elastic and transition form factors of baryons,
either with vector currents in view of electromagnetic form factors~\cite{Eichmann:2011vu,Sanchis-Alepuz:2015fcg,Sanchis-Alepuz:2017mir}, axial-vector currents for axial form factors~\cite{Eichmann:2011pv}, 
and also gravitational form factors~\cite{Yao:2024ixu}. 
There has also been much activity in the quark-diquark approach; if one assumes diquark clustering inside a baryon, the three-body equations reduce to two-body equations, which yield a similar current decomposition
and provide additional information on the various diquark contributions to the form factors~\cite{Eichmann:2016yit,Barabanov:2020jvn}.
This approach has recently been applied to axial-vector and pseudoscalar form factors of the nucleon and the $\Delta$-baryon~\cite{Chen:2021guo2,Chen:2022odn,Yin:2023kom},
demonstrating that the results match those of lattice QCD where there is overlap but also extending them to large $Q^2$.
Moreover, the $N\to\Delta$ axial and pseudoscalar transitions have been studied in~\cite{Mader:2011zf,Chen:2023zhh} and 
octet-baryon axial-vector and pseudoscalar form factors in Ref.~\cite{Cheng:2022jxe}.
Current efforts are also underway to extend such form factor calculations to the timelike region~\cite{Miramontes:2021xgn}.

\begin{figure}[!t]
	\centering
 \includegraphics[width=0.8\textwidth]{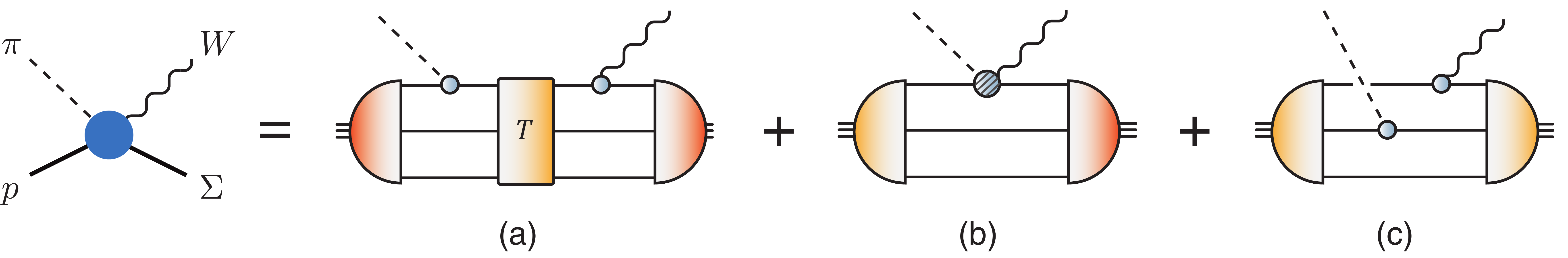}
    \caption{Decomposition of a four-point function, here exemplified for the weak $\Sigma^- \to p \pi^-$ transition.
             Additional permutations and diagrams where the currents couple to the two- and three-body kernels are not shown~\cite{Eichmann:2011ec}.}
    \label{fig:fm-4pt}
    \vspace{0mm}
\end{figure}

The approach has also been extended to four-body scattering amplitudes involving hadrons or external currents~\cite{Eichmann:2011ec}. The derivation proceeds along the same lines as for current-matrix elements and leads to the diagrams in Fig.~\ref{fig:fm-4pt}, where it is exemplified for the weak $\Sigma^- \to p\pi^-$ transition. Among the different classes of diagrams are those with an intermediate six-point function (a), which contain all possible baryon resonances; those with a four-point function (b), which contain handbag-like topologies and $t$-channel meson poles, and couplings to different quark lines (c). 
Compared to the extensive work on  form factors in the DSE/BSE approach, however, apart from a few applications~\cite{Eichmann:2012mp} this framework has remained largely unexplored so far. 

The apparent advantage of the approach with functional methods is that many physically different processes are all described by the same formalism with the same ingredients -- e.g., electromagnetic transitions of hyperons, semi-leptonic decays involving charmed baryons, or even processes involving flavor-changing neutral currents. To obtain quarks with different flavors, one solves the quark DSE for different current-quark masses. For the vertex that couples the current to a given quark flavor or mediates a flavor transition, one solves the respective vertex BSE using those quark propagators as input. To obtain baryons with different flavors, one solves the Faddeev equation where those quark propagators enter. In that way, observable phenomena can be related to and explained by the nonperturbative quark-gluon dynamics of QCD. This establishes an exciting prospect to explore a variety of electroweak transitions using functional methods.

\input{experiment-status}

\section{Selected \texorpdfstring{$u$-$d$}{u-d} semi-leptonic transitions }

\begin{figure}[h!]
	\centering
 \includegraphics[width=0.75\textwidth]{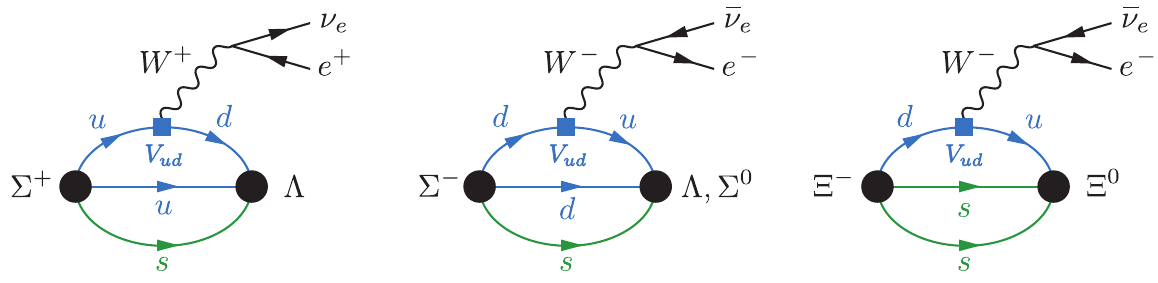}
    \caption{Relevant $u-d$ transitions involving hyperons.}
    \label{fig:ud-transitions}
    \vspace{0mm}
\end{figure}

The value of $|V_{ud}|$ governing such processes comes from the study of superallowed $0^+\to0^+$ nuclear beta decays, which are pure vector transitions, where the average of the most precise determinations gives $|V_{ud}|=0.97367(32)$, i.e., $3\times10^{-4}$ uncertainty. A less precise determination is obtained from the measurement of the neutron lifetime, but it is limited by the knowledge of the ratio of the axial-vector and vector couplings. Measurement of  $\pi^+\to \pi^0 e^+\nu$ branching ratio to 0.6\% from PIBETA  \cite{Pocanic:2003pf} is very clean theoretically as a pure vector transition and is free from nuclear-structure effects.  The precise knowledge of $|V_{ud}|$ allows one to use the 
semileptonic strangeness-conserving processes for precision studies of the form factors and for an unambiguous comparison with  lattice QCD calculations. One particular motivation is neutrino cross sections in  quasi-elastic nucleon interactions and the related $q^2$ dependent form factors, which are the necessary ingredients in neutrino detection~\cite{Formaggio:2012cpf,Meyer:2022mix}. Here, not only is the knowledge of  elastic transitions $\nu n\to p e^-$ needed, but also of processes such as  $\nu n\to p\pi^0 e^-$ as a function of energy. Related form factors could also be studied in the $u-d$ transitions in the hyperon sector, such as the $\Sigma\to \Lambda$ transition; see Fig.~\ref{fig:ud-transitions}. 

\subsection{Experimental status}

Since $u-d$ transitions have small excess energy, only electron decay modes are possible. Most measurements of the hyperon transitions branching fractions predate 1970 and suffer from limited statistics. Two results stand out: the 1982 CERN SPS measurement of $\Sigma^-{\to}\Lambda e^-\bar{\nu}_e$ with 1620 events and a substantially smaller uncertainty than any other~\cite{Bristol-Geneva-Heidelberg-Orsay-Rutherford-Strasbourg:1981uuv, Bristol-Geneva-Heidelberg-Orsay-Rutherford-Strasbourg:1983rna}, and the 2023 BESIII measurement of $\Sigma^+{\to}\Lambda e^+\nu_e$ based on $15.7\pm4.0$ events~\cite{BESIII:2022lsr}. The ratio of the $\Sigma^-\to\Lambda$ to $\Sigma^+\to\Lambda$ partial decay widths is $1.0(4)$  using these results. Alternatively, using the PDG averages, which include pre-1970 results, gives $1.4(2)$. For the ratios, the decay rates are obtained from the respective branching fractions using PDG lifetimes~\cite{ParticleDataGroup:2024cfk}. These two determinations are in tension, and both imply $10\%$-level isospin-breaking effects in the axial charge, yielding $\Delta_{g}$ of $21(10)\%$ and $10(8)$\%, respectively. For the decay $\Xi^-\to\Xi^0e^-\bar{\nu}_e$ the upper limit is set to be $2.59\cdot10^{-4}$ at 90\% confidence level~\cite{BESIII:2021emv}.

\subsection{Lattice QCD status} 

The charged-current transition $d \to u$ is realized in the baryon sector through neutron beta decay $n \to p\,\ell^- \bar{\nu}_\ell$,
which provides one of the most precise low-energy tests of the Standard Model. The hadronic structure of this process is encoded in the nucleon weak matrix elements, commonly expressed in terms of vector and axial-vector form factors or, in the zero–momentum-transfer limit, in terms of the nucleon charges. In particular, the isovector axial charge $g_1$ controls the dominant axial contribution to neutron $\beta$ decay and is a key input for determining the CKM matrix element $|V_{ud}|$.

Over the past decade, substantial progress in lattice QCD calculations has led to increasingly precise determinations of nucleon charges and form factors. Improvements in controlling the dominant systematic uncertainties—most notably excited-state contamination, discretization effects, and simulations performed directly at or near the physical pion mass—have enabled precision determinations of the nucleon axial charge, axial form factors, and electromagnetic form factors~\cite{Walker-Loud:2019cif,Bali:2023sdi,Djukanovic:2024krw,Alexandrou:2024ozj,FlavourLatticeAveragingGroupFLAG:2024oxs,Jang:2023zts,Djukanovic:2022wru,Alexandrou:2023qbg,Djukanovic:2023beb,Tsuji:2023llh,Alexandrou:2025vto,Alexandrou:2026cnj}. 

In addition to the axial coupling, lattice QCD has determined the scalar and tensor nucleon charges, which parameterize possible non-SM scalar and tensor interactions in beta decay~\cite{Bhattacharya:2016zcn,Hasan:2019noy,Park:2021ypf,Alexandrou:2020okk,FlavourLatticeAveragingGroupFLAG:2024oxs}. These quantities are essential inputs for low-energy searches for new physics using precision beta-decay experiments. The same framework also allows the computation of  weak nucleon form factors governing neutron beta decay, including the axial, induced pseudoscalar, and weak-magnetic contributions~\cite{Jang:2023zts, Park:2021ypf, Alexandrou:2023qbg, Djukanovic:2024krw, FlavourLatticeAveragingGroupFLAG:2024oxs}. Overall, lattice QCD calculations of $d \to u$ are extracting several key observables at the percent-level precision, establishing neutron beta decay and related baryon processes as powerful probes of both SM parameters and possible new physics.

Beyond the nucleon, recent lattice QCD  calculations have been extended to the corresponding isovector charges across the baryon octet, providing a unified nonperturbative framework for comparing the nucleon with the $\Sigma$ and $\Xi$ baryons and for quantifying SU(3)-flavor symmetry breaking in the axial, scalar, and tensor sectors~\cite{Alexandrou:2016xok,Bali:2023sdi}. Such studies are relevant both for baryon structure and for assessing the extent to which SU(3)-based relations can be used in phenomenological analyses of weak decays.

Transitions within isospin multiplets of hyperons and charmed baryons are another area in which lattice QCD is providing predictions for the corresponding branching fractions. Since the mass splitting $\delta$ is very small, the next-to-leading-order (NLO) expansion is sufficient.
Similarly, for the vector form factor $f_1$, one may safely use its SU(3)-symmetric value. The axial coupling $g_1$ has been computed in lattice QCD for the octet baryons~\cite{Alexandrou:2016xok} using gauge ensembles with pion masses heavier than the physical value and, more recently, including an ensemble at the physical point~\cite{Bali:2023sdi}.

To determine the decay rate, an accurate value of the mass splitting $\delta$ is required. Experimental values compiled by the PDG are available only for the octet baryons. For the remaining systems, we assume the scaling relation $\delta \simeq 5,\mathrm{MeV}/m_B$. The branching fraction also depends on the baryon lifetime. The values of these input parameters are summarized in Table~\ref{tab:couplings}.

The expectation is that the mass splitting decreases approximately as $1/m_B$ as the baryon mass increases. Since the decay rate scales as $\delta^5$, even a modest reduction in $\delta$ leads to significant suppression. In addition, the baryon lifetime generally decreases with increasing mass. Consequently,
$
\mathcal{B}_B(d\to u) \propto {\tau_B}/{m_B^5},
$
implying that the branching fraction falls rapidly with increasing baryon mass. As a result, weak transitions within isospin multiplets are expected to be highly suppressed.

\begin{table}[h!]
\caption{For each process given in the first row we give, form left to right,  the vector coupling $f_1^{\rm SU(3)}$, the axial coupling $g_1$ computed either from lattice QCD \cite{Bali:2023sdi} or by inverting Eq.~\eqref{eq:GNLO} using the branching fraction $\mathcal{B}$ as an input, the mass splitting $\Delta^{\rm PDG}$, the relative mass splitting $\delta_B^{\rm PDG}$, $\mathcal{B}^{\rm PDG}$ are from PDG \cite{ParticleDataGroup:2024cfk}. The values with an asterisk are computed from Eq.~\eqref{eq:GNLO} by using the known LQCD and PDG inputs.}\label{tab:couplings}
\begin{tabular}{cccccc}
\hline
    Process&$f_1^{\rm SU(3)}$& $g_1$&$\Delta^{\rm PDG}$ [MeV]& $\delta^{{\rm PDG}}$ [\%]&$\mathcal{B}$ [$\times10^{-4}$]\\
   \hline
     $\Sigma^+\to\Lambda e^+\nu_e$&0&0.7696(20)$^*$&73.687(60)&6.1955(51)&0.23(4)~\cite{ParticleDataGroup:2024cfk}\\
     $\Sigma^-\to\Lambda e^-\bar{\nu}_e$&0&0.6205(27)$^*$&81.69(7)&6.8220(58)&0.573(27)~\cite{ParticleDataGroup:2024cfk}\\
     $\Xi^-\to\Xi^0 e^-\bar{\nu}_e$&-1&$0.267(12)$ \cite{Bali:2023sdi}&6.3(7)&0.477(53)&$2.1(1.2)\cdot10^{-6}$$^*$\\
     $\Sigma^-\to\Sigma^0e^-\bar{\nu}_e$&$\sqrt{2}$&$0.875_{(39)}^{(30)}/\sqrt{2}$ \cite{Bali:2023sdi}&4.807(35)&0.4014(29)&1.278(57)$\cdot10^{-6}$$^*$\\      
    \hline
\end{tabular}
\end{table}

The ratio of the $\Sigma^+{\to}\Lambda e^+\nu_e$ and $\Sigma^-{\to}\Lambda e^-\bar{\nu}_e$ decay rates provides a clean probe of isospin breaking in the axial current, $g_{\Sigma\Lambda}^{u-d}$. Combining the Standard Model prediction with existing data indicates that radiative effects in $g_{\Sigma\Lambda}^{u-d}$ may be as large as $10\%$.  The Standard Model provides a prediction for this ratio.   Starting from the NLO expression for the decay rate \eqref{eq:GNLO} we find,
\begin{equation}\label{eq:cabibbo}
    \Gamma_{\Sigma^\pm\to\Lambda e^\pm\nu}\simeq \frac{G_F^2 |V_{ud}|^2  m_{\Sigma^{\pm}}^5\delta_{\pm}^5}{60 \pi^3}\bigg(1+\Delta_{\rm RC}\bigg)\!\bigg(1-\frac{3}{2}\delta_{\pm}\bigg)\!\bigg(f_{1,\pm}^2+3g_{1,\pm}^2\bigg)\!\bigg(1+\mathcal{O}(\delta_{\pm}^2)\bigg)\,,
\end{equation}
with $\delta_\pm=(m_{\Sigma^\pm}-m_\Lambda)/m_{\Sigma^\pm}$. The vector and axial charges $f_{1,\pm}$ and $g_{1,\pm}$ include inner radiative corrections, while $\Delta_{\rm RC}$ collects the outer corrections. Since real and virtual photon contributions are identical under charge conjugation, the Sirlin factor $\Delta_{\rm RC}$ is the same for $\Sigma^+$ and $\Sigma^-$. Taking the ratio gives
\begin{equation}
    \frac{\Gamma_{\Sigma^-\to\Lambda e^-\bar{\nu}_e}}{\Gamma_{\Sigma^+\to\Lambda e^+\nu_e}}= \bigg(\frac{m_{\Sigma^-}-m_\Lambda}{m_{\Sigma^+}-m_\Lambda}\bigg)^{\!\!5}\!\!\bigg(\frac{2-3\delta_{-}}{2-3\delta_{+}}\bigg)\!\bigg(\frac{f_{1,-}^2+3g_{1,-}^2}{f_{1,+}^2+3g_{1,+}^2}\bigg)\!\bigg(1+\mathcal{O}(\delta_{-}^2\!{-}\delta_+^2,\,\delta_{\pm}^2\alpha)\bigg)\,,
\end{equation}
where higher-order effects enter either as $\mathcal{O}(\delta_-^{2}-\delta_+^{2})\approx0.08\%$, since coefficients are identical up to $\mathcal{O}(\alpha)$, or as $\mathcal{O}(\delta_{\pm}^2\alpha)\approx0.05\%$ when accounting for isospin-breaking effects.
The Ademollo–Gatto theorem suppresses radiative corrections to $f_1$, implying $f_{1,+}\simeq f_{1,-}\simeq f_1\equiv f_{\Sigma\Lambda}^{u-d}$ up to $\mathcal{O}(\alpha^2)$, with deviations from the SU(3) limit only at $\mathcal{O}(\delta_\pm^2)$. Since $f_{\Sigma\Lambda}^{u-d}|_{\rm SU(3)}=0$, the ratio of charges reduces to
\begin{equation}
    \frac{f_{1,-}^2+3g_{1,-}^2}{f_{1,+}^2+3g_{1,+}^2} = \frac{f_{1}^2|_{\rm SU(3)}+3g_{1,-}^2+\mathcal{O}(2\delta_-^2)}{f_{1}^2|_{\rm SU(3)}+3g_{1,+}^2+\mathcal{O}(2\delta_+^2)} = \frac{g_{1,-}^2}{g_{1,+}^2}\bigg(1+\mathcal{O}(2(\delta_{-}^2\!{-}\delta_+^2),\,2\delta_{\pm}^2\alpha^2)\bigg)\,,
\end{equation}
Relating $g_\pm$ via a $\Delta_g$ and using PDG masses yields
\begin{equation}
    \frac{\Gamma_{\Sigma^-\to\Lambda e^-\bar{\nu}_e}}{\Gamma_{\Sigma^+\to\Lambda e^+\nu_e}}\bigg|_{\rm SM}\!\!\!\!\!= (1-\Delta_g)^2\times 1.665(7)_{\rm m_{PDG}}(4)_{\rm NNLO}\quad\text{with}\quad\Delta_g\equiv\frac{g_{1,+}-g_{1,-}}{g_{1,+}}\,,
\end{equation}
where various sources of higher-order terms are combined in quadrature, yielding a systematic uncertainty larger than two per mille due to NNLO corrections.

Since $\Delta_g$ isolates the relative radiative shift between $g_{1,+}$ and $g_{1,-}$, all charge-conjugation–symmetric contributions cancel. The surviving effects stem from quark-mass–induced isospin breaking and consist of two pieces: a pure-QCD term proportional to $m_u-m_d$ and a radiative term proportional to $m_{\pi^+}-m_{\pi^-}$, arising from sea-pion loops in the transition matrix element~\cite{Cirigliano:2022hob}, the latter enhanced by the strong coupling of the axial current to pion states. Both of these are expected to contribute at the percent-level, and thus 10\%-level isospin-breaking effects are unexpectedly large.

\subsection{Opportunities} 
\label{sec:u-d-opp}
New experimental measurements of the absolute decay rates with controlled systematics and meaningful event ratios $\Sigma^\pm\to\Lambda$. The opposite-sign $VA$ interference in the neutrino and antineutrino channels (see Sec.~\ref{sec:(anti)neutrino}) provides asymmetries with distinct theoretical systematics; measuring them would supply complementary probes beyond absolute decay rates. Such measurements are possible at intensity-upgraded tau-charm factories and at Belle~II.
At CBM, $\Sigma^+\rightarrow \Lambda$ transitions can be studied via the exclusive reaction channel $pp\to p\Sigma^+ K^0$, which has a cross section $\sigma \simeq 25\,\mu b$ at a proton kinetic beam energy of $6$~GeV~\cite{Sibirtsev:1998dh}. The $\Sigma^-\rightarrow \Lambda$ transitions can be measured in $p\Sigma^- K^+\pi^-$ final state. Measurements of semileptonic decays can be complemented by studies of the $\Sigma^0\to\Lambda e^+e^-$ Dalitz decay, which would provide an independent constraint on vector form factors of this transition (see Sec.~\ref{sec:EM_tran}) and measurements of hadronic decay modes $\Sigma\rightarrow N\pi$. The latter will provide normalization for the branching ratio estimations and model-independent information on the hyperon production amplitudes. 

On the theory side, lattice QCD can directly determine the quark-mass and radiative isospin-breaking contributions to $g_{\Sigma\Lambda}^{u-d}$, yielding a first-principles value of $\Delta_g$. Phenomenology can refine estimates of pion–loop–induced radiative effects following Ref.~\cite{Cirigliano:2022hob} for this transition and assess additional sources of the observed deviation, including potential non-standard interactions. Space-like form factors for the $u-d$ transitions are needed to predict cross sections and the angular distributions for quasi-free $\nu A$ interactions that are necessary to describe the properties of neutrino detectors.

\section{ Selected \texorpdfstring{$s\to u$}{s to u} semileptonic transitions }
\begin{figure}[!t]
	\centering
 \includegraphics[width=0.75\textwidth]{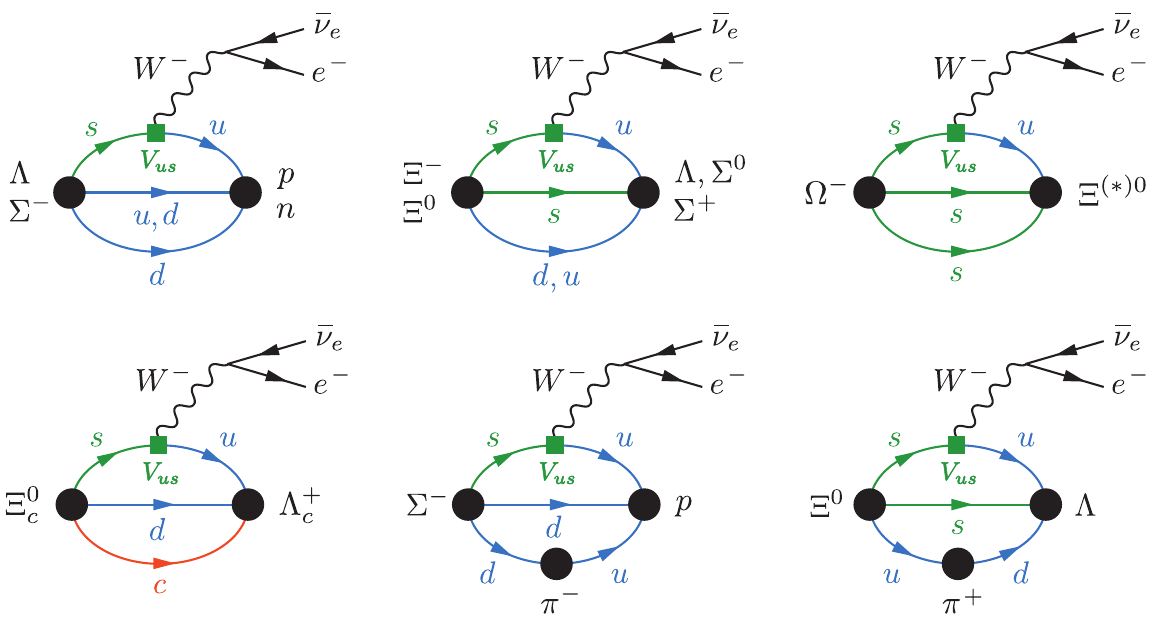}
    \caption{Baryon to baryon $u-s$ transitions}
    \label{fig:us-transitions}
    \vspace{0mm}
\end{figure}

\begin{figure}[htp]
    \begin{center}
    \includegraphics[width=0.8\linewidth]{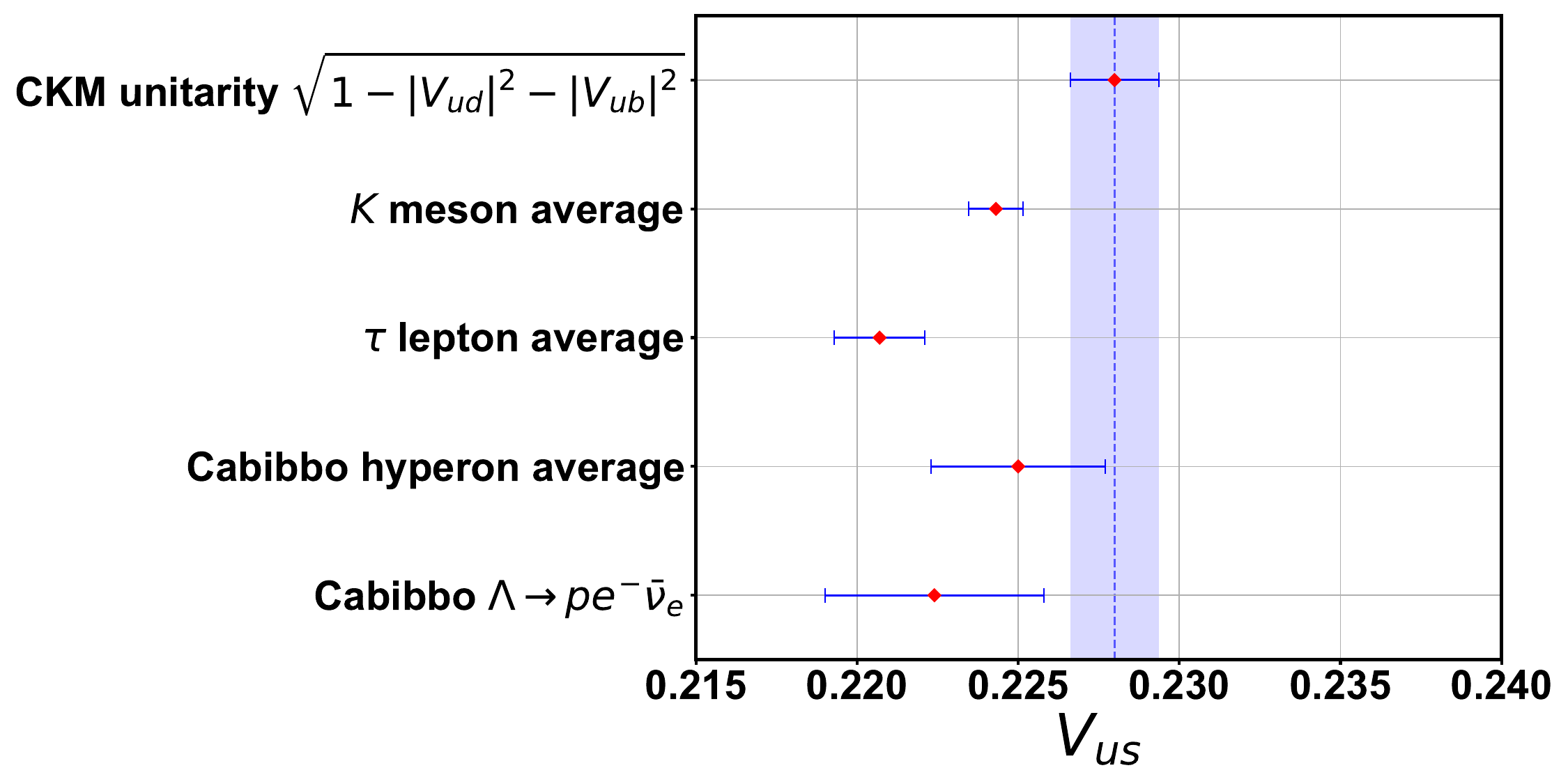}
    \caption{Results on ${|V_{us}|}$. The first-row CKM unitarity constraint and $K$ average are from Ref.~\cite{ParticleDataGroup:2024cfk}. The $\tau$ average is from the Heavy Flavor Averaging Group~\cite{HFLAV:2022esi}. The two hyperon results were evaluated by Cabibbo in~\cite{Cabibbo:2003ea, Cabibbo:2003cu}. 
    }  
      \label{fig:Vus}
      \end{center}
\end{figure}

Baryon semileptonic decays governed by $V_{us}$ are illustrated in Fig.~\ref{fig:us-transitions}. In addition to the transitions within the octet, they include $\Omega^-$ decays as well as $\Xi^0_c\to\Lambda_c^+$ transitions. There is also a limited number of transitions to excited states, represented in the figure by the $\Sigma^-\to p\pi^-$ transition and $\Omega^-\to\Xi^{*}$. A more comprehensive list of the decays is given in Table~\ref{tab:us-trans}.  The most fundamental aspect that could be studied in $s\to u$
processes is the determination of $|V_{us}|$. This value is known from kaon decays $|V_{us}| = 0.22431(85)$~\cite{ParticleDataGroup:2024cfk} and $\tau$ decays $|V_{us}| = 0.2207(14)$~\cite{HFLAV:2022esi}. These are not consistent with each other and are not compatible with $|V_{us}|$ calculated assuming CKM unitarity and using the values of the other first row elements $|V_{ud}|$ and $|V_{ub}|$~\cite{Cirigliano:2022yyo} (Fig.~\ref{fig:Vus}).  Semileptonic decays of strange baryons provide a complementary path to determine $|V_{us}|$. A prerequisite for extracting $|V_{us}|$ from such decays is, in addition to the measurement of branching fractions for the octet hyperon decays, a good knowledge of the form factors. The $f_1$ and $f_2$ form factors are analogous to the electromagnetic Dirac and Pauli form factors.  The value of $f_1$ in the SU(3) limit is $-\sqrt{3/2}$. Since, according to the Ademollo--Gatto theorem~\cite{Ademollo:1964sr}, the first order correction in $m_s-m_{ud}$ is zero, Cabibbo suggested extracting $|V_{us}|$ using the SU(3) value for $f_1$ and determining $g_{av}$ from the experiment~\cite{Cabibbo:2003cu,Cabibbo:2003ea}. The results of such analysis for combined hyperon decays are $|V_{us}|=0.2250(27)$, as shown in Fig.~\ref{fig:Vus}.  In addition, the result based only on the $\Lambda\to p$ electron transition is shown. The  measurement of $g_{av}$ using angular analysis  limits the precision of the determination of \cite{Cabibbo:2003cu}, contributing to as much as 80\% to the overall uncertainty.  With the input from the Lattice QCD~\cite{Bacchio:2025auj} for all form factors, it is possible to extract $V_{us}$ in a model-independent way solely from the $\Lambda\to p$ transition branching fraction measurements. In particular, the lattice calculations give values for $f_1=1.1856(6)$ and $g_{av}=0.690(4)$, which differ from the SU(3) prediction by $-3.3\%$ and $-5.5\%$, respectively. The resulting uncertainty is improved over the Cabibbo method in two ways: the value is shifted due to changes in the form factors, and the precision is increased by not relying on the experimental extraction of $g_{av}$.
 
The relative form factors, such as $g_{av}$ and $g_w$  \eqref{eq:g_q2},  are interesting in their own right, as they encode information on hyperon structure. They can be measured and calculated from first principles using lattice QCD formulation or derived using other theoretical approaches, such as functional methods (Dyson-Schwinger), quark models, or dispersion theories. 

\begin{figure}[htp]
  \begin{center}
      \includegraphics[width=0.95\textwidth]{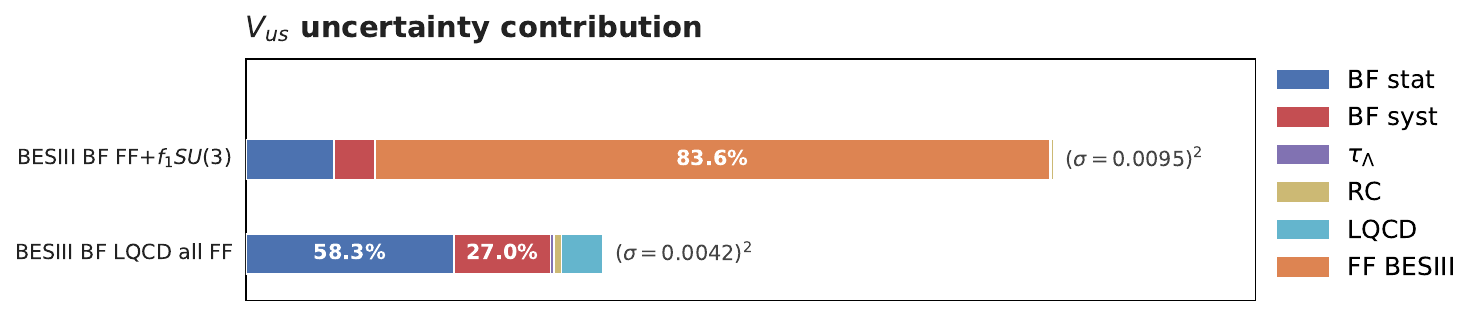}
      \caption{Two ways to extract $V_{us}$ from the BESIII data~\cite{BESIII:2026ehc}, either by using the BESIII form factors, assuming $f_{1}$ in the SU(3) limit (top), or by using form factors that have been determined from Lattice QCD~\cite{Bacchio:2025auj}. The contributing sources to the $|V_{us}|$ uncertainty are given as a fraction  of the total variance. 
      } 
      \label{fig:Vus_variance}
    \end{center}
\end{figure}

\subsection{Experimental status}

Pioneering measurements of $\Lambda\to pe^-\bar{\nu}_e$ were conducted in the 1970s using polarized beams~\cite{Argonne-Chicago-OhioState-WashingtonUniversity:1971hso,Lindquist:1975fc,Burnett:1976fr,Althoff:1971hxl,Althoff:1973my,Baggett:1972jnv,Canter:1971zj,Maloney:1969xd}, albeit with limited statistics. Subsequent experiments~\cite{Wise:1980iq, Dworkin:1990dd}, including the charged-hyperon beam experiment at SPS~\cite{Bristol-Geneva-Heidelberg-Orsay-Rutherford-Strasbourg:1983jzt}, collected larger event samples, enabling measurements of the axial-vector coupling $g_{av}$ for $\Lambda$ decays~\cite{ParticleDataGroup:2024cfk}. This experiment has gathered information on $\Xi^-\to\Lambda e^-\bar{\nu}_e$ and $\Xi^-\to\Sigma^0e^-\bar{\nu}_e$, providing the branching fraction and axial-vector coupling measurements~\cite{Bristol-Geneva-Heidelberg-Orsay-Rutherford-Strasbourg:1983jzt}. The SPS data sample has also been used to study the $\Sigma^-\to ne^-\bar{\nu}_e$ decay, allowing the extraction of the branching fraction and axial-vector coupling~\cite{Bristol-Geneva-Heidelberg-Orsay-Rutherford-Strasbourg:1983jpz}. More recently, the KTeV~\cite{KTeVE832E799:1999tte,KTeV:2001djr,KTeV:2005nvk} and NA48~\cite{NA48I:2006yat,NA481:2012dtx} collaborations studied $\Xi^0\to\Sigma^+e^-\bar{\nu}_e$, determining relative form factors $g_{av}$, $g_w=f_2/f_1$, and $g_{av2}$. Semi-leptonic $\Omega$ decays have been almost unexplored so far, there exists only a branching fraction measurement of $\Omega^-\to\Xi^{0}$ in the electron-mode, $\mathcal{B}(\Omega^-\to\Xi^0 e^-\bar{\nu}_e)=6(3)\times10^{-3}$~\cite{BRISTOL-GENEVA-HEIDELBERG-ORSAY-RUTHERFORD-STRASBOURG:1984jku}. All experiments before 2019 have relied on $\Lambda$ polarization via the decay asymmetry parameter $\alpha_{\Lambda}=0.642(13)$~\cite{ParticleDataGroup:2018ovx}, whereas the current value is 16\% larger $\alpha_{\Lambda}=0.746(8)$ \cite{ParticleDataGroup:2024cfk}.

\begin{table}[ht]
\caption{Semileptonic $u-s$ decays where $^*$ in the first column means that the channel is experimentally favorable since, except for the neutrino, there are only charged particles in the final state. The values for the form factor ratios are experimental. The corresponding muon modes are implied for the transitions where there is sufficient energy excess. $\Delta M$ refers to excess energy for the electron mode. Data for the branching fractions and form factors are taken from~\cite{ParticleDataGroup:2024cfk} except for the $\Lambda\to p$ transition$^\dagger$.}\label{tab:us-trans}
\begin{tabular}{llrllllrr}
\hline
    & Process&$\Delta M$  [MeV]&  & ${\cal B}\ (e$-mode) $\times10^{-3}$&$R^{\mu e}_{\rm th}$  \cite{Alexandrou:2026noh}& \hfill $g_{av}$ & $g_w$ & $g_{av2}$ \\
   & &&  &  & & & & \\
   \hline
     *&{$\Lambda\to pl^-\bar{\nu}_e$}&{$177$}&  & {$0.83(1)$}& $0.162$ & 0.719(14) & 0.97(34)& $-0.2(7)$\\
     &$\Sigma^-\to nl^-\bar{\nu}_e$&$258$& & 1.02(3) & $0.446$ &0.340(17) & 0.97(14) &\\      
     *&$\Xi^-\to\Lambda l^-\bar{\nu}_e$&$206$&  & 0.56(3) &  $0.273$& 0.25(5) & &\\
     &$\Xi^-\to\Sigma^0e^-\bar{\nu}_e$&$129$&  & 0.087(2)& 0.013& & & \\
     &$\Xi^0\to\Sigma^+e^-\bar{\nu}_e$&$126$&  &0.252(8)& $0.008$ &1.22(5) & 2.0(9) & 2(2)\\
    &$\Omega^-\to\Xi^0l^-\bar{\nu}_e$&$357$& & 6(3)& & && \\
    *    &$\Xi_c^0\to\Lambda_c^+e^-\bar{\nu}_e$&$184$&& & --&& &  \\
    \hline
    *  &$\Sigma^-\to p\pi^-e^-\bar\nu_e$&$120$&&&&  && --\\
     *&$\Xi^0\to\Lambda\pi^+e^-\bar\nu_e$&$64$&&&& & & --\\
 *        &$\Omega^-\to\Xi^{*0}e^-\bar{\nu}_e$&$142$&&&&& & --\\ 
         \hline
\end{tabular}
$^\dagger$ For the $\Lambda\to p$ transition: $g_{av2}$~\cite{BESIII:2026ehc}; $g_w$ is  average of~\cite{Althoff:1971mul,Althoff:1973my,BESIII:2026ehc,Bristol-Geneva-Heidelberg-Orsay-Rutherford-Strasbourg:1983jzt}; $g_{av}$ is  average~\cite{ParticleDataGroup:2024cfk,BESIII:2026ehc}.
\end{table}

 At BESIII, large background-free samples of $J/\psi$ decays have enabled dedicated measurements of semileptonic hyperon decays, including channels such as $\Lambda\to pe^-\bar{\nu}_e$~\cite{BESIII:2026ehc}, $\Lambda\to p \mu^-\bar{\nu}_{\mu}$~\cite{BESIII:2021ynj}, and $\Xi^-\to\Lambda e^-\bar{\nu}_e$~\cite{BESIII:2025wov}. These studies allow for absolute branching-fraction measurements and, where statistics permit, the extraction of the form factor ratios. The analyses exploit the well-defined initial state and full event reconstruction using a modular framework~\cite{Perotti:2018wxm, Batozskaya:2023rek}. This approach enables, in principle, reconstruction of the complete decay kinematics, providing excellent control of systematic uncertainties and access to decay dynamics relevant for form factor studies. Recent measurement of the $\Lambda\to pe^-\bar{\nu}_e$~\cite{BESIII:2026ehc} 
 branching fraction based on $1854(49)$ reconstructed events is
 \[
\mathcal{B}(\Lambda\to p e^-\bar{\nu}_e)=
(8.16 \pm 0.22_{\rm stat} \pm 0.15_{\rm syst})\times10^{-4}\ ,
\]
i.e., the overall accuracy is 3\%. From this analysis, $V_{us}$ was determined using both the Cabibbo method and the form factors calculated by lattice QCD~\cite{Bacchio:2025auj}. The extracted values are  0.2194(95) and 0.2339(42), respectively. Contributions to the total uncertainty of the two determinations are compared in Fig.~\ref{fig:Vus_variance}. 

LHCb has recently demonstrated the feasibility of precision semileptonic hyperon-decay measurements in the high-rate environment of a hadron collider by measuring  the branching fraction of $\Lambda\to p\mu^-\bar{\nu}_\mu$~\cite{LHCb:2025wld}. Using $5.4~\mathrm{fb}^{-1}$ of $13~\mathrm{TeV}$ $pp$ data collected between 2016 and 2018, one obtained
\[
\mathcal{B}(\Lambda\to p\mu^-\bar{\nu}_\mu)=
(1.462 \pm 0.016_{\rm stat} \pm 0.100_{\rm syst} \pm 0.011_{\rm norm})\times10^{-4},
\]
 is based on a sample of
$1.637(16)\times10^4$ events. The analysis uses $\Lambda\to p\pi^-$ as a normalization and control channel and introduces a kinematic correction for candidates compatible with $\Lambda\to p\pi^-$, followed by $\pi^-\to\mu^-\bar\nu_\mu$ decay in flight. 
The measured LFU ratio,
$
R^{\Lambda\to p}_{\mu e}
=0.175(12)$, is consistent with the Standard Model expectations and with recent lattice QCD predictions~\cite{Bacchio:2025auj,Chang:2014iba}. Combined with the lattice predictions for
$\Gamma(\Lambda\to p\ell\bar\nu_\ell)/|V_{us}|^2$, the same measurement gives an independent baryonic extraction of $|V_{us}|$. The result is limited by systematic uncertainty on the ${\cal B}$, which is 7\%, while the statistical uncertainty is only 1\%, compared to 2.7\% for the  BESIII result. The experience gained in this analysis allows LHCb to devise methods that will mitigate the dominant systematic effects and could significantly reduce systematic uncertainty.

The $\Xi^-\to\Lambda e^-\bar{\nu}_e$ has also been studied by BESIII~\cite{BESIII:2025wov}. With approximately 110 events, the measurement of the absolute branching fraction $\mathcal{B}=3.6(4)(1)\cdot10^{-4}$ is determined. This value is lower than the world average value~\cite{ParticleDataGroup:2024cfk} by $3.9\sigma$. The axial-vector coupling value $g_{av}=0.18(7)$ is also determined and is found to be consistent with the earlier measurement~\cite{Bristol-Geneva-Heidelberg-Orsay-Rutherford-Strasbourg:1983jzt}.

\subsection{Lattice QCD status} 

Lattice QCD is enabling {\it ab initio} determinations of the form factors governing hyperon semileptonic decays, mediated by the weak transition $s \to u \ell \bar{\nu}$.
Such studies have been carried out for several hyperon semileptonic decay channels , including $\Sigma^- \to n$, $\Xi^0 \to \Sigma^+$, and $\Lambda \to p$~\cite{Guadagnoli:2006gj,Sasaki:2008ha,Sasaki:2012ne,Sasaki:2017jue,Bacchio:2025auj}. Early quenched calculations determined the complete set of six vector and axial-vector form factors and investigated SU(3)-flavor-breaking effects, including the possible size of second-class currents. Subsequent simulations with dynamical quarks concentrated on high-precision determinations of the vector form factor at zero momentum transfer, $f_1(0)$, exploiting the Ademollo--Gatto theorem to isolate the second-order SU(3)-breaking corrections that govern its deviation from unity.

Recent advances in lattice QCD are enabling precise  determinations of the form factors governing hyperon semileptonic decays.  These calculations provide the essential non-perturbative input for the Standard Model predictions of decay rates and angular observables, allowing increasingly stringent comparisons with experiments and more precise determinations of fundamental parameters, most notably the CKM matrix element $|V_{us}|$. 
An example is the first complete lattice QCD determination of the $\Lambda \to p,\ell^- \bar{\nu}*\ell$ form factors obtained directly at the physical point, including the full $q^2$ dependence and the complete set of vector, axial-vector, and second-class form factors~\cite{Bacchio:2025auj}. These results provide fully non-perturbative SM predictions for decay rates, angular observables, and lepton-universality ratios. They also enable a competitive baryon determination of $|V_{us}|$, offering important and complementary input to first-row CKM unitarity tests while providing new opportunities to probe SU(3)-flavour symmetry breaking and the structure of weak baryon transitions.

\subsection{Opportunities }

The hyperon determinations of $|V_{us}|$ are consistent with first-row CKM unitarity, although they are currently less precise than kaon-based measurements. A 1\% statistical uncertainty on the branching ratio at LHCb translates into a sub-percent determination of $|V_{us}|$ using existing data. Since a significant fraction of the current 7\% total uncertainty arises from the limited size of the simulated data sample, there is considerable room for near-term improvement.
The $\Lambda\to p$ and $\Xi^-\to\Lambda$ transitions will be accessible to all experimental setups described here and can therefore be considered golden modes for $|V_{us}|$ extraction among octet to octet transitions.  Many independent measurements would solidly establish $V_{us}$ based on these two transitions. Other hyperon decays, like $\Xi^0\to\Sigma$ and $\Sigma^-\to n$, will be experimentally challenging for LHCb and also for CBM, as they involve neutral particles. The $\Sigma^-\to n$ decay will likely only be accessible in well-defined initial processes like $e^+e^-$, where it is possible to determine the $\Sigma$ kinematics independently of its decay products. The $\Sigma\to n$ decays provide complementary constraints on the $\epsilon_S$-$\epsilon_T$ plane; see Figure~\ref{Fig:WC}.

Transitions involving decuplet baryons provide access to additional form factors \cite{Holmberg:2019ltw,Mommers:2022dgw,Bertilsson:2023htb}.  Since the semileptonic  $\Omega^-$ are unexplored,  any result would be welcomed. As large datasets are needed, presently only LHCb and Belle~II can study such decays. The $\Omega^-\to\Xi^{0}$, reconstructed via $\Xi^0\to\Lambda\pi^0$, would likely only be accessible at Belle~II. A decay mode that would also be feasible at LHCb would be the $\Omega^-\to\Xi^-\pi^+\mu^-\bar\nu_\mu$, where the final states could include both 1/2 and 3/2 spin baryonic states. In particular, $\Omega^-\to \Xi^*(1530) \mu^- \bar{\nu}_\mu$ is the only possible weak process between spin-3/2 baryons~\cite{Mommers:2022dgw, Bertilsson:2023htb}. This would extend the semileptonic-hyperon program to the heaviest strange baryon and provide a unique probe of a decuplet-to-decuplet charged-current transition. Furthermore, the corresponding monitoring channel $\Omega^-\to\Xi^- \pi^+ \pi^-$ mode is poorly known and offers unique insight into non-leptonic transitions between spin 3/2 baryons. 

Other semileptonic transitions to a baryon-meson pair are $\Sigma^-\to p\pi^-e^-\bar\nu_e$ and $\Xi^0\to\Lambda\pi^+e^-\bar\nu_e$. They allow for meson–baryon interaction studies. In particular, the former is related to the unknown $\Lambda-\pi$ interaction, and the latter could be used to validate theoretical approaches.
 
\section{ Selected \texorpdfstring{$c\to s/d$}{c→s/d} semileptonic transitions }

Semileptonic decays of $1/2^+$ charmed baryons are of great interest both experimentally and theoretically. The large mass difference between the charm quark and the three lighter quarks provides substantial phase space for the semileptonic decays, leading to large branching fractions on the order of percent (compared to $\mathcal{O}(10^{-4})$ for hyperons), facilitating their observation and making high-statistics measurements achievable without excessive experimental effort. The mass gap allows for the study of the $q^2$ dependence of the  form factors, enabling comparison of different theoretical parameterizations, like $z$-expansion~\cite{Faustov:2016yza, Bristol-Geneva-Heidelberg-Orsay-Rutherford-Strasbourg:1983jzt}, light-front approach~\cite{Zhao:2018zcb}, pole-dominance model~\cite{Avila-Aoki:1989arc, Perez-Marcial:1989sch}, and other frameworks~\cite{Gutsche:2015rrt, Liu:2009sn, Pervin:2005ve}. In addition, several  decay scenarios to excited hyperons and multibody baryon--meson $S=-1,0$ systems, e.g., $\Lambda_c\to\Lambda^*l\bar{\nu}_l$ and $\Lambda_c\to p\pi^-l\bar{\nu}_l$. They are potentially one of the best places to study isolated baryon-meson systems. Furthermore, lattice QCD has proven that precise predictions can be made for $\Lambda_c$ semileptonic decays~\cite{Meinel:2016dqj}. 

\subsection{Experimental status }
\begin{figure}[ht!]
    \centering
    \includegraphics[width=1.\textwidth]{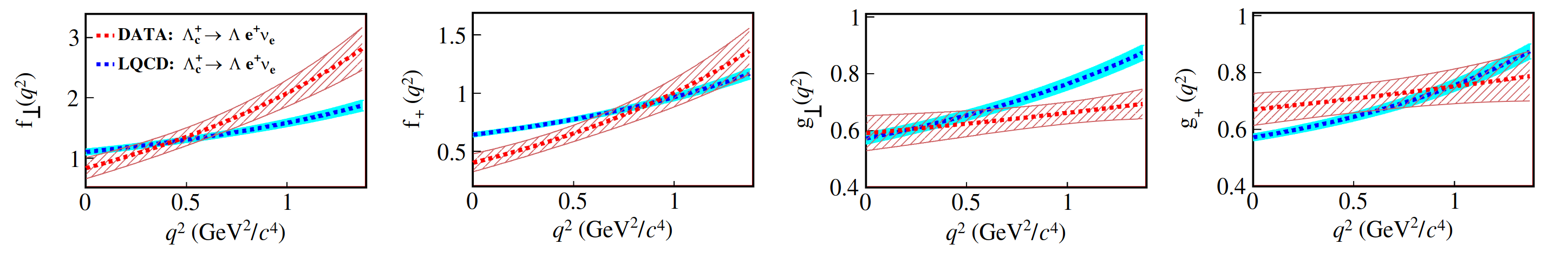}
    \caption{Comparison of vector (left) and axial-vector (right) form factors between BESIII measurement and lattice QCD calculations. The bands indicate the total uncertainties. The figure is taken from Ref.~\cite{BESIII:2022ysa}.}
    \label{fig:Lc2SL_FFs}
\end{figure}

The CKM matrix element $|V_{cs}|$ is predominantly extracted from mesonic processes, namely semileptonic $D\to Kl\nu$ and leptonic  $D_s\to l\nu$ decays, combined with lattice-QCD determinations of the relevant form factors and decay constants~\cite{FlavourLatticeAveragingGroupFLAG:2024oxs, HeavyFlavorAveragingGroupHFLAV:2024ctg}. Earlier averages gave $|V_{cs}|=0.987(11)$~\cite{ParticleDataGroup:2020ssz}, while recent global fits have substantially improved the precision, $|V_{cs}|=0.9691(45)$~\cite{HeavyFlavorAveragingGroupHFLAV:2024ctg}. A complementary but much less precise method is to determine $V_{cs}$ directly from flavor-tagged $W$-boson decays, $|V_{cs}|=0.94(3)$~\cite{DELPHI:1998hlc}. Experimentally, BESIII has accumulated a sample of $\Lambda_c^+\bar{\Lambda}_c^-$ pairs, from which pioneering studies of charmed baryon semileptonic decays have been conducted. A part of this dataset enabled the first form factor measurement in $\Lambda_c^+\to\Lambda l^+\nu_l$ and the initial $|V_{cs}|$ determination for the electron mode~\cite{BESIII:2022ysa, BESIII:2023jxv}. The first complete BESIII analysis of $\Lambda_c^+\to\Lambda e^+\nu_e$, including differential decay distributions and form-factor extraction, gave $|V_{cs}|=0.936(30)$~\cite{BESIII:2022ysa}, consistent with determinations from mesonic SLDs, $|V_{cs}|=0.972(7)$, with 1.2 standard deviations~\cite{ParticleDataGroup:2024cfk}. The analyses of the electron~\cite{BESIII:2022ysa} and muon~\cite{BESIII:2023jxv} modes of $\Lambda_c^+$ SLD reveal intriguing discrepancies with lattice QCD predictions~\cite{Meinel:2016dqj} in the $q^2$ dependence of FFs, as shown in Fig.~\ref{fig:Lc2SL_FFs}. A recent theoretical investigation~\cite{Geng:2022fsr} suggests that these differences may stem from overly restrictive assumptions for form factors, highlighting the need for improved approaches. 
The two BESIII branching measurements of the semileptonic $\Lambda_c^+\to\Lambda$ also allow for a lepton universality ratio, $R^{\mu e}=0.98(5)(3)$, which is consistent with the lattice QCD predicted value of 0.97~\cite{Meinel:2016dqj}. 

BESIII has also conducted searches for SLD transitions from charm to excited hyperon states, which subsequently decay to protons and kaons. A significant signal of the decay $\Lambda_c^+\to pK^-e^+\bar{\nu}_e$ is observed, and evidence for the decays $\Lambda_c^+\to\Lambda(1520)e^+\bar{\nu}_e$ and $\Lambda_c^+\to\Lambda(1405)e^+\bar{\nu}_e$~\cite{BESIII:2022qaf} (see Table~\ref{tab:cs-ground}). In particular, $\mathcal{B}(\Lambda_c^+\to pK^- e^+\nu_e)=2.1(4)\%\times\mathcal{B}(\Lambda_c^+\to Xe^+\nu_e)$ shows that  semileptonic $\Lambda_c^+$ decays are not saturated by the $\Lambda l^+\nu_l$ final state. 
By combining branching fractions of $\Lambda_c^+\to\Lambda(1520)e^+\nu_e$, $\tau_{\Lambda_c}$, and the $q^2$-integrated rate predicted by LQCD~\cite{Meinel:2021grq,Meinel:2021mdj}, one determines $|V_{cs}|=1.3(3)$. This is the first determination of $|V_{cs}|$ from data on baryonic SL decays that proceed via an excited $\Lambda$ state. The ratio of decay widths of inclusive semileptonic $\Lambda_c^+$ and charge-averaged nonstrange charmed mesons
has been measured by BESIII as $1.28(5)$~\cite{BESIII:2022cmg}. BESIII has also searched for the SLD transitions $\Lambda_c^+\to\Sigma^{\pm}\pi^{\mp}e^+\nu_e$~\cite{BESIII:2025kna}. Under the assumption that the two branching fractions are equal under isospin symmetry, evidence for $\Lambda_c^+\to\Sigma^{\pm}\pi^{\mp}e^+\nu_e$ is reported with a significance of $3.6\sigma$.

Several charm-to-strange semileptonic baryon decay results have been obtained by Belle/Belle II. While there is no measurement of $\Lambda^+_c\to \Lambda\ell^+\nu_\ell$ yet, Belle has instead measured the branching fractions of $\Xi_c^0\to \Xi^-e^+\nu_e$ and $\Xi_c^0\to \Xi^-\mu^+\nu_\mu$. This was done using data samples of 89.5 and 711~fb$^{-1}$ recorded at energies of $\sqrt{s}=10.52$~GeV and 10.58~GeV, respectively~\cite{Belle:2021crz}. In total, about $13\times 10^3\, (9\times 10^3)$ events are reconstructed in the electron (muon) mode, and the branching fractions are measured to be $\mathcal{B}(\Xi_c^0\to \Xi^-e^+\nu_e)=(1.31(4)(7)(38)\%$ and $\mathcal{B}(\Xi_c^0\to \Xi^-\mu^+\nu_\mu)=1.27(6)(10)(37)\%$, where uncertainties are statistical, systematic, and from the $\Xi^0_c\to \Xi^-\pi^+$ normalization mode. Also, the decay $\Omega^0_c\to\Omega^-\ell^+\nu_\ell$ has been observed for the first time in 89.5~fb$^{-1}$, 711~fb$^{-1}$, and 121.1~fb$^{-1}$ of Belle data collected at $\sqrt{s}=10.52$~GeV, 10.58~GeV, and 10.86~GeV, respectively~\cite{Belle:2021dgc}. The signal branching fractions relative to the normalization mode $\Omega^0_c\to \Omega^-\pi^+$ are found to be $\mathcal{B}(\Omega^0_c\to\Omega^-e^+\nu_e)/\mathcal{B}(\Omega^0_c\to \Omega^-\pi^+)=1.98(13)(8)$ and $\mathcal{B}(\Omega^0_c\to\Omega^-\mu^+\nu_\mu)/\mathcal{B}(\Omega^0_c\to \Omega^-\pi^+)=1.94(18)(10)$, where the first uncertainty is statistical and the second one is systematic.

Concerning $c$ to $d$ quark transitions, BESIII has been able to determine the branching fraction of the experimentally challenging decay mode $\Lambda_c^+\to n e^+ \nu_e$. An important component is the use of a Graph Neural Network to distinguish between the signal and the more dominant background channel $\Lambda_c^+\to\Lambda(\to n \pi^0) e^+ \nu_e$~\cite{BESIII:2024mgg}. Together with the LQCD calculation and lifetime measurement, the first baryon determination of  $|V_{cd}|$ was found to be  $\left\vert {V}_{cd}\right\vert=0.208(11)_{\rm{exp.}}(7)_{\rm{LQCD}}$, consistent with the world average value $0.221(4)$~\cite{ParticleDataGroup:2024cfk}.

   \begin{table}[h!]
  \caption{Selected semileptonic transitions between charm and strange baryons. }\label{tab:cs-ground}
\begin{tabular}{llrrr}
\hline
    & Process&$\Delta M$& ${\cal B}$ ($e$-mode)& Ref.\\
   & & [GeV]& $\times10^{-2}$&\\
   \hline
     &{$\Lambda_c^+\to \Lambda \ell^+\nu_\ell$}&{$1.17$}& {$3.6(1)$}& \cite{BESIII:2022ysa,BESIII:2023jxv}\\
     &$\Xi_c^0\to\Xi^-\ell^+\nu_\ell$&$1.15$& 1.0(2)&\\
     &$\Xi_c^+\to\Xi^0 \ell^+\nu_\ell$&$1.15$& 7(4)&\\
    \hline
      &{$\Lambda_c^+\to pK^- e^+\nu_e$}&  & {$0.088(18)$} & \multirow{4}{*}{\cite{BESIII:2022qaf}}\\
     &{$\Lambda_c^+\to \Lambda(1520)e^+\nu_e$}& 0.77  & 0.10(5) &\\
     &{$\Lambda_c^+\to \Lambda(1520)[\to pK^-] e^+\nu_e$}& & 0.023(12) &\\
      &{$\Lambda_c^+\to \Lambda(1405)[\to pK^-] e^+\nu_e$}&& 0.042(19) &\\
      \hline
      &{$\Lambda_c^+\to \Sigma^{\pm}\pi^{\mp} e^+\nu_e$}&  & {$0.077(28)$} & \cite{BESIII:2025kna}\\
     \hline
     &{$\Lambda_c^+\to X e^+\nu_e$}&& 4.06(13) & \cite{BESIII:2022cmg}\\
    \hline
    \hline
    &$\Omega_c^0\to\Omega^-e^+\nu_e$& 1.03 &$1.98(29)$& \cite{ParticleDataGroup:2024cfk}\\
    &$\Omega_c^0\to\Omega^-\mu^+\nu_\mu$&1.03&$1.94(21)$ & \cite{Belle:2021dgc}\\
    \hline
\end{tabular}
\end{table}

\subsection{Lattice QCD status} 

Charm semileptonic decays, where $c \to s \ell^+ \nu$,  have been studied in several baryon channels. The transition $\Lambda_c \to \Lambda \ell^+ \nu_\ell$ was computed with the full set of vector and axial form factors and, in some studies, including tensor structures, enabling precise SM predictions for decay rates and branching fractions for electron and muon modes~\cite{Meinel:2016dqj, Bahtiyar:2021voz}. These results provide a determination of $|V_{cs}|$ using baryons and, when combined with experimental data, allow detailed tests of lepton flavor universality in the charm sector.

Studies were further extended to excited states, computing the full set of form factors of $\Lambda_c \to \Lambda^*(1520)\ell^+ \nu_\ell$ and providing predictions for differential decay rates and angular observables~\cite{Meinel:2021mdj}. These results demonstrate that semileptonic charm decays to negative-parity baryons can be treated within a similar framework.

The analogous baryon transition $\Xi_c \to \Xi \ell^+ \nu_\ell$ has also been computed with the complete basis of six form factors governing the semileptonic decay~\cite{Zhang:2021oja, Farrell:2025gis}. These calculations yield SM predictions for branching fractions and lepton universality ratios and enable the extraction of $|V_{cs}|$. The most recent determinations achieve percent-level precision and highlight tensions between lattice QCD calculations and experimentally normalized branching fractions, emphasizing the importance of improved measurements for channels that determine normalization.

The decay $\Lambda_c \to n \ell^+ \nu_\ell$ provides access to the $c \to d$ charged-current transition in the baryon sector. The complete set of vector, axial-vector, and tensor form factors for this channel has been determined, enabling the Standard Model predictions for semileptonic decay rates and offering a determination of $|V_{cd}|$ in the baryon sector~\cite{Meinel:2017ggx}.

\subsection{Opportunities}

The transitions between ground states are the simplest to calculate within lattice QCD. These processes could provide information about  $|V_{cs}|$ and $|V_{cd}|$. Reaching a precision comparable to the benchmark, $\sigma(|V_{cs}|)\sim0.005$~\cite{HeavyFlavorAveragingGroupHFLAV:2024ctg} from the meson sector, will require improved measurements of differential decay rates and polarization observables, more precise lattice QCD calculations of the full set of baryon form factors, and refined theoretical control of SU(3)-flavor symmetry breaking and second-class current contributions. Such progress would establish charmed baryon semileptonic decays as a competitive and theoretically clean probe of the weak current in the baryon sector.
 The form factors for these processes can be computed in lattice QCD, providing a comparison to experiment. In contrast to the situation with hyperon decays, the new aspect is the large range of $q^2$, which requires detailed studies of the $q^2$ dependence of the form factors. Moreover, the integrated value of $R_{\mu e}$ is close to unity, meaning muonic and electronic rates are nearly the same. Concerning tests of lepton-flavor violation, it implies that for sensitive limits, one should study $R_{\mu e}$ as a function of $q^2$.

The main focus so far has been on $\Lambda_c$ and $\Xi_c$ decays, but one can also consider $\Omega_c$ and $\Sigma_c$ decays. Very little is known about the decays of these particles, even for hadronic decay modes. There are also extra  $dsc$ and $usc$ spin-1/2 states, labeled $\Xi_c^{0\prime}$ and $\Xi_c^{+\prime}$. This is because one of the $qq$ pairs can have spin 1 (symmetric), giving a total spin of 1/2 with the third quark. Since the mass difference between $\Xi_c^{\prime}$
and $\Xi_c^{\prime}$ is less than the pion mass, no hadronic decay is possible. This makes it possible to observe semileptonic decays since it is an analogous process to the observed vector meson $D^*_s$ purely leptonic decay $D^{*+}\to e^+\nu_e$\cite{BESIII:2023zjq}.
The large samples of charmed baryons collected by LHCb during Runs~1--3 allow for a broad experimental program for semileptonic charmed baryon decays.
The large excess energy in charm quark semileptonic decays allows for final states that contain mesons in addition to the baryon. In $|V_{cs}|$ transitions, the final state interactions of (multi-) strange baryon-meson systems could be studied free from other interactions with additional hadrons. 
For example, excited strange baryon resonances with spins 1/2, 3/2, and meson-baryon interactions in  $S=-1$ neutral baryon systems such as  $\Lambda-\pi^-$ in $\Lambda_c^+$ decays,  $S=-2$ baryon systems such as $\Xi^-\pi^+$ and   $\Xi^-\pi^+$ in   $\Xi_c^+$ and $\Xi_c^0$  decays, respectively. 
Finally, $S=-3$ baryon systems can be probed in $\Omega_c^0$ decays.

\section{Selected \texorpdfstring{$b\to c/u$}{b→c/u} semileptonic transitions}
Several earlier experiments, such as DELPHI, ALEPH, and CDF, have studied beauty baryons. Today, however, they are produced exclusively at the LHC. Currently, $\Lambda_b$ is the only baryon in which a CP asymmetry has been observed~\cite{LHCb:2025ray}, or for which evidence thereof exists~\cite{LHCb:2024yzj, LHCb:2025mnp}. Table~\ref{tab:bc-ground} lists the existing SLD beauty measurements. While semileptonic $b\to c$ transitions with $\ell=e,\mu$ are the dominant decay modes of bottom baryons~\cite{DELPHI:1995dtx, ALEPH:1997ake}, decay modes with $\ell=\tau$ are also kinematically available. 

  \begin{table}[h!]
  \caption{Semileptonic decays of beauty baryons.  }\label{tab:bc-ground}
\begin{tabular}{llrrr}
\hline
    & Process&$\Delta M$  [GeV]& {\cal B} $\times10^{-2}$ & Ref.\\
   \hline
     &{$\Lambda_b\to \Lambda_c^+ \ell^-\bar{\nu}_\ell X$}&\multirow{4}{*}{$3.33$}& {$10.9(2.2)$}& \cite{ParticleDataGroup:2024cfk}\\
     &{$\Lambda_b\to \Lambda_c^+ \ell^-\bar{\nu}_\ell$}&& {$6.2(1.4)$}& \cite{ParticleDataGroup:2024cfk}\\
     &{$\Lambda_b\to \Lambda_c^+ \tau^-\bar{\nu}_{\tau}$}&& {$1.9(5)$}& \cite{LHCb:2022piu}\\
     &{$\Lambda_b\to \Lambda_c^+\pi^+\pi^- l^-\bar{\nu}_l$}&& {$5.6(3.1)$}& \cite{DELPHI:2003qft}\\
     &{$\Lambda_b\to \Lambda_c(2595)^+l^-\bar{\nu}_l$}&{$3.03$}& {$0.8(4)$}& \cite{CDF:2008hqh}\\
     &{$\Lambda_b\to \Lambda_c(2625)^+l^-\bar{\nu}_l$}&{$2.99$}& {$1.3(6)$}& \cite{CDF:2008hqh}\\
     &{$\Lambda_b\to \Sigma_c(2455)^0\pi^+l^-\bar{\nu}_l$}&\multirow{2}{*}{$3.17$}& \multirow{2}{*}{$0.5(3)$}& \cite{CDF:2008hqh}\\
     &{$\Lambda_b\to \Sigma_c(2455)^{++}\pi^-l^-\bar{\nu}_l$}&&& \cite{CDF:2008hqh}\\
     \hline
     &{$\Lambda_b\to p\mu^-\bar{\nu}_{\mu}$}& {$4.68$}& {$0.04(1)$}& \cite{LHCb:2015eia}\\
    \hline
\end{tabular}
\end{table}

\subsection{Experimental status}
The first beauty baryon semileptonic decay involving a $\tau$ lepton was reported for $\Lambda_b^0\to \Lambda_c^+\tau^-\bar{\nu}_{{\tau}}$~\cite{LHCb:2022piu}. The analysis used 3 fb$^{-1}$ of data at collision energies of 7 and 8 TeV. The $\tau^-$ candidates were formed from their decay via three charged pions and included contributions from the $\tau\to\pi^-\pi^+\pi^-\nu_{\tau}$ and $\tau\to\pi^-\pi^+\pi^-\pi^0\nu_{\tau}$ decays. To cancel out detector-related systematic uncertainties, the selected normalization channel was chosen to be $\Lambda_b\to\Lambda_c^+\pi^-\pi^+\pi^-$. A lepton-flavor universality test was performed by comparing with the corresponding $\mu$ decay mode from the DELPHI experiment~\cite{DELPHI:2003qft}.  The ratio ${R}^{\Lambda_b\to\Lambda_c^+}_{\tau\mu}=0.242(76)$ was found to be consistent with the lattice QCD Standard Model prediction,  $0.3328\pm 0.0074\pm0.0070$~\cite{Detmold:2015aaa}.

From the branching fraction ratios of $\Lambda_b^0\to p^+\mu^-\bar{\nu}_{{\mu}}$ and $\Lambda_b^0\to \Lambda_c^+\mu^-\bar{\nu}_{{\mu}}$ LHCb determined the ratio $|V_{ub}/V_{cb}|$ ~\cite{LHCb:2015eia}. The result  $|V_{ub}/V_{cb}| = 0.083\pm0.004\pm0.004$ was based on 2.0 fb$^{-1}$. The first and second uncertainties arise from the experimental and LQCD predictions, respectively.  The limiting systematic factor comes from the branching fraction determination of the secondary decay, $\Lambda_c^+\to p K^- \pi^+$ from Belle~\cite{Belle:2013jfq}. 

LHCb has also measured the differential decay rate for $\Lambda_b^0\to\Lambda_c^+\mu^-\bar{\nu}_{\mu}$ using 3 fb$^{-1}$ of Run~1 data~\cite{LHCb:2017ruc}. The $\Lambda_c^+$ baryon was reconstructed in the $pK^-\pi^+$ final state, and the measured $q^2$ dependence was used to determine the slope of the Isgur--Wise function~\cite{Isgur:1989vq}. In the same analysis, the contribution from excited charm baryons was constrained using $\Lambda_b^0\to\Lambda_c^+\pi^+\pi^-\mu^-\bar{\nu}_{\mu}$ candidates, providing experimental information on feed-down from $\Lambda_b^0\to\Lambda_c^{*+}\mu^-\bar{\nu}_{\mu}$ decays.

Semileptonic beauty-baryon decays have also been used by LHCb as inputs to production-fraction measurements. In particular, using 13 TeV data, LHCb measured the ratio of the $\Lambda_b^0$ production fraction to the sum of the $B^-$ and $\bar{B}^0$ production fractions with semileptonic decays, finding a sizable $\Lambda_b^0$ component in the LHCb acceptance and a strong dependence on the transverse momentum of the $b$ hadron~\cite{LHCb:2019fns}.

\subsection{Lattice QCD status}
{\bf\boldmath Semileptonic transitions involving $b \to c \ell \bar{\nu}$} in the baryon sector have been studied in lattice QCD for a long time, first using heavy-quark effective theory  (HQET). The first investigations of $\Lambda_b \to \Lambda_c \ell \bar{\nu}$ and $\Xi_b \to \Xi_c \ell \bar{\nu}$ were performed in the quenched approximation within the HQET framework~\cite{Bowler:1997ej}. These works computed the full set of semileptonic form factors and expressed them in terms of the baryon Isgur–Wise function $\xi(\omega)$~\cite{Isgur:1989vq}, which governs the transition near zero recoil. Using this framework, partially integrated decay rates in the region $\omega \in [1,1.2]$ were evaluated, and the slope of $\xi(\omega)$ was extracted to characterize the shape of the differential decay rate. A subsequent quenched study of $\Lambda_b \to \Lambda_c \ell \bar{\nu}$ refined the determination of the dominant vector and axial structures and confirmed the central role of $\xi(\omega)$ as the key nonperturbative input controlling the decay amplitude near zero recoil~\cite{Gottlieb:2003yb}.

With the advent of relativistic heavy-quark actions and dynamical simulations, lattice QCD moved beyond the HQET limit and provided the complete set of QCD form factors for $\Lambda_b \to \Lambda_c \ell^- \bar{\nu}$ over the full kinematic range~\cite{Detmold:2015aaa}. These calculations enabled precise SM predictions for electron, muon, and tau modes separately, allowing the construction of lepton-flavor-universality ratios $R_{\tau\mu}$ for $\Lambda_b \to \Lambda_c \ell \bar{\nu}_\ell$. Building on these results, phenomenological analyses incorporated tensor operators and produced predictions for LFU ratios as well as differential and angular observables, such as forward–backward asymmetries, providing a baryonic probe of possible new physics in $b \to c \tau \nu$ transitions~\cite{Datta:2017aue}.

Lattice QCD has also extended this program to semileptonic decays into negative parity excited charm baryons, specifically $\Lambda_b \to \Lambda_c(2595),\ell^- \bar{\nu}$ and $\Lambda_b \to \Lambda_c(2625),\ell^- \bar{\nu}$~\cite{Meinel:2021rbm}. In these channels, the full set of vector, axial-vector, and tensor form factors was determined near zero recoil and used to predict differential decay rates and angular observables, including forward–backward asymmetries and symmetric  terms. These results provide the first nonperturbative description of $b \to c$ transitions to excited baryon final states and highlight the importance of subleading heavy-quark corrections in describing their kinematic behavior.

{\bf\boldmath Semileptonic baryon transitions mediated by $b \to u \ell \bar{\nu}$}  have been studied primarily through $\Lambda_b \to p \ell^- \bar{\nu}_\ell$. The earliest lattice investigation was performed in the static limit, where one assumes infinite mass for the $b$-quark. In this leading-order HQET limit,  the transition was parameterized in terms of two independent HQET form factors governing the dominant vector and axial structures~\cite{Detmold:2013nia}. The calculation provided the first nonperturbative determination of the differential decay rate in the high-$q^2$ region and established $\Lambda_b \to p$ as a viable baryonic channel for extracting the $|V{ub}|$ CKM matrix element, complementary to mesonic $B \to \pi$ decays.

Subsequently, a fully relativistic heavy-quark calculation determined the complete set of QCD form factors over the entire kinematic range~\cite{Detmold:2015aaa}. These results enabled precise Standard Model predictions for differential and integrated decay rates for electron, muon, and tau modes. Combined with experimental data, they provide an independent determination of $|V_{ub}|$ in the baryon sector and of the ratio $|V_{ub}/V_{cb}|$, contributing to the long-standing inclusive–exclusive tension in $|V_{ub}|$ determinations.

\subsection{Opportunities}

For $\Lambda_b^0\to \Lambda_c^+\mu^-\bar{\nu}_{{\mu}}$, improved determinations of the absolute $\Lambda_c$ branching fraction from BESIII~\cite{BESIII:2015bjk, BESIII:2026qbp} and Belle~II would reduce uncertainties on CKM elements in baryonic $b$-quark transitions. Large beauty-baryon samples from LHCb Runs~1--3 and the Upgrade enable an extensive semileptonic beauty-baryon program.

More precise measurements of the Cabibbo-suppressed decay $\Lambda_b^0\to p\mu^-\bar{\nu}_\mu$ will improve determinations of $|V_{ub}|$ and $|V_{ub}/V_{cb}|$, providing an independent probe of the inclusive--exclusive CKM tension.

Semileptonic beauty-baryon decays with electron, muon, and tau final states allow lepton-flavor-universality (LFU) tests across all three generations within the same hadronic transition. The measurement of $R_{\Lambda_b\to\Lambda_c}^{\tau\mu}$ is currently statistics-limited; the full Run~2 and Run~3 samples will significantly improve its precision. Angular analyses of this spin-1/2 baryon transition complement analogous $B$-meson measurements, probing LFU and possible New Physics~\cite{Ferrillo:2019owd}. Decays to excited charm baryons, such as $\Lambda_b^0\to\Lambda_c(2595)^+\tau^-\bar{\nu}_\tau$ and $\Lambda_b^0\to\Lambda_c(2625)^+\tau^-\bar{\nu}_\tau$, may also become accessible~\cite{Bernlochner:2021vlv,Lupato:2017zow}.


Beyond the $\Lambda_b^0$ sector, semileptonic decays of strange beauty baryons, such as
$\Xi_b\to\Xi_c\ell^-\bar{\nu}_\ell$ and
$\Omega_b\to\Omega_c\ell^-\bar{\nu}_\ell$, remain largely unexplored experimentally. Both $\Lambda_b\to \Lambda_c$ and $\Xi_b\to\Xi_c$ SLD data have been used to determine the polarimeter fields of $\Lambda_c\to p K\pi$ and $\Xi_c\to p K\pi$, respectively~\cite{LHCb:2022sck, LHCb:2025hul}.  
The large data samples available at LHCb offer the possibility to establish a comprehensive program of semileptonic measurements for the full beauty-baryon sector but will require precisely determined branching fractions from the charm and strange sectors used for normalization.

\section{ Electromagnetic transitions of hyperons  }
\label{sec:EM_tran}

\begin{figure}[!htb]
	\centering
 \includegraphics[width=0.25\textwidth]{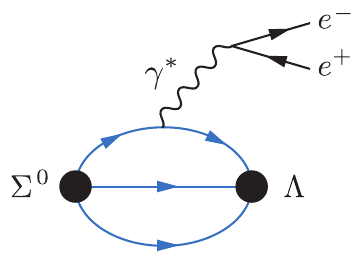}
    \caption{$\Sigma^0\to \Lambda e^+ e^-$ Dalitz decay. }
    \label{fig:dalitz-decay}
    \vspace{0mm}
\end{figure}

A complementary way to extract information on the form factors is to use electromagnetic processes. These processes are historically electron scattering on nucleons $e^- N\to e^- B_2$, providing access to the spacelike range $q^2<0$ and radiative decays $B_1\to B_2\gamma$ where $q^2=0$.  One has two timelike regions: Dalitz decays $B_1\to\ell^+\ell^-$, $4m_\ell^2<q^2<(M_1-M_2)^2$, and electron–positron annihilation $e^+e^-\to B_1\overline{B}_2$, $(M_1+M_2)^2<q^2$. For parity-conserving processes, the form factors given by Eqs. \eqref{eq:matrixelem_semil}, 
\eqref{eq:def-transFF13}, and \eqref{eq:def-transFF13} show that the axial-vector form factors vanish. The weak vector form factors are closely related to the electromagnetic transition form factors of baryons by the conserved vector current (CVC) theorem. For example, the vector transition form factor $f_1$  of the semileptonic transition $\Sigma^+\rightarrow \Lambda e^+ \bar{\nu}$ is directly connected to the electromagnetic transition form factor of the dilepton Dalitz decay $\Sigma^0 \rightarrow \Lambda e^+e^-$, Fig~\ref{fig:dalitz-decay}.  Here we focus mainly on the Dalitz decays, but lattice QCD could also provide information about the spacelike range.  

Dalitz decays of the type $B^*\rightarrow Bl^+l^-$ into dileptons and radiative decays $B^*\rightarrow B\gamma$ probe the electromagnetic structure of baryons 
via their interactions with a virtual or real photon, respectively.
Dalitz decays  have an advantage over the radiative decays because they provide additional information on the $q^2$ dependence of the respective transition form factors in the timelike region. However, the branching ratios are suppressed by $\alpha$ with respect to radiative decays and, consequently, measurements require large production rates. The $q^2$ dependence, provided by the invariant mass of dileptons, is also sensitive to meson-cloud effects, which, in the case of hyperons, are predicted to be governed by pions and kaons~\cite{Granados:2017cib, Husek:2019wmt,Junker:2019vvy,Ramalho:2011pp, Ramalho:2020tnn,Lin:2022dyu}. 
Expressions of differential decay rates for Dalitz decays of $J=1/2$ and $J=3/2$ baryons can be found in \cite {Krivoruchenko:2001jk,Husek:2019wmt,Aung:2024qmf,An:2024pip}. 
They encode vector form factors that are related to electric ($G_E$), magnetic ($G_M$), and, for $J=3/2$, also Coulomb $G_C$ form factors.
For elastic processes, the analogous electromagnetic form factors carry  information on the charge and magnetization distributions inside the hadron,
and one can extract the mean square radii $\langle r_{E,M}^2\rangle$ by measuring their slopes 
in the limit $q^2\rightarrow 0$. 

    \begin{figure}[!b]
	\centering
      \includegraphics[width=0.8\textwidth]{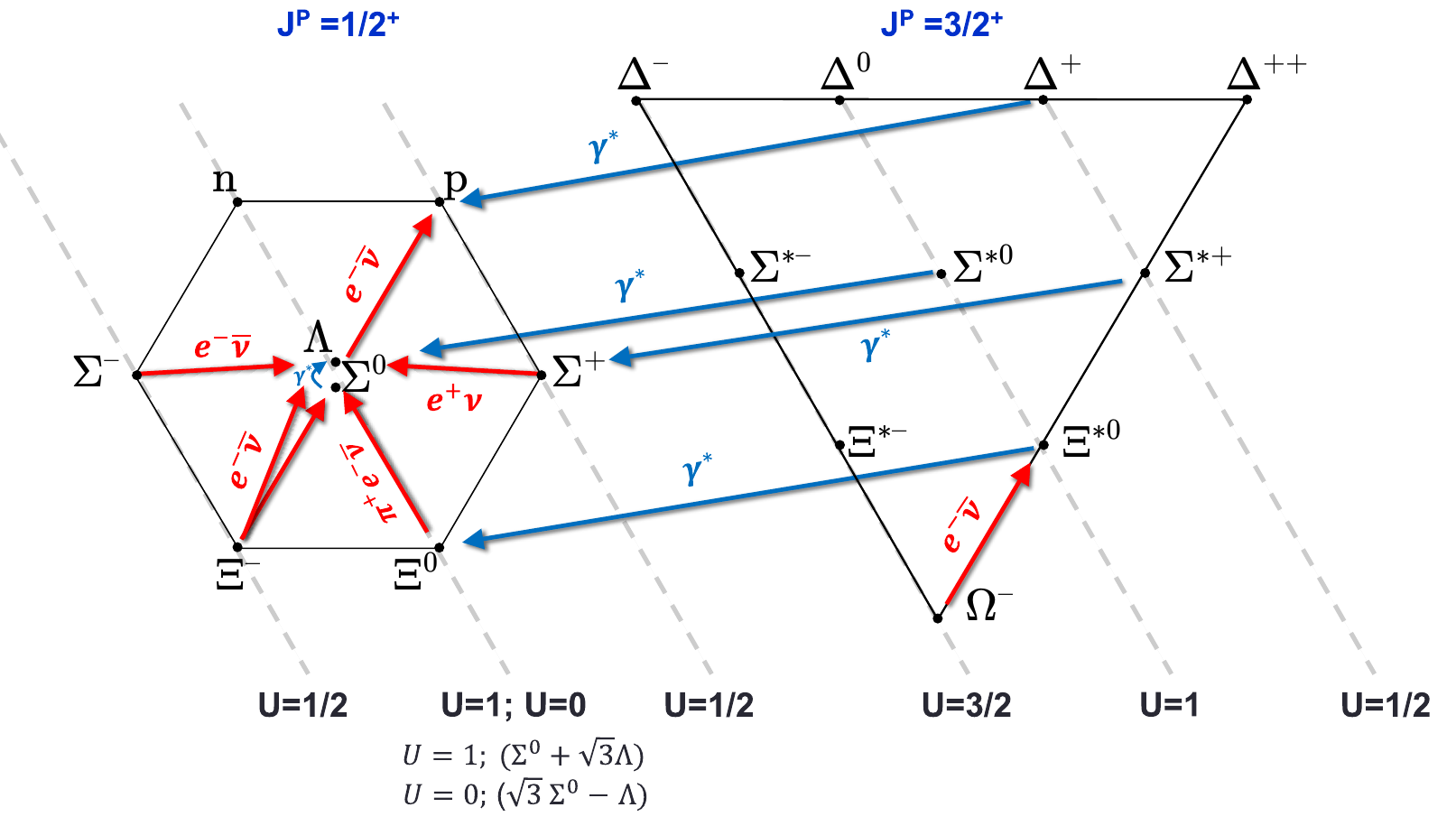}
    \vspace{5mm}
    \caption{Semi-leptonic and Dalitz transitions in octet and decuplet baryons that can be measured in proton-proton collisions in exclusive final states, with all charged states detected and a missing neutrino only (red lines). U spin is indicated by dashed lines and U-spin-allowed electromagnetic transitions by blue arrows.}
    \label{fig:ODtrans}
    \vspace{0mm}
\end{figure} 

A useful concept for hyperons belonging to the ground-state $SU(3)$ multiplets is U-spin symmetry.
U-spin describes the  $SU(2)$ subgroup of $SU(3)$ that refers to the interchange of $d$ and $s$ quarks with the same electric charge
(in contrast to isospin, which swaps $u$ and $d$ quarks). 
For example, as illustrated in Fig.~\ref{fig:ODtrans}, the proton and $\Sigma^+$ in the octet form a U-spin doublet with $U=1/2$, as do the $\Delta^+$ and $\Sigma^{\ast +}(1385)$  in the decuplet.
While isospin symmetry is approximately realized in nature, U spin is useful because the photon is a pure U-spin scalar (it carries $U=0$).
This implies that electromagnetic transitions only between states belonging to the same U-spin multiplets (those with $\Delta U=0$) are allowed,
 while all others are forbidden in the $SU(3)$ limit. With $SU(3)$ breaking, those are still strongly suppressed, and the suppression is governed by the difference in the masses of the $d$ and $s$ quarks. Electromagnetic transitions between the doublets $\{ p, \Sigma^+\}$ and $\{ \Delta^+, \Sigma^{\ast +}\}$ are therefore allowed and also related to each other,
 so one can infer information on the decay ${\Sigma^\ast}^+\to \Sigma^+ \gamma^\ast$ from its partner process $\Delta \to p \gamma^\ast$.
 Similar relations hold for the $U=1$ triplets $\{ \Xi^{\ast 0},\Sigma^{\ast 0},\Delta^0\}$ in the decuplet and $\{\Xi^0,\Lambda/\Sigma^0,n\}$ in the octet  
 (note that $\Sigma^0$ and $\Lambda$ are mixed $U=1$, $U=0$ states).

 \subsection{Experimental status}
 
 The HADES experiment at FAIR/GSI pioneered measurements of Dalitz transitions for non-strange baryons. The $\Delta \rightarrow N e^+e^-$ transition was measured for the first time, determining the branching ratio of $4.2(7)\cdot 10^{-5}$ and the angular distributions that show a dominant magnetic M1-contribution~\cite{HADES:2017jkt}. Meson-cloud effects are clearly visible in the mass-dependent form factor and agree with model predictions assuming contributions from the pion vector form factor. This effect is even more visible for the $N^\ast$ resonances in the second resonance region, with the dominant contribution coming from the $N^\ast(1520)$ state. The mass dependence of the effective form factor has been measured up to a dilepton invariant mass of 500 MeV/c$^2$ and reveals a very strong contribution from the $\rho$ meson that can be related to the pion electromagnetic form factor \cite{AbouYassine:2022hmt}, in agreement with predictions from quark models~\cite{Ramalho:2016zgc} and calculations based on effective field theories~\cite{Zetenyi:2020kon}  
 and dispersion theory~\cite{An:2024pip}.  

Regarding Dalitz decays of hyperons, initial pioneering measurements were achieved by the HADES experiment at SIS18, which identified dilepton production tagged by $\Lambda$ in $pp$ interactions at a kinetic beam energy of 4.5~GeV~\cite{Sumara:2024stj, Rieger:2026cxs}. In particular, the Dalitz decay $\Sigma^0\to\Lambda e^+e^-$ was reconstructed for the first time by BESIII, with the absolute branching fraction determined to be $\mathcal{B} = 6.34(34)\times 10^{-3}$~\cite{BESIII:2026ubj}.  The partial decay width is proportional to the magnetic form factor squared, $|G_{M}(0)|^2$.
Therefore, it is possible to determine the magnetic moment of the $\Sigma^0\to \Lambda$ transition as $\mu_{\Sigma-\Lambda} = 1.74(9)\mu_{N}$, where $\mu_N$ is the nucleon magnetic moment.

\subsection{Functional methods status}   

In contrast to semileptonic baryon decays, electromagnetic transitions of baryons have been extensively studied using functional methods; see, e.g., the reviews~\cite{Eichmann:2016yit,Burkert:2017djo,Barabanov:2020jvn,Ding:2022ows,Ramalho:2023hqd,Achenbach:2025kfx,Burkert:2025coj}.
The underlying diagrams are the same as in Fig.~\ref{fig:fm-current}, with the $W$ boson replaced by a photon.
While  the photon virtuality $Q^2 = -q^2$ can, in principle, be arbitrary, most  calculations have focused on spacelike values $Q^2 \geq 0$ or moderate timelike values to avoid the need for contour deformations in the integrations.
Because the photon can fluctuate into vector mesons, the timelike behavior is dominated by vector meson bumps.
The corresponding vector-meson poles are dynamically generated when the quark-photon vertex is calculated self-consistently~\cite{Maris:1999bh},
including the resonance behavior of the $\rho$ meson~\cite{Miramontes:2021xgn}, which makes vector-meson dominance an automatic feature. 
Concurrently, the nonperturbative physics of three valence quarks interacting by gluons is also implicitly captured; it is responsible for 
anomalous magnetic moments and charge radii,  potential zero crossings in excited states, and the power-law falloff of form factors at large $Q^2$.
Although the decomposition in Fig.~\ref{fig:fm-current} is general, in most truncations considered so far, the quark-gluon topologies that are responsible for creating meson-cloud effects are absent. These  can be significant at low $Q^2$, which suggests interpreting the results so far
as a `quark core' without meson-baryon effects.

Examples include calculations for the $\Delta\to N\gamma^\ast$ transition~\cite{Eichmann:2011aa,Segovia:2014aza,Lu:2019bjs}, which indicate
that the magnetic dipole ($G_M$) form factor is sensitive to meson-cloud effects, while the electric ($G_E$) and Coulomb ($G_C$)
quadrupole ratios are well described owing to relativistic orbital angular momentum in the nucleon and $\Delta$ wave functions~\cite{Eichmann:2011aa}. 
Calculations for the nucleon-to-Roper transition successfully predict a zero-crossing in the form factor~\cite{Segovia:2015hra,Chen:2018nsg},
which is a feature also seen in experiments and is characteristic of a first radial excitation.
A variety of other nucleon-to-resonance transitions have been investigated within this framework~\cite{Barabanov:2020jvn,Ding:2022ows,Achenbach:2025kfx,Burkert:2025coj,Raya:2021pyr}, including axial transition form factors~\cite{Chen:2023zhh}. 
While these are mostly quark-diquark calculations, the results for elastic and transition form factors are typically very similar to three-quark calculations~\cite{Eichmann:2016yit}.
Three-quark calculations are available for electromagnetic transitions of hyperons including $\Sigma^\ast \to \Sigma \gamma^\ast$, $\Sigma^{\ast,0}\to\Lambda\gamma^\ast$, $\Sigma^0 \to \Lambda \gamma^\ast$, and $\Xi^\ast \to \Xi \gamma^\ast$ for spacelike momenta~\cite{Sanchis-Alepuz:2015fcg,Sanchis-Alepuz:2017mir}.
A wide array of electromagnetic transitions has also been studied in the
covariant spectator approach, including timelike form factors~\cite{Ramalho:2023hqd}.

\subsection{Lattice QCD status}

The first dedicated lattice QCD studies of baryon electromagnetic transitions considered SU(3)-flavor octet-to-decuplet transition moments in the quenched approximation, including the $N\gamma \to \Delta$ transition and hyperon octet-to-decuplet radiative transitions~\cite{Leinweber:1992hy,Draper:1992bn}. These works implemented the methodology to extract transition matrix elements from three-point correlation functions and determined the dominant magnetic-dipole transition amplitudes, together with estimates for the electric-quadrupole to magnetic-dipole ratios. This  provided the first  lattice QCD study  of baryon radiative transitions of these systems.

The most extensively studied electromagnetic baryon transition in lattice QCD is $N\gamma^\ast \to \Delta$. An approach was developed to extract, beyond the dominant dipole form factor $G_{M1}(Q^2)$, the subleading electric-quadrupole and Coulomb-quadrupole form factors $G_{E2}(Q^2)$ and $G_{C2}(Q^2)$~\cite{Alexandrou:2003ea,Alexandrou:2004xn,Alexandrou:2007dt}. The dominant magnetic form factor controls the radiative strength of the transition, while the quadrupole form factors, usually expressed through the ratios $R_{EM}$ and $R_{SM}$, probe possible deformation in the nucleon-$\Delta$ system. Lattice QCD  calculations found nonzero quadrupole amplitudes with the same qualitative sign as the experiment, providing evidence that such deformation arises from nonperturbative QCD dynamics.

The same methodology developed for the transition $N \to \Delta$ form factors has also been applied to radiative transitions of charmed baryons. Lattice QCD calculations of $\Omega_c\gamma \to \Omega_c^\ast$ and $\Xi_c\gamma \to \Xi_c^\prime$ determined the magnetic transition form factors, transition moments, helicity amplitudes, and radiative decay rates, providing first-principles input for electromagnetic decays in the singly charmed sector~\cite{Bahtiyar:2015sga,Bahtiyar:2016dom}. The program has further been extended to doubly charmed baryons, where spin-$3/2 \to 1/2$ radiative transitions were studied through the magnetic-dipole and electric-quadrupole form factors~\cite{Bahtiyar:2018vub}.

\subsection{Opportunities}\label{sec:EM_tran-opportunities}

Future opportunities to study electromagnetic transitions of hyperons will expand significantly with the upcoming Compressed Baryonic Matter (CBM) experiment at FAIR, while continuing to utilize the ongoing capabilities of the HADES detector. In the forthcoming years,   
 Dalitz transitions of hyperons with $S=-1$ can also be studied by HADES using the pion beam facility at FAIR in reactions $\pi^-p\rightarrow \Lambda/\Sigma^0 K^0$ \cite{Messchendorp:2025men}.
At SIS100,  measurements of Dalitz decays of hyperons will benefit from substantially larger production cross sections, 
larger counting rates, and the vertexing capabilities of the CBM detector. Furthermore, the characteristically narrow widths of hyperons facilitate the clean identification of the decaying states that can be reconstructed in the invariant mass distributions of the decay products (for example $\Lambda e^+e^-$ or $\Xi e^+e^-$),
bypassing the need for complicated partial-wave analyses that are mandatory to disentangle contributions from overlapping, broad non-strange resonances ($N^\ast/\Delta$). In addition, the excellent dielectron mass resolution and acceptance of CBM down to very low masses will allow for a precise determination of the form factor slopes, 
thereby enabling direct measurements of the transition radii $\langle r^2 \rangle$ (as demonstrated for the $\Lambda(1520)\rightarrow \Lambda e^+e^-$ transition in \cite{Messchendorp:2025men}). Furthermore, a set of measurements of Dalitz and semileptonic transitions for the octet and decuplet states, indicated in Fig.~\ref{fig:ODtrans}, would provide a unique opportunity to extract electromagnetic and weak form factors, which, in the case of vector form factors, are interconnected by the Current Vector Conservation. 

Another connection between electromagnetic transitions is provided by the aforementioned U-spin symmetry.         
 According to U-spin symmetry, electromagnetic transitions between $U=1$ multiplets are related to each other. Hence, the $\Delta\rightarrow N$ transition is related to  the $\Sigma(1385)$ and $\Xi^*$ Dalitz decays mentioned above (see Fig. \ref{fig:ODtrans}). The calculations in Ref.~\cite{Ramalho:2020tnn} predict a  $\Sigma(1385)$ branching ratio comparable to that of the $\Delta$ ($\simeq 4.2 \cdot10^{-5}$), but with smaller pion cloud effects; however, no kaon cloud effects were considered. It would be interesting to compare the $\Sigma(1385)^{0,+}\rightarrow \Sigma^{0,+}/\Lambda$  and $\Xi^{*,0}\rightarrow \Xi^0$ transitions allowed by the U-spin symmetry to the forbidden ones $\Sigma(1385)^-\rightarrow \Sigma^{-}$ and $\Xi^{*-}\rightarrow \Xi^-$. Observations of the latter would provide information on SU(3) breaking effects driven by the $d-s$ mass difference, which are also important for semi-leptonic decays.  
 
 On the theory side, calculations of electromagnetic hyperon transitions using functional methods must be developed further to extend their reach into the timelike region.
 This will require advanced contour-deformation techniques, alongside more sophisticated truncations to accommodate the resonant nature of baryons and to capture meson-cloud effects. Recent progress in this direction has been made in the meson sector~\cite{Huber:2021yfy, Miramontes:2021xgn, Miramontes:2025ofw, Ferreira:2025wpu} and can be systematically extended to baryons. 
 Such developments will also provide a quantitative handle on deviations from U-spin symmetry driven by the $d-s$ quark mass difference. 
 As outlined in Sec.~\ref{sec:function-methods}, the same tools can be employed to calculate semileptonic decays, which can provide a unified description of both electromagnetic and weak baryon transitions.

Lattice QCD is now in a good position to compute 
the $\Sigma^0\to\Lambda$ transition that has not yet been investigated. For spin-$3/2$ to spin-$1/2$ transitions, it is essential to verify that the spin-$3/2$ baryon lies below the strong-decay threshold. This threshold depends on both the pion mass used in the simulation and the lattice volume, which determines the discrete momentum spectrum. Whenever the spin-$3/2$ baryon is stable, the matrix elements can be computed assuming a stable particle.  Transitions of hyperons and the $\Delta\to N\pi$  can be considered at physical pion mass and for  moderately large volumes, which for hyperons can be larger than for the $\Delta$.  Recent  advances~\cite{Lozano:2022kfz,Agadjanov:2014kha} have made it possible to study such resonant states  either through the L\"uscher formalism~\cite{Luscher:1990ux} or using the spectral density method~\cite{Patella:2024cto}. Combining these recent complementary advancements, there are now opportunities for lattice QCD  to extract transition form factors and provide valuable input for experimental analyses.

\section{ Weak radiative and dileptonic decays}

  \begin{table}[h!]
  \caption{Weak radiative and dileptonic baryon decays}
  \label{tab:fcnc}
\begin{tabular}{llrrrr}
\hline\hline
    & Process&$\Delta M$& {\cal B} & $\alpha_\gamma $& Ref.\\
   & & [GeV]& $\times10^{-4}$& &\\
   \hline
     &{$\Lambda \to n \gamma $}&{$0.16$}& {$8.3(7)$}&-0.16(11) &\cite{ParticleDataGroup:2024cfk}\\
     &{$\Sigma^+\to p \gamma $}&{$0.23$}& {$10.4(6)$}& -0.69(5)&\cite{ParticleDataGroup:2024cfk}\\
     &{$\Xi^0\to \Lambda \gamma $}&{$0.18$}& {$12.4(7)$}& --&\cite{ParticleDataGroup:2024cfk}\\
     &{$\Xi^0\to \Sigma^0 \gamma $}&{$0.12$}& {$33.3(10)$}& -0.72(5)&\cite{ParticleDataGroup:2024cfk}\\
     &{$\Xi^-\to \Sigma^- \gamma $}&{$0.18$}& {$1.27(23)$}& -0.69(6)&\cite{ParticleDataGroup:2024cfk}\\ 
     &{$\Lambda_b\to \Lambda \gamma $}&{$-$}& {$0.07(2)$}& $0.8^{+0.2}_{-0.3}$&\cite{LHCb:2021byf}\\
     &{$\Lambda_b\to \Lambda^*(\to p K^-) \gamma $}&{--}& {--}&&\cite{LHCb:2024vtc}\\
     \hline
 &      $\Lambda_c^+\to p\mu^+\mu^-$~ & & & & \cite{LHCb:2024hju,LHCb:2025bfy}\\
     &{$\Lambda_b\to \Lambda \mu^+ \mu^- $}&{$-$}& $0.011(3)$&&\cite{ParticleDataGroup:2024cfk}\\
    & $\Lambda_b\to\Lambda(1520)\mu^+\mu^-$& - &  & &\cite{LHCb:2023ptw}\\
     \hline\hline
\end{tabular}
\end{table}

Flavor-changing neutral-current (FCNC) decays are strongly suppressed in the Standard Model since they are forbidden at tree level and further suppressed by the Glashow-Iliopoulos-Maiani mechanism~\cite{Glashow:1970gm}, proceeding only through higher-order electroweak loop amplitudes. In contrast, weak radiative and dileptonic transitions of strange and charmed hadrons are generally dominated by long-distance, non-local contributions that combine the electromagnetic current with four-quark operators responsible for non-leptonic weak interactions~\cite{Jenkins:1992ab,Neufeld:1992np,Borasoy:1999nt,Lach:1995we}. For example, the processes $\Lambda\to n\gamma$ and $\Lambda\to n e^+e^-$ can proceed through intermediate $p\pi^-$ or $n\pi^0$ states, followed by the rescattering process $p\pi^- \to n\gamma$. A unitarity bound for these transitions can be estimated using experimental information on the photoproduction reaction $\gamma n\to p\pi^-$. Radiative and dileptonic decays of beauty hadrons, on the other hand, are predominantly governed by short-distance electroweak penguin and box amplitudes, while non-local effects become important only in specific regions of dilepton invariant mass, particularly near charmonium resonances. Consequently, these transitions are regarded as sensitive probes of physics beyond the Standard Model. In particular, comparisons of electron and muon modes provide stringent tests of lepton-flavor universality. Owing to the spin of both the initial- and final-state baryons, processes of the type $B_1\to B_2\ell^+\ell^-$ provide access to combinations of Wilson coefficients that are not available in mesonic decays~\cite{Boer:2014kda,Golz:2021imq}.

In view of the non-local contributions to the weak radiative transitions, it is not possible to formulate the pertinent matrix elements as local quark operators sandwiched between baryon states. Instead, we employ a general transition operator $T$ that takes the initial state to the final state, with the latter also containing the photon:
\begin{align}
 \langle B_2 \gamma|T |B_1\rangle 
& = \epsilon^*_\mu(q) \bar{u}(p_2) (a + b \gamma_5) \sigma^{\mu\nu} q_\nu u(p_1) 
\label{eq:WRBD}
\end{align}
with the photon polarization vector $\epsilon$ and $q=p_1-p_2$. 
Here we have restricted ourselves to spin-1/2 baryons. The helicity of the photon can only be $\pm 1$. Therefore, the helicity of the initial baryon uniquely determines the helicities of the final baryon and photon. As a consequence, there are only two (complex) numbers $a$ and $b$ in (\ref{eq:WRBD}). The transitions where one baryon has spin 3/2 and the other one 1/2 have four amplitudes. If both baryons have spin 3/2, one has to deal with six amplitudes. The parity violating aspects of the radiative decays are characterized by the decay asymmetry parameter $\alpha_\gamma$, given as
\begin{equation}
    \alpha_\gamma=\frac{2 {\rm Re}(a b^*)}{\vert a \vert^2 + \vert b \vert^2} \ .
\end{equation}

Weak dileptonic decays with a charged lepton-antilepton pair (dilepton) instead of a real photon might be generated by a virtual photon or $Z$ boson, but also by loop effects where the pair does not emerge from the same vertex. Therefore, simply assigning a $q^2$ dependence to the parameters of \eqref{eq:WRBD} is not adequate. We will also refrain from writing down the most general expression for these three-body decays involving dileptons.

\subsection{Experimental status}

Concerning weak radiative decays of hyperons, BESIII has measured absolute branching fractions and decay asymmetries of $\Lambda\to n\gamma$~\cite{BESIII:2022rgl}, $\Sigma^+\to p\gamma$~\cite{BESIII:2023fhs}, and $\Xi^0\to\Lambda\gamma$~\cite{BESIII:2024lio} as shown in table~\ref{tab:fcnc}. The branching fraction of $\Sigma^+\to p\gamma$~ was found to be lower than the world average of earlier results by 4.2$\sigma$. The related rare decay $\Sigma^+\to p\mu^+\mu^-$ has a notable experimental history. In 2005, the HyperCP experiment reported the first evidence, based on three candidates
clustered around the dimuon invariant mass of 214 MeV, suggesting a narrow exotic state labeled $X^0(214)$ contributing to the branching fraction at the level of about $10^{-7}$~\cite{HyperCP:2005mvo}.
Recently, LHCb has reported the most precise measurement to date of the $\Sigma^+\to p\mu^+\mu^-$ branching fraction of $1.08(17)\times10^{-8}$, making this the rarest baryon decay observed to date~\cite{LHCb:2025evf}.  The branching fraction and the shape of the dimuon spectrum are compatible with the SM expectations~\cite{LHCb:2017rdd}. Therefore, the LHCb measurement finds no evidence for the $X^0(214)$ state and sets the limit $\mathcal{B}\bigl(\Sigma^+\to pX^0(214)\bigr)<1.4\times10^{-8}$ at a 90\% C.L.  

Weak radiative decays of charmed baryons have not yet been observed; there are only upper limits for $\Lambda_c^+\to \Sigma^+\gamma$ and $\Xi^0_c\to\Xi^0\gamma$ of $2.5\times10^{-4}$ and $1.7\times10^{-4}$ at 90\% C.L., respectively,  based on the results from BESIII~\cite{BESIII:2022aok} and Belle~\cite{Belle:2022raw}. The $c\to u\ell^+\ell^-$ FCNC transitions were studied in $\Lambda_c^+\to p\mu^+\mu^-$ and 
$\Xi^0_c\to \Xi^-\ell^+\ell^-$ at LHCb~\cite{LHCb:2017yqf,LHCb:2024hju} and Belle~II~\cite{Belle:2023ngs}, respectively. In $\Lambda_c^+\to p\mu^+\mu^-$, the highly suppressed short-distance contribution from the FCNC transition $c\to u\mu^+\mu^-$ is expected to contribute to the branching fraction below $10^{-8}$ in the SM. Instead, the long-distance four-quark transitions, proceeding predominantly via intermediate  mesons $V=\rho,\omega,\phi$ through $\Lambda_c^+\to pV(\to\mu^+\mu^-)$, contribute at the level of $\mathcal{O}(10^{-6})$. LHCb has studied the process in different regions of dimuon mass, thereby examining both resonant and nonresonant contributions~\cite{LHCb:2024hju}. In addition to $\Lambda_c^+\to p\phi(\to\mu^+\mu^-)$, signals with significance above $5\sigma$  are observed in the regions dominated by the $\rho$ and $\omega$ resonances. Under the assumption that each region is dominated by the corresponding resonance, the branching fractions $\mathcal{B}(\Lambda_c^+\to pV)$ are determined to be of $\mathcal{O}(10^{-3})$, while stringent limits are set on the nonresonant FCNC contribution at the level of $\mathcal{O}(10^{-8})$. In addition to the vector meson dominant contributions, LHCb sees evidence for 
$\Lambda_c^+\to p\eta$, followed by the rare decay $\eta\to\mu^+\mu^-$, governed by a two photon intermediate state. The Belle experiment has searched for the decay $\Xi^0_c\to \Xi^-\ell^+\ell^-$. The full Belle data sample of 980~fb$^{-1}$ is analyzed, and upper limits at 90\% C.L. with respect to the normalization mode $\Xi^0_c\to \Xi^-\pi^+$ are set for $\Xi^0_c\to\Xi^0 e^+e^-<6.7\cdot 10^{-3}$ and $\Xi^0_c\to\Xi^0 \mu^+\mu^-<4.3\cdot 10^{-3}$~\cite{Belle:2023ngs}. These searches are limited by the available statistics, which could be doubled by including Belle~II data.

The radiative decay $\Lambda_b^0\to\Lambda\gamma$ provides direct access to the electromagnetic-penguin transition $b\to s\gamma$ in the baryon sector. LHCb reported the first observation of this decay and measured its branching fraction to be $\mathcal{B}(\Lambda_b^0\to\Lambda\gamma)=7.1(1.8)\times10^{-6}$~\cite{LHCb:2019wwi}. In spite of the use of an unpolarized $\Lambda_b^0$ sample, it was possible to determine  $\alpha_\gamma$ from $\Lambda$ polarization using the properties of weak decay $\Lambda\to p\pi^-$, in agreement with the SM prediction~\cite{LHCb:2021byf}.

The multibody decay $\Lambda_b^0\to pK^-\ell^+\ell^-$ is dominated by the $\Lambda_b^0\to pK^-J/\psi(\to\mu^+\mu^-)$ resonant process, which was first observed by LHCb in an analysis motivated by CP violation studies~\cite{LHCb:2017slr}. No evidence for CP violation was found in the analysis normalized as a control channel to reduce production and detection asymmetries.  Using the full Run~1 and Run~2 data samples, LHCb subsequently measured the differential branching fraction and angular coefficients of $\Lambda_b^0\to pK^-\mu^+\mu^-$ decays in bins of both $q^2$ and $m(pK^-)$. The results are consistent with contributions from a mixture of $\Lambda$ resonances with different spin-parity quantum numbers~\cite{LHCb:2024ick}.

The first LFU test in beauty-baryon decays was performed using $\Lambda_b^0\to pK^-\ell^+\ell^-$, with $\ell=e,\mu$, in the region $0.1<q^2<6~\mathrm{GeV}^2$, and for $m(pK^-)<2.6~\mathrm{GeV}$~\cite{LHCb:2019efc}. The LFU observable was measured as a double ratio in which the electron and muon modes were normalized separately to the corresponding $\Lambda_b^0\to pK^-J/\psi(\to\ell^+\ell^-)$ channels, reducing experimental systematic uncertainties. The ratio  $R_{e\mu}^{\Lambda_b\to pK}=1.17(18)$ (with the electron mode in the numerator) is consistent with LFU. 
The decay $\Lambda_b^0\to\Lambda(1520)\mu^+\mu^-$ has emerged as an important rare beauty-baryon channel. LHCb measured its differential branching fraction in five intervals of dimuon invariant-mass squared, excluding the $J/\psi$ and $\psi(2S)$ resonances~\cite{LHCb:2023ptw}.  
While the $\Lambda(1520)$ is a prominent contribution to the $pK^-$ final state, several other $\Lambda$ resonances also contribute. LHCb has performed an amplitude analysis of the $\Lambda_b^0\to pK^-\gamma$ mode, which also constituted the first official observation of this decay~\cite{LHCb:2024vtc}, with a model comprising all thirteen well-established $\Lambda$ resonances, from $\Lambda(1405)$ to $\Lambda(2350)$, plus a non-resonant component. The largest contributions, in decreasing order, arise from the $\Lambda(1800)$, $\Lambda(1600)$, $\Lambda(1890)$, and $\Lambda(1520)$ states, while the largest interference term arises between the $\Lambda(1405)$ and $\Lambda(1800)$ states.

The decay $\Lambda_b^0\to\Lambda\mu^+\mu^-$ is the hallmark FCNC transition in the beauty-baryon sector.   Due to the parity-violating decay $\Lambda\to p\pi^-$, the polarization of the $\Lambda$ can be determined. Therefore, it provides information about form factors  that is not accessible in mesonic $b\to s\ell^+\ell^-$ transitions. LHCb has measured the differential branching fraction and the lepton- and hadron-side forward--backward asymmetries as functions of $q^2$~\cite{LHCb:2015tgy}. A subsequent moment analysis determined a larger set of angular observables in the low-hadronic-recoil region, $15<q^2<20~\mathrm{GeV}^2$~\cite{LHCb:2018jna}. Together with the available lattice-QCD form factors, these measurements provide complementary constraints on the short-distance couplings governing $b\to s\mu^+\mu^-$ transitions.

\subsection{Lattice QCD status} 
Rare decay $\Lambda_c \to p \mu^+ \mu^-$, which proceeds via the flavor-changing neutral current $c \to u \ell^+ \ell^-$, was studied in ref.~\cite{Meinel:2017ggx}. Lattice input allows Standard Model predictions for differential branching fractions and angular observables, including null-test quantities such as the forward–backward asymmetry, which vanishes in the Standard Model and thus provides a clean probe of possible new physics contributions.

Rare baryonic $b \to s \ell^+ \ell^-$ transitions have been extensively studied via $\Lambda_b \to \Lambda \ell^+ \ell^-$. The first lattice calculation in this channel was performed in the static-$b$ approximation, reducing the full QCD structure to two HQET form factors and enabling SM predictions for differential branching fractions in the high-$q^2$ region~\cite{Detmold:2012vy}.

Later work using relativistic heavy quarks provided the full set of ten QCD form factors (vector, axial-vector, and tensor currents), allowing for precision SM predictions of differential rates and complete angular observables in $\Lambda_b \to \Lambda(\to p\pi^-)\ell^+\ell^-$~\cite{Detmold:2016pkz}. These results supply essential nonperturbative input for global fits of Wilson coefficients and establish baryonic rare decays as theoretically clean probes of short-distance physics.

This program has also been extended to excited final states. The decay $\Lambda_b \to \Lambda^*(1520)\ell^+\ell^-$ has been computed with the full set of vector, axial-vector, and tensor form factors~\cite{Meinel:2020owd}. Using these, lattice QCD provides SM predictions for differential branching fractions and angular observables in the high-$q^2$ region, offering an additional baryonic channel to test the flavor anomalies observed in mesonic $b \to s \ell^+\ell^-$ processes.

\subsection{Opportunities}

Numerous rare radiative and conversion hyperon decays studied at BESIII remain statistics-limited, but many could be investigated with significantly greater sensitivity at Belle~II and CBM. At the same time, channels such as $\Sigma^+\to p\gamma^{(*)}$ (including the dielectron mode) and $\Xi^0\to\Lambda\gamma^{(*)}$ are being, or are expected to be, investigated at LHCb. In the charm-baryon sector, studies of $\Lambda_c^+\to p\ell^+\ell^-$ at LHCb could be extended to $\Xi_c^0\to\Lambda\ell^+\ell^-$ using the large available data sets.

Both radiative and lepton-antilepton  flavor-changing neutral-current decays of beauty baryons provide a rich and largely unexplored laboratory to probe short-distance BSM physics. Owing to the spin structure of the initial and final state baryons, these decays provide access to a wider set of angular observables than their mesonic counterparts, offering complementary ways to search for new physics. The decay $\Lambda_b^0\to\Lambda\mu^+\mu^-$ is already experimentally established and can benefit from increasingly precise measurements of differential branching fractions and angular observables with the full Run~2 and Run~3 data sets of LHCb. Decays involving spin 3/2 excited hyperons, such as $\Lambda_b^0\to\Lambda(1520)\mu^+\mu^-$, provide additional observables and complementary sensitivity to possible modifications of the effective Wilson coefficients governing $b\to s\ell^+\ell^-$ transitions. LHCb has recently reported the first angular analysis of this decay using $9\,\mathrm{fb}^{-1}$ of Run~1 and Run~2 data \cite{Felicia}. The leptonic forward--backward asymmetry $A_{\mathrm{FB},3/2}^{\ell}$ and the CP-averaged angular observable $S_{1cc}$ were measured in several intervals of $q^2$ and found to be compatible with SM predictions. Since the measurement is statistically limited, the larger Run~3 data sample should enable substantial improvements in precision and potentially allow a broader set of angular observables to be determined. Furthermore, as demonstrated in ref.~\cite{LHCb:2024vtc}, there are many more $\Lambda^*$ baryon resonances that can be used for probing weak radiative decays. In addition, weak radiative $\Xi_b$ decays have not been explored at all. So far, the only $\mu^+\mu^-$ modes have been studied, but it has been demonstrated that all lepton flavors are accessible.

Rare beauty-baryon decays involving a $\tau^+\tau^-$ pair are 
particularly attractive targets. An analysis of the decay
$\Lambda_b^0 \to pK^-\tau^+\tau^-$ is already being actively pursued
within the LHCb collaboration. The ongoing program considers hadron--hadron, hadron--muon, and muon--muon $\tau$-decay topologies. Looking further ahead, the decay $\Lambda_b^0 \to \Lambda\tau^+\tau^-$ is worth highlighting. Measurements of these channels would extend existing $b \to s\ell^+\ell^-$ studies to tau final states and open the way to tests of lepton flavor universality involving electron, muon, and $\tau$ modes.

Separately, in charged-current semileptonic
$\Lambda_b \to \Lambda_c\ell\bar{\nu}_{\ell}$ transitions,
measurements of the $q^2$-dependent ratios
$R_{\ell\ell'}^{\Lambda_b\to\Lambda_c}(q^2)$, with
$\ell,\ell'=e,\mu,\tau$, would provide complementary tests of lepton
flavor universality in a baryonic environment. Tau modes in these
charged-current transitions provide enhanced sensitivity to
non-SM scalar and tensor interactions and complement
corresponding studies in mesonic semileptonic decays. The rare
$\Lambda_b^0 \to \Lambda\tau^+\tau^-$ channel would, in turn,
complement searches such as $B_s^0 \to \tau^+\tau^-$ while providing
access to a richer set of observables arising from the spin structure
of the baryonic decay.

\section{Prospects}

This final section is intentionally selective: we outline projects in
which the authors are directly involved or for which we can provide
quantitative projections, rather than attempting a comprehensive
catalog of all the opportunities identified in the preceding sections.
The omission of a channel should therefore not be interpreted as a
judgment on its scientific interest or feasibility.

\subsection{Experimental prospects}

Determination of $V_{us}$ from semileptonic decays offers a complementary method to kaon and tau decays.
A model-independent approach should be based on LQCD calculations and precise measurements. 
Although most hyperons can be used to probe $V_{us}$, the best precision is expected for the decay modes $\Lambda\to p\ell^-\bar\nu_\ell$ and $\Xi^-\to\Lambda\ell^-\bar\nu_\ell$. They have relatively large branching fractions (see table \ref{tab:us-trans}) and exclusively charged particles in the final state (sans neutrinos ). Finally, their long life-times allow for a clean separation from combinatorial proton-pion background, which is especially needed for hadronic experiments like LHCb and CBM. Furthermore, the determination of the branching ratios for these decays can be improved by measurements of hadronic and semileptonic decays using the same detector configuration.

\subsubsection{LHCb}

The integrated luminosity of Run~3 is $26.7~\mathrm{fb}^{-1}$, compared to the 5.4~$\mathrm{fb}^{-1}$ collected in Run~2. This implies a relative statistical uncertainty on $\mathcal{B}(\Lambda\!\to p\mu^-\bar{\nu}_\mu)$ of $\sim0.5\%$. The dominant Run 2 systematic uncertainties came from signal modeling, trigger efficiency corrections, and calibration uncertainties stemming from particle identification and tracking. If these were to be reduced to the 2\% level in Run 3, made possible by dedicated trigger lines and larger control samples, the total branching-fraction uncertainty would realistically reach the $\mathcal{O}(2)\%$ level. Combined with the corresponding PDG-value for the electron-mode branching fraction, $\mathcal{B}(\Lambda\!\to pe^-\bar{\nu}_e)=(8.34\pm0.14)\times10^{-4}$~\cite{ParticleDataGroup:2024cfk}, this would imply $\delta R_{\Lambda}^{\mu e}/R_{\Lambda}^{\mu e}\simeq2.6\%$. The same measurement can be translated into an extraction of $|V_{us}|$ using lattice-QCD input~\cite{Bacchio:2025auj}. With a 2\% relative uncertainty of $\mathcal{B}_\mu$, it would give the relative error on $|V_{us}|$ of about 1.0\%. On top of that, the LHCb collaboration can also experimentally determine the semi-leptonic transition form factors from the differential distributions of the $\Lambda\!\to p\mu^-\bar{\nu}_\mu$ channel.

For a competitive \emph{single-experiment} lepton flavor universality test, a precise measurement of the electron mode
$\Lambda\!\to pe^-\bar{\nu}_e$ at LHCb would also be desirable. While electron reconstruction and identification are intrinsically more challenging for electrons compared to muons for LHCb, this channel also benefits from substantially reduced backgrounds with respect to the muon mode. In particular, the dominant peaking contribution from $\Lambda\!\to p\pi^-$ with a pion decaying in flight is strongly suppressed in the electron case by
the helicity suppression of $\pi^-\!\to e^-\bar{\nu}_e$. Moreover, an LHCb determination of both semileptonic modes removes the reliance on $\Lambda\!\to p\pi^-$ as a normalization channel and enables the use of closely aligned trigger and selection requirements for the two final states. In such a strategy, many detector- and analysis-related systematic effects (including production, reconstruction, and common selection efficiencies) would largely cancel in the ratio, making a competitive single-experiment determination of $R^{\Lambda\to p}_{\mu e}$ feasible with Run~3 data.\\
While the $\Xi^-\!\to\Lambda e^-\bar{\nu}_e$ branching fraction has been measured as 
$\mathcal{B}(\Xi^-\!\to\Lambda e^-\bar{\nu}_e)=5.6(3)\times 10^{-4}$~\cite{BRISTOL-GENEVA-HEIDELBERG-ORSAY-RUTHERFORD-STRASBOURG:1984jku} and $3.6(4)\times 10^{-4}$, i.e., 3.9. standard deviations different~\cite{BESIII:2025wov}, the uncertainty in the muon mode is at the $100\%$ level,
$\mathcal{B}(\Xi^-\!\to\Lambda\mu^-\bar{\nu}_\mu)=(4(4)\times 10^{-4}$ based on just one observed event in a bubble chamber experiment dating back to 1974~\cite{Yeh:1974wv}. The muonic decay mode  can be substantially improved with LHCb data and would lead to a more precise constraint on LFU in a hyperon channel other than $\Lambda\to p$, which is important as BSM operators are  strongly channel-dependent~\cite{Chang:2014iba}. The expected number of $\Xi^-\!\to\Lambda\mu^-\bar{\nu}_\mu$ events in Run~3  is $2\times10^{3}$. From the strange physics prospects study at LHCb~\cite{AlvesJunior:2018ldo}, one obtains (i) the production rates in the LHCb acceptance relative to $K_S^0$ and
(ii) the acceptance/selection efficiencies for partially reconstructed semileptonic modes, separately for long (L) and downstream (D) reconstruction. Assuming  that the dominant reducible systematic uncertainties (fit/yield modeling,
trigger-efficiency corrections, and PID/tracking calibrations) can be reduced to the $\sim2\%$ level with dedicated trigger lines and
larger calibration samples (as motivated by the $\Lambda\!\to p\mu^-\bar{\nu}_\mu$ experience)~\cite{LHCb:2025wld}, the total Run~3 precision is expected to reach 3\%.

For an LFU test, one can either use the existing electron-mode branching fraction as external input or determine it with Run 3 data. The LFU ratio is determined by
$R^{\Xi\to\Lambda}_{\mu e}\simeq \mathcal{B}_\mu/\mathcal{B}_e$
The relative uncertainty for ${R^{\Xi\to\Lambda}_{\mu e}}$  will be set by the lepton mode with  the dominant relative uncertainty. Even if $\delta\mathcal{B}_\mu/\mathcal{B}_\mu\simeq3\%$ can be achieved with Run~3 data, the uncertainty remains limited by the current
$\Xi^-\!\to\Lambda e^-\bar{\nu}_e$ precision of  $\delta\mathcal{B}_e/\mathcal{B}_e\simeq 5.5\%$~\cite{ParticleDataGroup:2024cfk}, yielding $\delta R_{\Xi}^{\mu e}/R_{\Xi}^{\mu e}\simeq6\%$.  In contrast, the SM expectation is much more precise: $0.275(14)$~\cite{Chang:2014iba}. Therefore, reaching a relative uncertainty on ${R^{\Xi\to\Lambda}_{\mu e}}$ of $6\%$ with Run~3 data (using the available electron-mode input) would already strongly tighten the LFU constraint in this channel and provide a sensitive and complementary test of LFU in semileptonic hyperon decays. This also calls for a dedicated LHCb electron mode measurement with aligned trigger/selection requirements, similar to the muon mode. Such a measurement would enable a competitive single-experiment LFU test in the $\Xi$ system, representing a substantial improvement.

The feasibility of this program is supported by dedicated simulation and trigger developments. A dedicated model for semileptonic hyperon decays, based on theoretical differential decay rates, has been integrated into the LHCb simulation chain and validated against the theoretical distributions. Dedicated Run~3 trigger lines for $\Lambda\to p\mu^-\bar\nu_\mu$, $\Lambda\to pe^-\bar\nu_e$, and $\Xi^-\to\Lambda\mu^-\bar\nu_\mu$ are available, ensuring that the key ingredients from simulation to online selection are in place. In addition, LHCb plans to exploit differential distributions in $\Lambda\to p\mu^-\bar\nu_\mu$ to obtain the experimental constraints on the form factors.

\subsubsection{BESIII and STCF}
In the perspective of a few years with BEPCII-CW, a four- and ten fold increase in statistics could be expected at $J/\psi$ and $\psi(2S)$, respectively.  This will allow an increase in the precision of SL decays with neutral hadrons, in particular $\Omega^-$ decays. The projected uncertainties on the form-factor ratios $g_{av}^{\Lambda}$ and $g_w^{\Lambda}$ in the decay $\Lambda\to p e \nu$ are given in table~\ref{tab:FF_SLHD}.

\subsubsection{CBM}

\begin{figure}[h!]
    \includegraphics[width=1\textwidth]{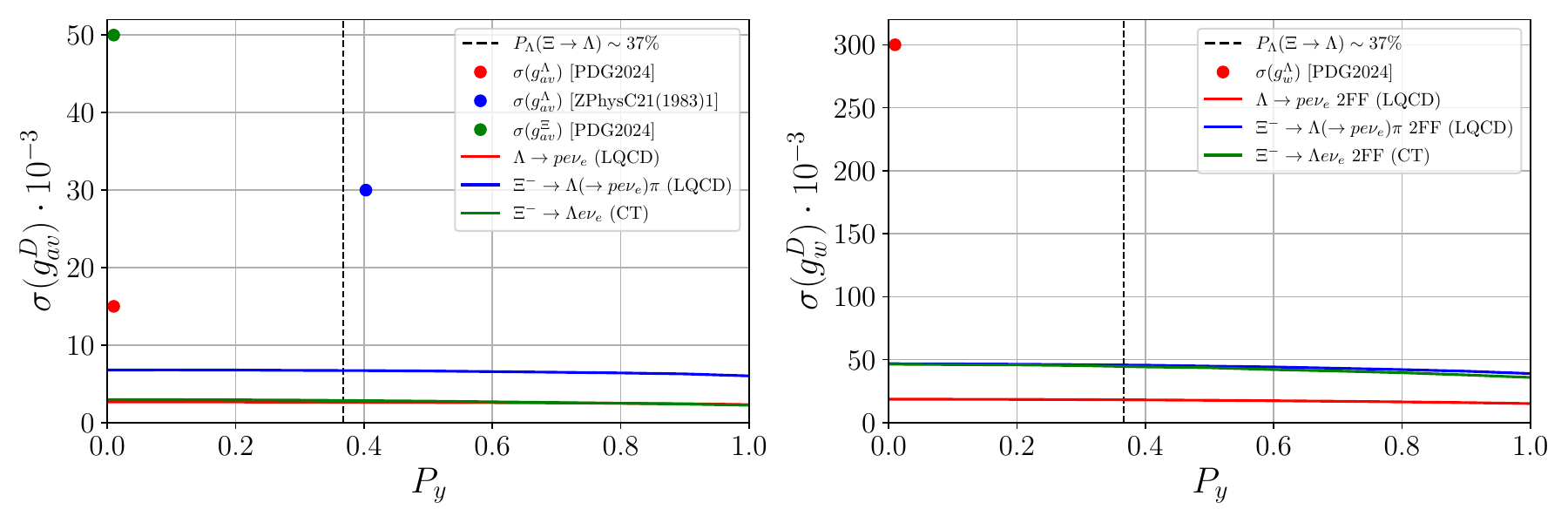}
    \caption{Statistical uncertainties (a) $\sigma(g_{av})$ and (b) $\sigma(g_w)$ for semi-leptonic $\Lambda$ and $\Xi^-$ decays as a function of the initial baryon polarization. LQCD FFs and radii are used as input~\cite{Bacchio:2025auj}. }
    \label{fig:Py-CBM}
\end{figure}

Building upon the modular approach introduced in~\cite{Batozskaya:2023rek}, a dedicated project will be conducted to perform a comprehensive feasibility study of the expected sensitivity to FFs for the "golden" hyperon SLDs: $\Lambda \rightarrow p e^-\bar{\nu_e}$ and $\Xi^-\rightarrow \Lambda e^-\bar{\nu_e}$ under CBM conditions.  To illustrate the predictions of the approach, we present in Figure~\ref{fig:Py-CBM} the statistical uncertainty for $g_{av}$ and $g_w$  for the $\Lambda\rightarrow pe^-\bar{\nu_e}$, $\Xi\rightarrow \Lambda e^-\bar{\nu_e}$ and $\Xi\rightarrow (\Lambda\rightarrow pe^-\bar{\nu_e})\pi^-$ transitions. The estimation is based on the modular approach for the decay of a spin-polarized initial baryon~\cite{Batozskaya:2023rek} and signal statistics of $4\cdot 10^5$, $4\cdot 10^4$, and $6\cdot 10^4$, respectively. Effects of the detector acceptance, reconstruction efficiency, and the background are not yet included. The LQCD prediction for $\Lambda$ FFs and $q^2$ parametrization is used as an input value~\cite{Bacchio:2025auj}, while the Cabibbo prediction~\cite{Cabibbo:2003cu} and experimental results~\cite{ParticleDataGroup:2024cfk} are used for other decays.

At CBM, SLD decays of $\Lambda$ can be measured in proton-proton collisions with a beam of kinetic energy in the range of $5-6$ GeV, where the cross section for the exclusive final state $p \Lambda K^+$ is maximal, with a value of $\sigma=70 \mu b$ \cite{Sibirtsev:1998dh}. For the on-line event selection, $\Lambda$ reconstruction associated with an additional two charged tracks could be used to reduce prompt background from inelastic proton-proton collisions. The respective cross sections for the inclusive $\Lambda$ production and inelastic collisions amount to $\sigma\simeq 0.3-0.4 m b$ and $\sigma_{in}\simeq 30 mb$. Hence, on-line $\Lambda$ tagging should provide at least one order of magnitude of suppression. Cascade production at FAIR energies is very poorly known \cite{HADES:2020pcx}, and the excitation function of the production has to be investigated with CBM to find the maximal cross section energy for the exclusive $pp\Xi KK$ channel. The lowest energy at which $\Xi^-$ production was observed in the exclusive channel $p\Xi^- K^+K^+ $ is at a beam momentum of $10$ GeV/c with a cross section of $\simeq7 \mu b$.

 As explained in sec. \ref{sec:u-d-opp} studies of $u\rightarrow d$ transitions in SL decays of $\Sigma^{\pm}\rightarrow \Lambda$ can be  studied for CBM in combination with the electromagnetic (Dalitz) $\Sigma^0\rightarrow \Lambda e^+e^-$ decays for the extraction of vector form factors. All considered SLD and electromagnetic transitions that can be targeted by CBM are summarized in Fig. \ref{fig:ODtrans}. The strategy of their reconstruction is based on exclusive final states with the reconstruction of all charged particles, except for one neutral particle (neutrino or photon in the case of Dalitz transitions to $\Sigma^0$). It is also important to emphasize that measurements of the leptonic decays of hyperons will be combined with the measurements of hadronic decay channels in the same experiment. Therefore, besides normalization for the branching ratio determination , measurements of hadronic decay channels will provide model independent information on the production of hyperons (i.e., kinematic distributions, polarization) necessary for the form factor extraction.

In summary, the LHCb $s\to u$ hyperon program offers three near-term deliverables: percent-level branching-fraction measurements in the $\Lambda\to p$ and $\Xi^-\to\Lambda$ modes, single-experiment LFU tests through matched electron and muon measurements, and experimental constraints on semileptonic hyperon form factors from differential distributions. In longer perspectives, the CBM experiment at FAIR can contribute to these measurements and complement them with studies of electromagnetic transitions using exclusive production channels in $pp$ interactions. 
Together with lattice QCD input, these measurements provide an independent  route to $|V_{us}|$ and a complementary probe of BSM scalar and tensor charged-current interactions.

\subsection{Lattice QCD prospects}
\subsubsection{Beyond semileptonic decays: axial structure and rare hyperon transitions}

Lattice QCD has determined intrinsic baryon axial couplings, such as $g_{\Sigma\Sigma}$ and $g_{\Xi\Xi}$, providing direct input for SU(3) axial low-energy constants and quantifying SU(3)-flavor symmetry breaking in baryon axial currents~\cite{Savanur:2018jrb}.

In addition, rare hyperon decays, such as $\Sigma^+ \to p \ell^+ \ell^-$, have been recently explored through four-point correlation functions, allowing first-principles access to the long-distance amplitude structure of flavor-changing neutral-current processes in the strange sector~\cite{Erben:2025jkq}. These studies represent an important step toward fully non-perturbative control of rare baryon decays beyond the semileptonic framework.

\subsubsection{Beyond semileptonic decays: coupled-channel scattering and hyperon resonances}

Lattice QCD has recently reached the stage where it can determine hyperon resonance properties from first principles through coupled-channel scattering analyses. In the $I=0$, $J^P=\frac{1}{2}^-$ channel relevant for the $\Lambda(1405)$, the authors of Ref.~\cite{BaryonScatteringBaSc:2023zvt} extracted the coupled $\pi\Sigma$–$\bar{K}N$ scattering amplitude from finite-volume spectra using a coupled-channel formalism. This study determined the pole content of the amplitude and its channel couplings, offering a direct lattice resolution of the long-standing question of the $\Lambda(1405)$ structure. These results demonstrate that lattice QCD can access not only single hadron matrix elements but also multi-hadron dynamics and near-threshold resonance behavior that are essential ingredients of hyperon phenomenology.

Beyond the L\"uscher formalism for treating exclusive decays, new theoretical developments are enabling the study of inclusive processes using spectral densities \cite{Hansen:2017mnd,Bulava:2021fre,Bulava:2021fre,Gambino:2020crt}.
This approach has recently been successfully applied in  tau decays and D-meson decays and mixing~\cite{Evangelista:2023fmt,ExtendedTwistedMass:2024myu,Gambino:2022dvu,DeSantis:2025qbb,DeSantis:2025yfm,DiCarlo:2025mnm}.
Applying this method to the baryon sector is an interesting direction, with applications to inclusive hyperon rates, radiative corrections, and multi-hadron final states. Spectral densities can also include radiative corrections to semi-leptonic decays, such as $\beta$-decay, even from isospin-symmetric simulations~\cite{Seng:2021syx,Feng:2020zdc,Ma:2023kfr,DiCarlo:2026kpv}.

\subsubsection{Beyond SU(2) isospin symmetry}

The vast majority of results produced so far by lattice QCD are based on the assumption of exact isospin symmetry, i.e., degenerate masses for the up and down quarks, and no QED effects. Isospin breaking due to the difference in the masses of the up and down quarks $\Delta m_{ud}$ and QED effects enters at the same order and must ultimately be treated together. In the isospin limit, the light flavor octet reduces to the four independent states $N$, $\Lambda$, $\Sigma$, and $\Xi$, and many electroweak transitions are related by isospin rotations. See Ref. \cite{Bickerton:2019nyz} for relations among isospin multiplets.

Using the spectral density approach can enable us to take into account radiative corrections that can be important for semi-leptonic decays such as $beta$-decay \cite{Seng:2021syx,Feng:2020zdc,Ma:2023kfr,DiCarlo:2026kpv}. This can be achieved by performing simulations in the isospin limit.
Similarly, one can applied the Rome123 \cite{deDivitiis:2013xla,Giusti:2017dwk,Frezzotti:2020bfa} approach to expand to first order in $\Delta m_{ud}$ and the electromagnetic coupling and include isospin corrections to quantities computed using simulations in the isospin limit.

The most consistent way to include isospin breaking is to simulate QCD with $\Delta m_{ud}\neq 0$ plus QED. Various formulations exist to do this \cite{Hayakawa:2008an,Endres:2015gda,Lucini:2015hfa,Hansen:2017zly,RCstar:2022yjz}. Recently, it has been demonstrated that the formalism using $C^*$ boundary conditions yields promising results when using a gauge ensemble with a pion mass that is heavier than physical~\cite{RC:2025zoa}.

\subsection{Prospects with functional methods}

As outlined in Sec.~\ref{sec:function-methods}, a body of baryon form factor calculations has been established using functional methods, including
elastic and transition form factors with electromagnetic, axial, and pseudoscalar currents. 
In parallel, systematic improvements of the underlying building blocks through the quark and gluon correlation functions are underway.
Together, this enables calculations of semi-leptonic decays of a wide range of baryons under the same umbrella,
especially since changing the flavor content of a baryon and tuning the current-quark masses from the chiral limit up to heavy quarks is, in principle, straightforward.
Studies in this direction are  underway and will complement efforts with experiment, phenomenology, and lattice QCD
in our understanding of electroweak transitions of hyperons and charmed baryons.

\section*{Acknowledgments}
  C.~Alexandrou, S.~Bacchio and A.~Konstantinou acknowledge partial support from the projects  IMAGE-N (EXCELLENCE/0524/0459), MuonHVP (EXCELLENCE/0524/0017), PulseQCD (EXCELLENCE/0524/0269), StrongILA (EXCELLENCE/0524/0001), RobustSigma (EXCELLENCE/0925/0325), and partonWF (VISION ERC/0525/0010) funded by the European Regional Development Fund and the Republic of Cyprus through the Cyprus Research and Innovation Foundation. C.A. is partly supported by European Union’s HORIZON MSCA Doctoral Networks programme, under Grant Agreement No. 101072344, project AQTIVATE (Advanced computing, QuanTum algorIthms and data-driVen Approaches for science, Technology and Engineering). A.Kupsc and V.Batozskaya acknowledge partial support from the Polish National Science Center through Grants Nos. 2024/53/B/ST2/00975 and 2025/58/E/ST2/00051. V.B. is partly supported by the Beijing Natural Science Foundation of China (BNSF) under Contract No.~IS23014. G.E. acknowledges support from the Austrian Science Fund FWF under grant number 10.55776/PAT2089624. We thank the conveners and members of the LHCb Charm, Semileptonic, and Rare Decays working groups for discussions and helpful feedback on the experimental status and prospects presented in this white paper.

\printbibliography[category={reviews},title={Reviews}]

\printbibliography[notcategory={reviews},title={References}]


\end{document}

%% file: experiment-status.tex
\section{Experiments}
Measurements of semileptonic hyperon decays were performed primarily during the second half of the twentieth century, often with limited event yields. Owing to their relatively small branching fractions and the absence of direct information about the neutrino in the final state, these decays are experimentally challenging, particularly in high-background environments. The most precise measurement of $\Lambda$ semileptonic decays dates back more than three decades \cite{Dworkin:1990dd}, while the most recent results were obtained by the KTeV and NA48 collaborations, which measured the form factors and branching fraction of the $\Xi^0 \to \Sigma^+ e \nu$ transition \cite{KTeV:2001djr, NA481:2012dtx}.

In recent years, both strange- and charm-baryon semileptonic decays have attracted renewed interest at $e^+e^-$ colliders and hadron experiments. This resurgence has been driven by large data samples collected by BESIII, LHCb, and Belle~II, together with the development of novel reconstruction strategies and analysis techniques. Looking ahead, FAIR may also provide opportunities to study semileptonic decays as well as electromagnetic transitions, including Dalitz decays.
\subsection{BESIII and STCF}

The Beijing-based BESIII detector~\cite{BESIII:2009fln} collects data from nearly symmetric collisions of electron and positron beams provided by the BEPCII accelerator~\cite{BES:1994bjo,BES:2001vqx}. Charged tracks are identified and measured with high efficiency in the mini drift chamber, and their flight times are recorded by the time-of-flight detector. Neutral particles, including photons and (anti)neutrons, are reconstructed by the electromagnetic calorimeter. Large samples at  the $J/\psi$ and $\psi(2S)$ resonances are used for studying hyperon decays, while data with $\sqrt{s}$ above the $\Lambda_c^{+}\bar{\Lambda}_c^{-}$ production threshold is used for charm baryon decays. The hyperon semileptonic decays have been made possible by the $10^{10}$ $J/\psi$ data set collected between 2009 and 2018~\cite{BESIII:2021cxx}. The BESIII physics program post 2024~\cite{BESIII:2020nme} focuses on $\sqrt{s}$ above the charm baryon-antibaryon threshold. The upgraded BEPCII-U~\cite{Lyu:2021biq} achieved a threefold increase in peak luminosity, reaching $1.06 \times 10^{33}$ cm$^{-2}$s$^{-1}$ at $e^{\pm}$ at 4.7 GeV, which will provide large data samples of charmed baryons. 
For all these studies, one can exploit the fact that baryon--antibaryon pairs are produced by using the double-tag technique~\cite{MARK-III:1985hbd}. Here, one reconstructs one of the baryons in its dominant non-leptonic decay modes (for charm baryons, often several modes are used). This constitutes the single-tag event sample. From this sample, one then searches for the decay mode of interest. There are three advantages to using this technique. First, it allows for the determination of the absolute branching fractions $\Gamma_i/\Gamma_\text{total}$ by determining only the yields and efficiencies of the single- and double-tag samples. Second, the momentum and polarization of the baryon decaying via the semileptonic mode are known; therefore, it is possible to determine the neutrino momentum. Third, due to its event topology -- two baryons reconstructed back-to-back—it is possible to suppress background and determine the branching fraction of experimentally challenging modes, e.g., decays like $\Lambda_c^+\to n l^+\nu_l$, which contain both a neutron and a neutrino in their final state~\cite{BESIII:2024mgg}. 

Recently, a nonlinear focusing of colliding bunches, called the Crab Waist (CW) 
collision scheme~\cite{Zobov:2016sxm, Raimondi:2007vi}, was proposed for the next BEPCII upgrade (BEPCII-CW). It holds the promise of increasing the luminosity of the storage-ring colliders by 1-2 orders of magnitude, without any significant increase in beam current and without a reduction in bunch length. The design luminosity goal of the BEPCII upgrade project is $10^{34}$~cm$^{-2}$s$^{-1}$ at energies less than $3.77$~GeV, i.e., a factor of 10 higher than the record peak  luminosity of $10^{33}$~cm$^{-2}$s$^{-1}$ achieved so far. 

Looking beyond the BESIII physics program, the increase in yields is expected to come from the next-generation Super Tau-Charm factory (STCF) that will operate within a CMS energy range of 2.0 to 7.0~GeV~\cite{Ai:2025xop}.The full physics program of BESIII can be carried out with a significantly expanded reach. According to the STCF conceptual design report \cite{Achasov:2023gey}, it is projected to collect $3.4\times10^{12}$ $J/\psi$ events, which would amount to a production of $10^{9}$ $\Lambda\bar{\Lambda}$ pairs. This will enable precise measurements of branching fractions and form factors of semileptonic hyperon decays.  The design luminosity is $0.5\cdot10^{35}$~cm$^{-2}$s$^{-1}$ or higher at the optimal CMS energy of 4.0~GeV. With this luminosity, it will be possible to collect 1~ab$^{-1}$ of data per year. Relevant for hyperon-antihyperon production are the expected $J/\psi$ yields, which correspond to $3.4\cdot 10^{12}$ events per year, given a beam energy spread equal to that of BEPCII. This opens up different high precision studies in the decays of hyperon-antihyperon pairs. Projections of the precision can be obtained by dividing the statistical uncertainties achieved with BESIII by at least a factor of ten, due to two orders of magnitude larger data sets expected with the STCF. The results are presented in Table~\ref{tab:FF_SLHD}.

\begin{table}[h!]
  \caption{Illustration of the present and the expected statistical uncertainty for the branching fraction and axial-vector coupling $g_{av}$ for $\Lambda$ and $\Xi$ SLDs at BESIII for current and proposed BEPCII upgrade, and the proposed SCTF electron--positron collider. The results of the published BESIII measurements are given in the first row~\cite{BESIII:2026ehc,BESIII:2025wov}. The uncertainties given in the second and third rows are straightforward re-scaling based on the expected number of events. \label{tab:FF_SLHD}
     }
\begin{tabular}{lllllll}
\hline
    &$\sigma(\mathcal{B}^{\Lambda})$&$\sigma(g_{av}^{\Lambda})$&$\sigma(\mathcal{B}^{\Xi^-})$&$\sigma(g_{av}^{\Xi^-})$&&Comment\\
    \hline
BESIII&$2.2\times10^{-5}$&$4.8\times10^{-2}$&$4.0\times10^{-5}$&$7.0\times10^{-2}$&&$1.0\times10^{10}$ $J/\psi$\\
BESIII&$1.1\times10^{-5}$&$2.4\times10^{-2}$&$2.0\times10^{-5}$&$3.5\times10^{-2}$&&$4.0\times10^{10}$ $J/\psi$ (BEPCII-CW projection)\\
STCF&$1.2\times10^{-6}$&$2.6\times10^{-3}$&$2.2\times10^{-6}$&$3.8\times10^{-3}$&&$3.4\times10^{12}$ $J/\psi$ (projection)\\
\hline
  \end{tabular}
\end{table}

\subsection{LHCb}
The Large Hadron Collider beauty (LHCb) experiment~\cite{LHCb:2008vvz, LHCb:2014set} is a single-arm forward spectrometer at the CERN Large Hadron Collider, originally optimized for precision studies of beauty- and charm-hadron decays. Its forward geometry, excellent vertexing, tracking, and particle-identification capabilities also make it a powerful experiment for long-lived strange baryons. It is a premier facility to reconstruct hadrons decaying into charged final states, although final states with neutral particles remain experimentally challenging. High collision energies give access to the complete spectrum of hadron flavors, while high luminosity yields datasets of unmatched size. Initially, the LHCb detector collected data during Runs 1 and 2, from 2010 to 2018. After the first major detector upgrade (Upgrade~I), LHCb resumed data taking in Run~3 with a fully software-based trigger, which ended in July~2026. According to the current CERN schedule, the long shutdown will end around 2030, with Run~4 expected to begin in the high-luminosity LHC, HL-LHC, era. A second detector upgrade, Upgrade~II, is planned for the fourth longer shutdown, enabling LHCb to exploit the high-luminosity phase of the LHC~\cite{LHCbcollaboration:2903094}. 

The high boost of particles produced in the forward region, combined with the precision vertex detector, allows LHCb to separate primary production vertices from secondary and tertiary decay vertices. This is essential for reconstructing weakly decaying beauty, charm, and strange hadrons. A key quantity is the track impact parameter, defined as the minimum distance between a reconstructed track and its primary proton--proton collision vertex. In Run~1/2 performance studies, the impact-parameter resolution was approximately $(15+29/p_T)\,\mu\mathrm{m}$, with $p_T$ expressed in $\mathrm{GeV}/c$~\cite{LHCb:2024wvl}. For hyperon physics, this displaced-vertex capability is especially important because particles such as $\Lambda$, $\Xi^-$, and $\Omega^-$ have measurable flight distances before decay, enabling strong suppression of combinatorial background. Although both beauty and charm baryons decay inside the vertex detector, long-lived hyperons can decay outside its geometrical acceptance. These events are still reconstructable, but with a slightly worse resolution. Experimentally, hyperons benefit from the large strange-baryon production rate at LHCb. Since the strange particles like $\Omega^-$, $\Xi^-$, and $\Lambda$ are long-lived, they decay into charged particles relatively far from the collision vertex. This means that combinatorial background from spurious charged tracks can be suppressed to negligible levels. Furthermore, both $\Omega^-$ and $\Xi^-$ are charged particles; hence, they also leave footprints in the vertex locator, which can be used to improve the track resolution, enabling a more precise momentum determination and helping to distinguish whether the hyperon was produced promptly or as a descendant of a charm or beauty hadron~\cite{LHCb:2024wvl}. This allows for improved resolution and a means to remove background from combinatorial muons.
In that case, events can still be reconstructed, but with a slightly worse resolution. 

Although LHCb was optimized for beauty and charm physics, it can also perform high-precision measurements of strange hyperons. 
In 2018, the report \textit{Prospects for measurements with strange hadrons at LHCb}~\cite{AlvesJunior:2018ldo} was published. In that work, acceptance efficiencies and invariant-mass resolutions for several strange-hadron decays were computed, and production rates in the LHCb acceptance relative to $K_S^0$ were also evaluated. A method based on fitting in the plane of reconstructed mass versus estimated missing transverse momentum with respect to the baryon
flight direction was established as promising for extracting semileptonic hyperon decay yields. This cleanly separates $\Lambda\!\to p\mu^-\bar{\nu}_\mu$ from the peaking $\Lambda\!\to p\pi^-$ background,
and likewise $\Xi^-\!\to\Lambda\mu^-\bar{\nu}_\mu$ from $\Xi^-\!\to\Lambda\pi^-$. The prospects study demonstrates this
separation explicitly with LHCb Upgrade-like kinematics, validating the use of partially reconstructed observables to control these dominant backgrounds. 
\subsection{Belle and Belle~II }
The Belle experiment~\cite{Belle:2012iwr} at the KEKB~collider~\cite{Abe:2013kxa} located at KEK in Tsukuba, Japan, recorded asymmetric $e^+e^-$~collisions between the years 1999 and 2010, equivalent to an integrated luminosity of about 1~ab$^{-1}$. Most of these data (711~fb$^{-1}$) were recorded at $\sqrt{s}=10.58$~GeV, which corresponds to the energy of the $\Upsilon(4S)$~resonance and yields the production of a $B\bar B$~meson pair at threshold. However, smaller datasets were recorded below and above this energy. From 2011 onward, both the machine and the detector underwent major upgrades to allow for a higher instantaneous luminosity and ultimately to aim for an integrated luminosity of 50~ab$^{-1}$. The new facility, the Belle~II experiment~\cite{Abe:2010gxa} at SuperKEKB~\cite{Akai:2018mbz}, started operating in 2018.

The Belle~II detector records the collisions of 4~GeV positrons and 7~GeV electrons. It consists of nested detector subsystems arranged around the beam pipe in a cylindrical geometry. The innermost subsystem is the vertex detector, which includes two layers of silicon pixel detectors and four outer layers of silicon strip detectors. Most of the tracking volume consists of a helium and ethane-based small-cell drift chamber (CDC). Outside the drift chamber, a Cherenkov-light imaging and time-of-propagation detector (TOP) provides charged-particle identification in the barrel region. In the forward endcap, this function is provided by a proximity-focusing, ring-imaging Cherenkov detector with an aerogel radiator (ARICH). Further out is the electromagnetic
calorimeter (ECL), consisting of a barrel and two endcap sections made of CsI(Tl) crystals. A uniform 1.5~T magnetic field is provided by a superconducting solenoid situated outside the calorimeter. Multiple layers of scintillators and resistive-plate chambers, located between the magnetic flux-return iron plates, constitute the $K^0_L$ and muon identification system (KLM). As of May 2026, Belle II has recorded a data sample equivalent to 840~fb$^{-1}$, most of it at the center-of-mass energy of the $\Upsilon(4S)$~resonance.

Despite the experiment being optimized for the study of $B$~mesons, charmed baryons are copiously produced, and the clean $e^+e^-$~events are well suited for the study of final states with invisible particles. The weakly decaying single charm baryon states $\Lambda^+_c$, $\Xi^+_c$, $\Xi^0_c$, and $\Omega^0_c$ are produced either in $B$~meson decays or through charm fragmentation from the hadronic continuum $e^+e^-\to q\bar q$. The charm baryon signal in fragmentation is usually cleaner, while the production from $B$~decays allows for fully resolving the decay kinematics of semileptonic charm baryon decays. The latter is possible as the $B$~meson four-momentum can be inferred from the second $B$~meson in the event (a procedure called \emph{tagging}). The neutrino 4-momentum is thus merely the difference between $p_B$ and the 4-momentum sum of the visible decay products. The total production cross-section of $\Lambda^+_c$ at the $\Upsilon(4S)$ resonance, which includes both production mechanisms, has been measured to be $(189\pm 93)$~pb~\cite{Belle:2005mtx}. In the Belle and Belle~II datasets currently available on tape, about 283 million $\Lambda^+_c$ baryons have been produced. The number of $\Xi^0_c$ and $\Xi^+_c$ states is expected to be lower by an order of magnitude. While these numbers are small compared to LHCb, it should be noted that the Belle~II detector is nearly hermetic and that $\Upsilon(4S)$~events contain only about 10 charged and 10 neutral particles. This allows for excellent reconstruction of final states containing neutral or missing particles and provides a complementary reach to the LHCb experiment.

Another major upgrade of Belle~II is planned for the year 2032~\cite{Aihara:2024zds}, which also includes an improvement to the injector system. This will allow raising the center-of-mass energy above the $\Lambda_b$~pair threshold (11.24~GeV/c$^2$), but in the current data, the study of $b$-baryon transitions is out of scope.
\subsection{HADES and CBM at FAIR}
The High Acceptance DiElectron Spectrometer (HADES) experiment ~\cite{HADES:2009aat} has been in operation at SIS18 at GSI/FAIR for more than 20 years. Proton beams with a kinetic energy of up to 4.5 GeV and secondary pion beams with momenta up to 2.5 GeV/c are used to study hadron production and decays on proton and nuclear targets. One of the highlights of the experiment are the first measurements of Dalitz decays for non-strange $N^*/\Delta$ baryons \cite{HADES:2017jkt, AbouYassine:2022hmt, HADES:2023jaj}. Complementarily, the Compressed Baryonic Matter (CBM) experiment, currently under construction at FAIR~\cite{Senger:2018zax}, can extend these studies to the hyperon sector and combine them with measurements of SLD decays ~\cite{Messchendorp:2025men}. This will become possible thanks to the higher proton beam energies from the new SIS100 machine, ranging up to 30 GeV   ($\sqrt{s}=7.6$ GeV), higher beam intensities ($10^{11} p/s$), along with data taking combined with on-line event selection, allowing for measurements with a 10 MHz interaction rate \cite{Collaboration:2026jdg}.

HADES combines a large angular acceptance (almost complete in azimuthal angle and covering polar angles from $18^\circ$ to $80^\circ$) with excellent proton,kaon, pion, and lepton (electron) identification. The latter is assured by a RICH detector and electromagnetic shower detection in a lead glass calorimeter, providing an electron/pion suppression factor exceeding $10^4$, which is sufficient for complete hadron rejection and efficient di-electron reconstruction with invariant masses up to $1.5$ GeV/c$^2$. A low mass tracking system based on drift chambers combined with toroidal magnetic fields assures a mass resolution of about $8$ MeV/c$^2$ in the low mass vector meson region. 
Using proton beams at maximal kinetic energies  $E_{\rm kin}=4.5$~GeV, HADES also targets the first measurements of hyperon Dalitz-decays $\Sigma^0 \rightarrow \Lambda e^+e^-$, 
$\Sigma(1385)^0 \to \Lambda e^+e^-$, and $\Lambda(1520) \to \Lambda e^+e^-$~\cite{HADES:2020pcx}, which are experimentally unexplored processes. Such transitions probe the electromagnetic transition form factors of strange baryons in the timelike region. Present studies show that hyperon decays can be reconstructed inclusively, tagged by the $\Lambda \to p\pi^-$ weak decay, with total efficiencies of about $0.5\%$ \cite{Rieger:2026cxs}. Clear signals of the hyperon decays can be expected above the combinatorial and $\Delta$~Dalitz-decay backgrounds for dilepton masses above the $\pi^0$ mass, but with limited statistics allowing for the estimation of the branching ratios . In addition, the radiative decays $\Sigma(1385)^0 \to \Lambda\gamma$ and $\Lambda(1520) \to \Lambda\gamma$ can be measured through exclusive reconstruction using the calorimeter~\cite{HADES:2020pcx}. 
Measurements of the production of ground $\Sigma^0/\Lambda$ \cite{HADES:2023dti,Munzer:2017hbl} and the excited $\Lambda(1405)$ \cite{HADES:2012csk,Wladyszewska:2026dwq}, $\Sigma(1385)$,$\Lambda(1520)$ \cite{Sumara:2024stj} hyperon states, as well as dielectron invariant mass distributions in elementary and nuclear collisions~\cite{HADES:2019auv}, have provided both the physics motivation and the experimental know-how on which the CBM program related to Dalitz and SLD decays is built.

CBM \cite{Collaboration:2026jdg} will extend these capabilities to significantly higher beam energies and, crucially, to interaction rates of about 10 MHz, which will result in a number of reconstructed hyperon decays that are orders of magnitude beyond what is achievable with HADES. CBM provides excellent particle identification, allowing for electron-$\pi/K/p$ separation by combining information from a RICH, a Transition Radiation Detector, and a TOF system based on multi-gap resistive plate chambers (MRPCs) counters. Alternatively to the electron channel, muon measurements are possible with a dedicated setup composed of five hadron absorbers interleaved with four tracking stations. Charged hadron identification is achieved in MRPCs featuring a time resolution of about $80$ ps which provides efficient $K/\pi/p$ separation up to a few $GeV/c$ at flight paths of about $10 m$.
A precise vertex reconstruction is assured by the Micro-Vertex Detector (MVD), based on large-area monolithic pixel sensors organized in four planar layers, with a spatial resolution of $5 \mu$m and a vertex resolution of about $100 \mu$m measured along the beam axis. A small material budget per station of 0.3–0.5$\%X0$ only minimizes multiple scattering and photon conversion. The proximity of the MVD to the target (the first station is located 8 cm downstream of the target) region enables the reconstruction of secondary vertices of charm and hyperon weak decays, as well as low-momentum track fragments and short track segments resulting from photon conversion, which contributes to combinatorial background.  Eight planar Silicon Tracking Stations (STS), built from highly segmented double-sided silicon micro-strip detectors with a spatial resolution of $15 \mu m$ and a time resolution of $5ns$ and efficiencies above $95\%$, comprise a tracking system. STS is located in the large gap of a superconducting dipole magnet with a vertical acceptance of $\pm 25^{\circ}$ and a horizontal acceptance of $\pm 30^{\circ}$, providing a 1~Tm magnetic field integral  along straight trajectories within $\pm0.5$m around the magnet center. Momentum resolution in track reconstructions amounts to  $\Delta p/p\simeq 1.5\%$ . Simulation studies performed with Au+Au collisions demonstrate a clean identification of hyperon states like $\Lambda,\Xi,\Omega$ in a high multiplicity environment of heavy ion events. 

In proton-proton collisions, the multiplicity of tracks is lower, and running with even higher interaction rates could be possible. However, even assuming 10 MHz interaction rates and hyperon production cross sections of the order of tens of microbarns, production rates of hyperons (i.e., $\Lambda,\Sigma$) reach several times $10^{8}$/s,
exceeding by orders of magnitude those available in any previous fixed-target experiment devoted to semileptonic hyperon decays. Feasibility studies of the reconstruction of semileptonic and Dalitz decays in proton-proton interactions in CBM have been started and are focused on various exclusive channels. First results on Dalitz decays in the dielectron channel of hyperons are presented in~\cite{Messchendorp:2025men} and demonstrate excellent performance with dilepton mass resolution $\sigma \simeq 4$ MeV, which is sufficient to map the $q^2$ dependence of the electromagnetic
transition form factors over the accessible kinematic range. For the measurements of semileptonic decays, a strategy based on the complete reconstruction of exclusive channels, where all particles except the neutrino are reconstructed in a given event, is envisaged.  Kinematical constraints resulting from the exclusive reconstruction, combined with very good electron/pion separation provided by RICH and TRD ($\sim 10^{4}$), are sufficient for the reconstruction of electron decays. The separation of secondary vertices of the weak decays of hyperons from prompt background channels is provided by a very good vertex resolution of STS. Data acquisition based on continuous trigger-less sampling with online data processing permits the implementation of various trigger types based on the topology of decays (as, for example, hyperon weak decays).